\documentclass[fleqn,usenatbib]{mnras}
\usepackage[normalem]{ulem}
\usepackage[T1]{fontenc}
\usepackage{ae,aecompl}
\usepackage{gensymb}
\usepackage[spanish,es-minimal,english]{babel}
\usepackage[utf8]{inputenc}
\usepackage{graphicx}	
\usepackage{amsmath}	
\usepackage{amssymb}	
\usepackage{xcolor}
\usepackage{hyperref}
\hypersetup{colorlinks=True, linkcolor=blue!50!., citecolor=.,
  urlcolor=blue!50!.}
\usepackage{hyphenat}
\usepackage{capt-of} 
\usepackage{xfrac} 
\usepackage{etoolbox}
\robustify\bfseries
\robustify\itshape
\makeatletter
\let\orig@footnote\footnote
\renewcommand{\footnote}{%
  \@ifnextchar[
    \my@footnote@opt
    \my@footnote@noopt
}
\newcommand{\my@footnote@opt}[2][]{%
  \orig@footnote[#1]{\color{.}#2}%
}
\newcommand{\my@footnote@noopt}[1]{%
  \orig@footnote{\color{.}#1}%
}
\makeatother
\usepackage{newtxtext}
\usepackage[varg,varvw,smallerops]{newtxmath}
\usepackage{booktabs}
\usepackage{enumerate}
\usepackage{array}   
\newcolumntype{L}{>{$}l<{$}} 
\newcolumntype{R}{>{$}r<{$}} 
\newcolumntype{C}{>{$}c<{$}}
\usepackage{ragged2e}
\newcolumntype{P}[1]{>{\RaggedRight\arraybackslash}p{#1}}

\usepackage[english]{isodate}
\usepackage{siunitx}
\DeclareSIUnit\msun{\text{M\ensuremath{_\odot}}}
\DeclareSIUnit\lsun{\text{L\ensuremath{_\odot}}}
\DeclareSIUnit\rsun{\text{R\ensuremath{_\odot}}}
\DeclareSIUnit\angstrom{\text {E}}
\DeclareSIUnit\kms{\kilo\meter\per\second}
\DeclareSIUnit\pcc{\per\centi\meter\cubed}
\DeclareSIUnit\erg{\text{erg}}
\DeclareSIUnit\cgsflux{\erg\per\second\per\centi\meter\squared}

\ProvideDocumentCommand\unit{om}{\si[#1]{#2}}
\ProvideDocumentCommand\qty{omm}{\SI[#1]{#2}{#3}}
\ProvideDocumentCommand\qtylist{omm}{\SIlist[#1]{#2}{#3}}
\ProvideDocumentCommand\qtyrange{ommm}{\SIrange[#1]{#2}{#3}{#4}}

\definecolor{NEWcolor}{rgb}{0.7,0.1,0.4}
\newcommand\startNEW{\color{black}}
\newcommand\stopNEW{\color{black}}
\newcommand\NEW[1]{\startNEW #1\stopNEW\relax}

\usepackage{newtxtext,newtxmath}



\def\apj{ApJ}
\def\aj{AJ}
\def\mnras{MNRAS}
\def\pasp{PASP}
\def\aap{A\&A}
\def\apjs{ApJS}

\newcommand*\chem[1]{\ensuremath{\mathrm{#1}}}
\newcommand\Wav[1]{\ensuremath{\lambda #1}}
\newcommand\Wavv[1]{\ensuremath{\lambda\lambda \mbox{#1}}}
\newcommand{\Sub}[1]{_\mathrm{#1}}

\newcommand{\nii}{[\ion{N}{II}]}
\newcommand{\sii}{[\ion{S}{II}]}
\newcommand{\oiii}{[\ion{O}{III}]}
\newcommand{\oii}{\ion{O}{II}}
\newcommand{\heii}{\ion{He}{II}}

\newcommand\hi{\ion{H}{I}}
\newcommand\hii{\ion{H}{II}}
\newcommand\heiilam{\heii{}\ \Wav{4686}}
\newcommand\heiiredlam{\heii{}\ \Wav{6560}}
\newcommand\oiiilam{\oiii{}\ \Wav{5007}}
\newcommand\oiiialam{\oiii{}\ \Wav{4363}}
\newcommand\niilam{\nii{}\ \Wav{6584}}
\newcommand\niilamlam{\nii{}\ \Wavv{6548,84}}
\newcommand\Halam{\Ha{}\ \Wav{6563}}
\newcommand\Hblam{\Hb{}\ \Wav{4861}}
\newcommand\Ha{\ensuremath{\mathrm{H}\alpha}}
\newcommand\Hb{\ensuremath{\mathrm{H}\beta}}

\newcommand{\vsys}{\ensuremath{V_\mathrm{sys}}}
\newcommand{\vexp}{\ensuremath{V_\mathrm{exp}}}

\newcommand{\Teff}{\ensuremath{T_\mathrm{eff}}}
\newcommand{\PA}{\ensuremath{\mathrm{PA}}}
\newcommand\SN{\ensuremath{\mathrm{S/N}}}
\newcommand\elec{\ensuremath{_\mathrm{e}}}
\newcommand\splitting{\ensuremath{_\mathrm{split}}}
\newcommand\blue{\ensuremath{_\mathrm{blue}}}
\newcommand\red{\ensuremath{_\mathrm{red}}}
\newcommand\email[1]{\href{mailto:#1}{\nolinkurl{#1}}}

\title[Planetary nebula Ou~5]{Kinematics and physical structure of Ou~5: A bipolar planetary nebula with extreme abundance discrepancy and an eclipsing binary core\thanks{%
Based on observations carried out at the Observatorio Astron\'omico
Nacional, Sierra San Pedro M\'artir (OAN-SPM), Baja
California, Mexico, operated by the Universidad Nacional Aut\'onoma de M\'exico.}}

\author[Henney et al.]{%
  William J. Henney,\textsuperscript{1}%
  \thanks{Email: \email{w.henney@irya.unam.mx} (WJH), \email{jal@astro.unam.mx} (JAL), \email{tere@astro.unam.mx} (MTGD).}%
  \thanks{These three authors contributed equally to this work.}
  J. A. L\'opez,\textsuperscript{2}\footnotemark[2]\footnotemark[3]
  Ma.~T. Garc\'ia-D\'iaz,\textsuperscript{2}\footnotemark[2]\footnotemark[3]
  M. G. Richer,\textsuperscript{2}
  D. H. Gonz\'alez-Buitrago,\textsuperscript{2}
  \newauthor
  S. V. Zharikov,\textsuperscript{2}
  S. Noroozi,\textsuperscript{3}
  F. Pozo Nu\~nez\textsuperscript{4}\\
  \textsuperscript{1}Instituto de Radioastronomía y Astrofísica,
  Universidad Nacional Autónoma de México,
  Antigua Carretera a Pátzcuaro 8701,
  58089 Morelia, Michoacán, Mexico\\
  \textsuperscript{2}Instituto de Astronom\'ia, Universidad Nacional
  Aut\'onoma de M\'exico, Sede Ensenada. Km 103 Carretera Tijuana-Ensenada,
  22860 Ensenada,\\ Baja California, M\'exico\\
  \textsuperscript{3}Pre-Degree Department, Constructor University, Campus Ring 1, 28759 Bremen, Germany\\
  \textsuperscript{4}Astroinformatics, Heidelberg Institute for Theoretical Studies, Schloss-Wolfsbrunnenweg 35, 69118 Heidelberg, Germany\\
}

\date{Accepted XXX. Received YYY; in original form ZZZ}

\pubyear{2025}

\begin{document}

\date{Submitted: \today}

\pagerange{\pageref{firstpage}--\pageref{lastpage}} \pubyear{2025}

\maketitle

\label{firstpage}


\begin{abstract}
We present new deep imaging and high-resolution spectroscopy of the
extreme-abundance-discrepancy planetary nebula Ou~5, together with
photoionization modelling aimed at probing its unusual thermal and
chemical structure. The nebula exhibits a nested bipolar morphology,
including inner and outer shells, faint outer lobes, and polar knots.
Remarkably, all these components share a dynamical age of order
\(\qty{1e4}{yr}\).
Thermal broadening of the \Ha{} line relative to heavier ions implies a
hydrogen temperature \qtyrange{3000}{6000}{\kelvin}, in contrast to the
\(\sim \qty{1e4}{\kelvin}\) derived from collisionally excited line
diagnostics. This provides independent support for the presence of at least two
distinct temperature/metallicity phases, as previously proposed to
explain extreme abundance discrepancies.
Photoionization models with sinusoidally varying metallicity successfully reproduce
the observed nebular spectrum and morphology.
A mixture of fluctuations with both extreme and moderate
metallicity contrasts is required to simultaneously
fit the \oii{} and the \oiii{} observations.
The nebular \heii{} emission demands a hotter and more
luminous central star than previously inferred, consistent with
a \(\approx \qty{0.58}{\msun}\) post-AGB progenitor evolving toward a
CO white dwarf. Ou~5 thus reinforces the link between close-binary
nuclei and extreme abundance discrepancies, and provides a valuable benchmark
for understanding how common-envelope ejections give rise
to the thermal and abundance inhomogeneities observed in planetary nebulae.
\end{abstract}


\section{Introduction}
\label{sec:introduction}

Planetary nebulae (PNe) are formed when low- to intermediate-mass stars
eject their envelopes during late stages of evolution.
In the canonical picture, a dense slow wind from the asymptotic giant branch
(AGB) is swept up by a subsequent fast wind, leading to an ionized shell
\citep{Kwok:1978a, Kahn:1985a}.
While this framework has been broadly successful, many observed features
remain challenging to explain, including the wide diversity of morphologies
and the prevalence of highly collimated outflows
\citep[for a review, see][]{Lopez:2022a}.
Mounting evidence suggests that binary interactions,
and in particular common-envelope (CE) evolution,
play a key role in producing these structures
\citep[e.g.][]{Nordhaus:2006, De-Marco:2009a, Jones:2017b}.

Ou~5 (IPHASXJ211420.0+434136) is a striking example of a PN with a
post-CE binary central star.
It was first identified in the IPHAS survey \citep{Drew:2005a}
and independently rediscovered by Nicolas Outters \citep{Outtters2012}.
The central binary, confirmed as an eclipsing system with a
\(\qty{8.74}{\hour}\) orbital period, was studied by
\citet{Corradi:2014a} and \citet{Jones:2022a}.
Ou~5’s main shell shows a complex bipolar morphology with truncated inner lobes,
faint outer extensions, and filamentary cusps,
making it an excellent candidate for probing the dynamical effects
of a CE ejection.

Spectroscopic studies have also revealed that Ou~5 has an
extreme abundance discrepancy factor (ADF),
with recombination lines indicating much higher abundances than
collisionally excited lines
\citep{Corradi:2015a,Wesson:2018a}.
Such high-ADF nebulae are frequently associated with post-CE binaries,
but the physical origin of the discrepancy remains uncertain
\citep[see review by][]{Garcia-Rojas:2019a}.
Ou~5 therefore provides an important test case for exploring both the
morphological imprint of CE evolution and the thermal/chemical
inhomogeneities implied by its spectrum.

In this paper we present new imaging and high-resolution spectroscopy
of Ou~5, together with photoionization models that explore its
thermal and chemical structure.
Section~\ref{sec:observations} describes the observations and \NEW{data reduction}.
In Section~\ref{sec:dist-determ} we estimate the distance.
Section~\ref{sec:nebular-morphology} analyzes the nebular morphology and kinematics,
while Section~\ref{sec:physical-parameters} derives physical parameters of the
nebula and its central star.
Section~\ref{sec:cloudy-phot-models} presents photoionization models with
spatially varying metallicity and compares them with the observations.
Section~\ref{sec:discussion} discusses the implications of these results,
and Section~\ref{sec:conclusions} summarizes our conclusions.

\section{Observations and data reduction}
\label{sec:observations}

The Observations were carried out over multiple observing runs, as summarized in Table~\ref{tab:observing-log}. The data reduction was performed  using standard procedures with IRAF\footnote{
IRAF is distributed by the National Optical Astronomy Observatories, which is operated by the Association of Universities for Research in Astronomy, Inc. under cooperative agreement with the National Science foundation.} \citep[][]{Tody1986, Tody1993, Fitzpatricketal2024}. The reduction process included bias subtraction, cosmic ray removal, and wavelength calibration of the two-dimensional spectra.

\subsection{Long-slit Spectroscopic Observations}
\label{sec:long-slit-spectr}
The planetary nebula Ou~5 was observed at the \emph{Observatorio Astronómico Nacional} in \emph{San Pedro Mártir}, Baja California, Mexico.
Observations were conducted using the Manchester Echelle Spectrometer \citep[MES-SPM:][]{Meaburn:2003} mounted on the \qty{2.1}{\meter} telescope in its f/7.5 configuration.
The MES-SPM, was operated in its long-slit echelle mode without a cross-disperser, employing narrow-band interference filters to isolate individual echelle orders.
Specifically, a \qty{90}{\angstrom} bandwidth filter  was used to isolate the 87th order, covering the \Halam{} and \niilamlam{} emission lines, while a \qty{50}{\angstrom} bandwidth filter covered the 114th order, corresponding to the  \oiiilam{} emission line.
The detector was a Marconi 2 CCD  featuring a \(2048 \times 2048\) pixels array, with each pixel measuring \qty{13.6}{\micron} on a side. A two-by-two binning was applied during the observations.
The standard slit width was \qty{150}{\micron}, resulting in a projected slit width
\qty{1.9}{arcsec}, a slit length \qty{5.47}{arcmin}
and a velocity resolution of \NEW{\qty{11.8}{\kms} at \Ha{}}.
Additionally, for \NEW{slit positions~b, c and d} a narrower \qty{70}{\micron} wide slit was used,
resulting in a projected slit width  \qty{0.95}{arcsec} and a velocity resolution of \qty{5.9}{\kms}
\NEW{at \Ha{}}. The pixel scale along the slit was \qty{0.312}{arcsec} per pixel.
The full-width half maximum of the seeing was  measured as \qty{1.6 \pm 0.1}{arcsec}.
Wavelength calibration was performed using Th-Ar arc lamp spectra, achieving a radial velocity accuracy of approximately \qty{\pm 1}{\kms}.
In this study, all presented spectra were corrected to heliocentric velocity. 
 
\begin{figure}
	\includegraphics[width=\columnwidth]{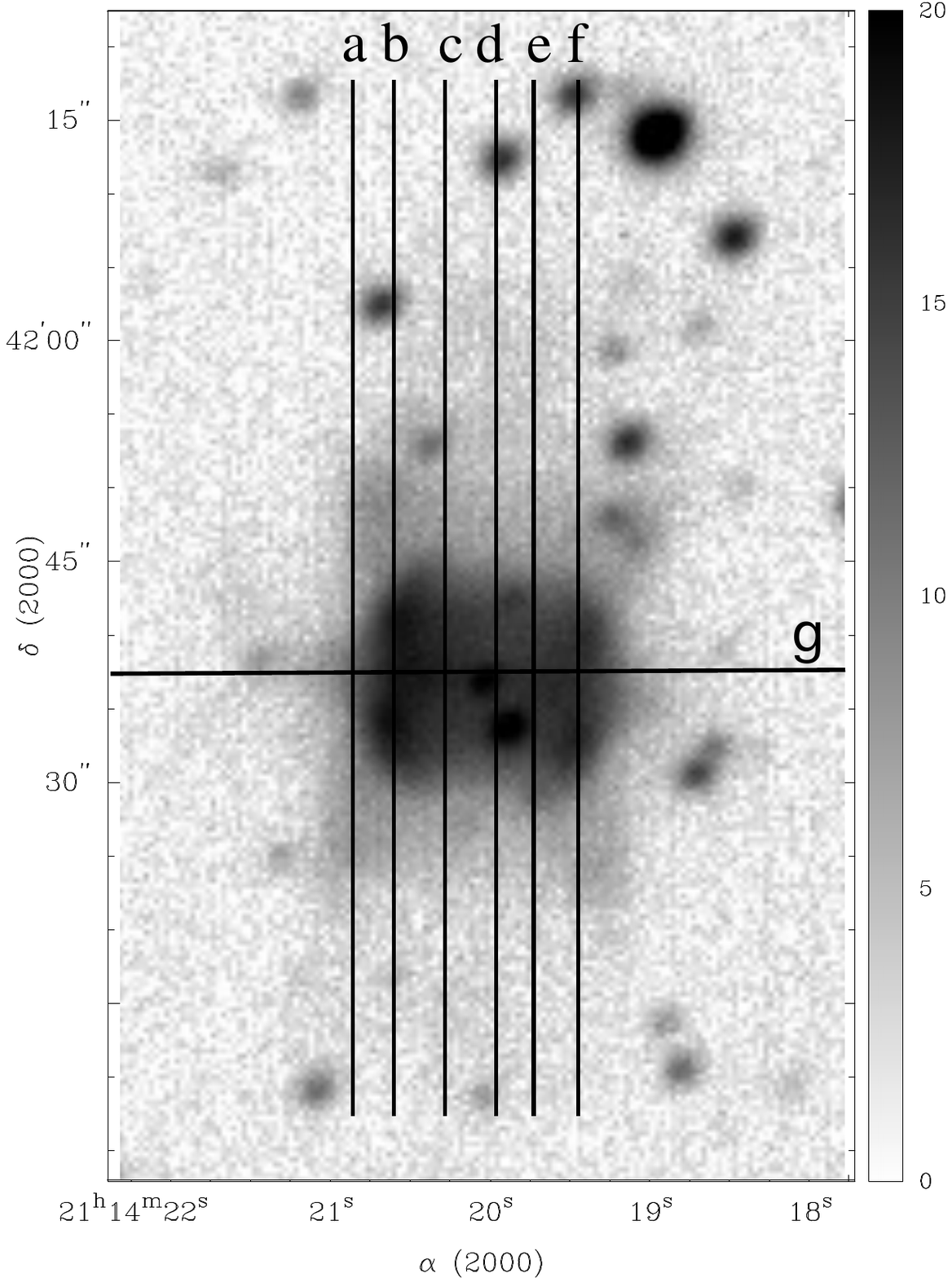}
    \caption{The 7 positions of the long-slits are indicated and labeled over an \oiii{} image of Ou~5}
    \label{fig:01}
\end{figure}

The spectroscopic observations covered seven distinct slit positions. Six consecutive North-South positions (a--f) were stepped approximately \qty{\sim 3}{arcsec} apart across the nebula \NEW{(position angle \(\PA = \ang{0}\))}. An additional east-west-oriented slit position \NEW{(slit g, \(\PA = \ang{90}\))} was placed centrally across the nebula (see Fig.~\ref{fig:01}). Each on-target spectrum was acquired with exposure times of \qty{1800}{\second}. 
Table~\ref{tab:observing-log} summarizes these spectroscopic observations.

\begin{table*}
\centering
\caption{Summary log of \NEW{OAN \qty{2.1}{\meter} MES-SPM long-slit echelle spectroscopic} observations of Ou~5.}
\label{tab:observing-log}
\isodate
\NEW{
\begin{tabular}{l ccccc} \toprule
  Dates & Exposure time & Filter (width) & Slit width & Velocity Resolution   & Slit Positions          \\
  20YY-MM-DD&    \unit{\second}   &  \unit{\angstrom}   & \unit{\micron} & \unit{\kms}   &  \\
  \midrule
  \daterange{2017-08-29}{2017-08-30} & 1800 & \Ha{} (90),  \oiii{} (60) & 70 & 6 & b, c, d\\
  \daterange{2018-05-02}{2018-05-04} & 1800 & \Ha{} (90), \oiii{} (60) & 150  &  12 & a, c, d, e\\
  \daterange{2019-05-04}{2019-05-18} & 1800 & \Ha{} (90) & 150  &  12 & b, d, f, g\\
  \bottomrule
\end{tabular}
}
\label{table:kn61}
\end{table*}

In addition, MES-SPM was used in its imaging configuration to obtain two deep \qty{1800}{\second} images of the Ou~5 field employing the  \Halam{} + \niilamlam{}  and \oiiilam{} filters.  Fig.~\ref{fig:01} shows the positions of the long slits projected onto the \oiii{} image obtained on August 28, 2017, which reveals the overall morphology of the nebula.


\begin{figure}
	\includegraphics
    [width=\linewidth] 
    {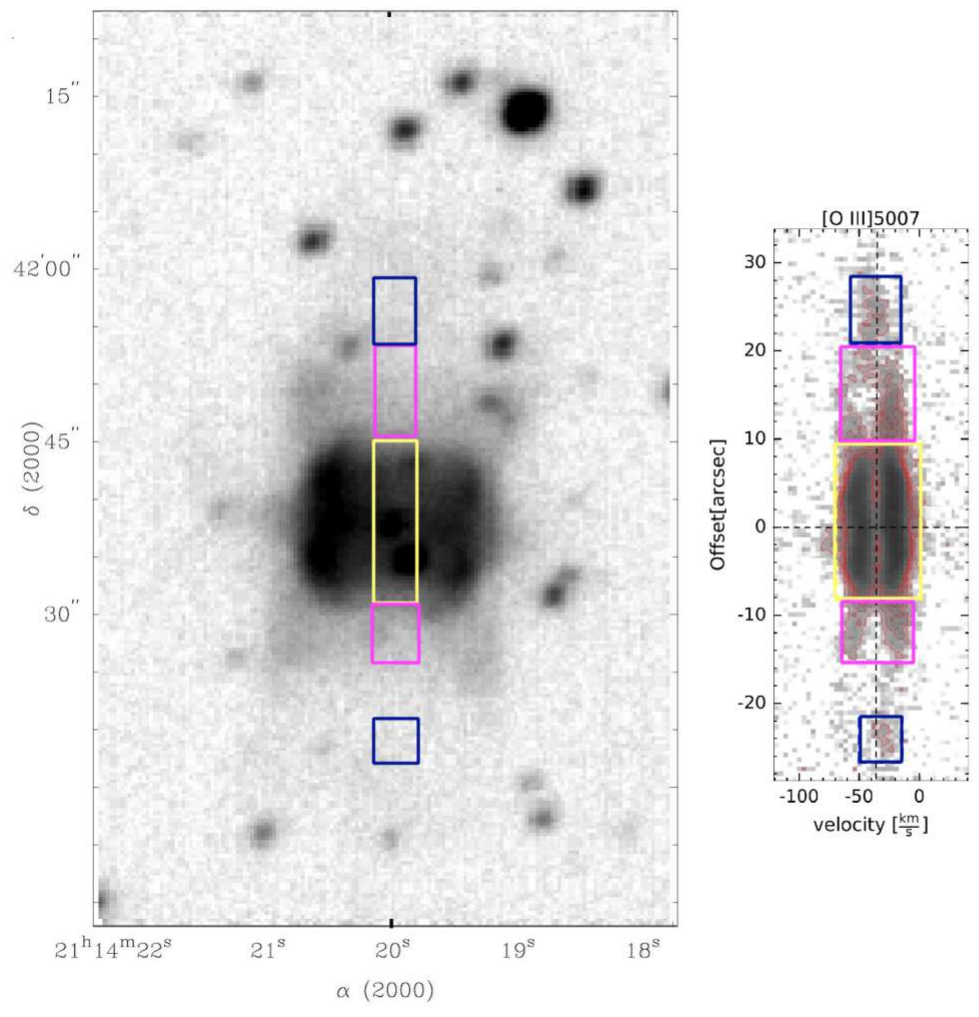}
    \caption{a. The central~d slit is projected over the \oiii{} image of Ou~5. Some regions of interest are indicated. The blue squares indicate the positions of the polar knots. The pink regions indicate the approximate length of the outer lobes and the central yellow region indicates the extent of the inner lobes. b. The P-V array in the right corresponds to the bi-dimensional \oiii~ spectrum from slit d.   
    }
    \label{fig:02}
\end{figure}

\subsection{Narrow-band imaging}
\label{sec:imaging-observations}

In addition to our longslit spectroscopy, we analyze archival
optical narrow-band imaging observations from the
Alhambra Faint Object Spectrograph and Camera (ALFOSC)
of the Nordic Optical Telescope (NOT)
in three filters.
Images in filters targeting \Halam{} and \oiiilam{} from
2013 November 7
were previously described in \citet{Corradi:2014a}, while
images targeting \heiilam{} from 2016 May 26 are presented here for the first time.
The central wavelengths (widths) are respectively
\qty{6577}{\angstrom} (\qty{180}{\angstrom}),
\qty{5007}{\angstrom} (\qty{30}{\angstrom}),
and \qty{4687}{\angstrom} (\qty{35}{\angstrom}).

\begin{figure*}
  \centering
  \includegraphics[width=0.97\linewidth]{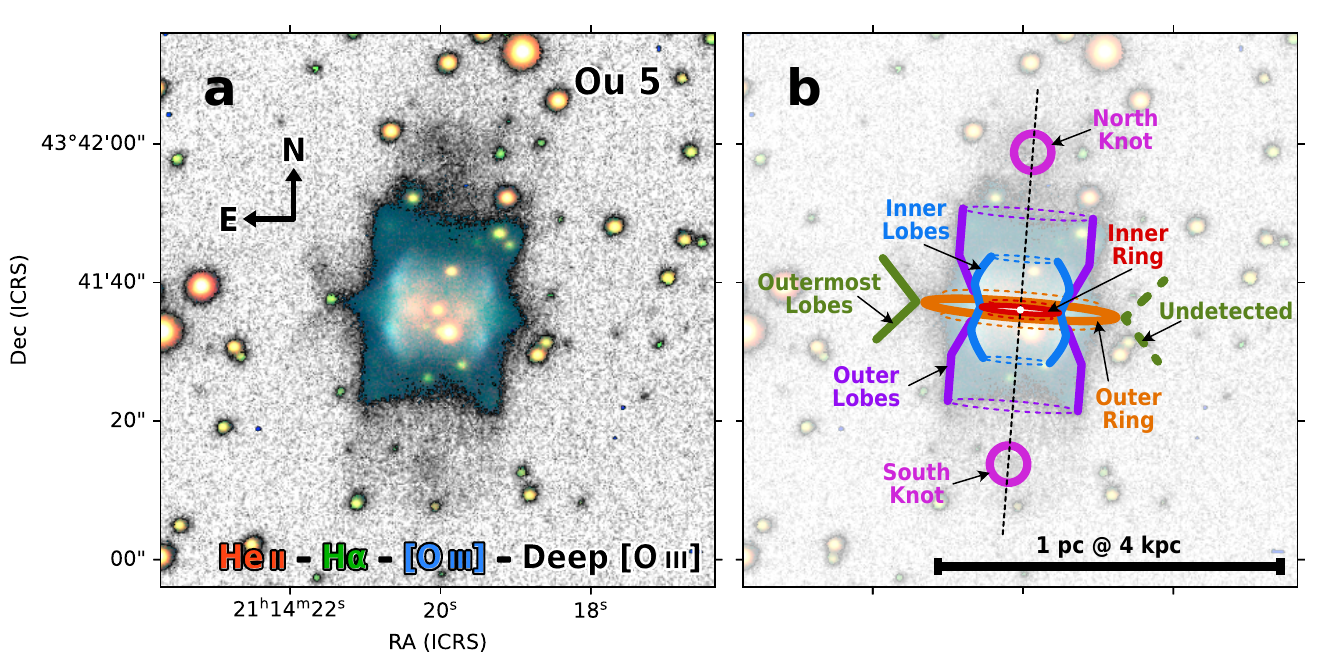}
  \caption{
    (a)~Three-color RGB image
    of Ou~5 from ALFOSC-NOT
    in filters of \heiilam{} (red), \Halam{} (green), and \oiiilam{} (blue),
    superimposed on a deep SPM + NOT co-added \oiii{} image
    (inverse grayscale).
    (b)~Schematic diagram of nebular components.
  }
  \label{fig:rgb-image}
\end{figure*}

All images were bias-subtracted, astrometrically aligned, and resampled
to a common pixel grid. 
Only one usable exposure (\qty{900}{\second}) was available for \oiii{},
but we combined two \qty{900}{\second} exposures for \Ha{}
and three \qty{1200}{\second} exposures for \heii{},
with outlier rejection to eliminate cosmic rays. 
Fig.~\ref{fig:rgb-image}a shows an RGB image of the bright core of the nebula that combines the
final images in the three filters,
with \heii{} in red,
\Ha{} in green,
and \oiii{} in blue.
The \Ha{} filter also admits the \nii{} nebular lines,
but their combined brightness is only 6\% of \Ha{}.
Contamination by continuum emission from stars is small for the \oiii{} filter
but is more significant for the wider \Ha{} filter
and for the \heii{} filter where the nebula is intrinsically fainter.

To obtain the deepest possible exposure of the faint
regions of the nebula, we co-add the NOT and SPM
\oiii{} observations.
The SPM image has worse seeing, but this is not important
for the more diffuse outer regions.
The result is shown as inverse grayscale in Fig.~\ref{fig:rgb-image}a
for the fainter parts of the image
below the 1\% \oiii{} brightness contour.
 
\section{Distance determination}
\label{sec:dist-determ}

Gaia parallax measurements for the
central star of Ou~5 have very large uncertainties,
which means that no usable direct parallax distance is available
\citep{Hernandez-Juarez:2024a}.
The most reliable distance can therefore be obtained
by statistical methods based on the mean trend of
surface brightness versus size \citetext{e.g., \citealp{Smith:2015a}}, which are anchored by reference samples of
nebulae with well-determined distances \citep{Frew:2016a}.
Based on earlier calibrations of the distance scale,
\citet{Corradi:2014a} found \(D = \qty{5.0 \pm 1.0}{kpc}\),
but we repeat the calculation here using the most recent calibrations,
which are based on Gaia DR3 parallaxes
\citep{Gaia-Collaboration:2016a, Gaia-Collaboration:2023a},
using the surface brightness of optical recombination lines
\citep{Bucciarelli:2023a}
or radio free-free emission
\citep{Ali:2022a}.

For the nominal angular radius of the nebula, \(R\),
we follow \citet{Frew:2016a} by taking the geometric mean
of the minor and major axes of the \Ha{} isophote contour at
10\% of the maximum brightness,
which yields \(R = \qty{11.4 \pm 1.0}{arcsec}\).
We combine this with the observed \Ha{} flux
and Balmer decrement from
\citet{Corradi:2014a}
to find the mean extinction-corrected \Ha{} surface brightness
as \(S(\Ha) = \qty{4.0(2.0)e-4}{erg.s^{-1}.cm^{-2}.sr^{-1}}\).
Taking the average of recent distance scale calibrations
\citep{Frew:2016a, Ali:2022a, Bucciarelli:2023a}
yields a physical radius \(R = \qty{0.22 \pm 0.04}{pc}\)
and a distance \(D = \qty{4.0 \pm 0.7}{kpc}\), which is the value that we adopt throughout this paper.
The uncertainty in the derived distance has roughly equal contributions from observational uncertainties and from
the dispersion of individual nebulae around the
mean radius--brightness trend.

\citet{Jones:2022a} found a smaller distance of \(D = \qty{3.1 \pm 0.3}{kpc}\)
from simultaneous fitting of the eclipsing binary light curve and radial velocity curve.
Note, however, that this is still marginally consistent with our adopted value.
\NEW{Ou~5 is a post-common-envelope system and there is some evidence \citep[Fig.~7.10]{Frew:2008a} that such nebulae are
  systematically offset from the mean radius--brightness trend, which would bring the two methods into closer agreement.
  However, preliminary analysis of a much larger sample of post-CE PNe shows no significant offset
  from the mean trend for single-star nebulae (Luna-Niño et~al., in preparation).}

\section{Nebular morphology and kinematics}
\label{sec:nebular-morphology}

\subsection{Morphological features}
\label{sec:morph-feat}

From the image presented in Fig.~\ref{fig:rgb-image}a
we can extend the analysis of \citet{Corradi:2014a, Corradi:2015a} to
arrive at a more complete description of the
different morphological features of the nebula,
as illustrated in Fig.~\ref{fig:rgb-image}b.
The bright inner nebula is revealed as having a composite structure,
with an inner equatorial ring in addition to the
inner bipolar lobes. The \heii{} emission is relatively brighter
in the ring,
while \oiii{} is brighter in the lobes,
with \Ha{} showing an intermediate behaviour.
The fainter outer lobes and outer equatorial ring combine
to produce a six-pointed star shape for the 1\% isophote of \oiii{}
(boundary between colour and greyscale image in Fig.~\ref{fig:rgb-image}a).
\heii{} is not detected from either the outer lobes or the outer ring.
The very faint ``fishtail'' structure on the \NEW{East} side \citep{Corradi:2014a}
is evidence for a third set of lobes, which we identify as ``outermost lobes''.
It is notable that no corresponding structure is seen to the \NEW{West},
but it is possible that it is too faint to detect,
given that the inner and outer lobes also show mild lateral asymmetries,
with the \NEW{East} side being the brighter.
The deep image also shows the morphology of the faint polar knots,
which we detect in our slit spectroscopy.
The South knot is relatively compact, whereas the North knot is much more diffuse.
There is an indication of a faint extension of the outer lobes around the location of the knots,
especially to the South,
but this is close to the noise limit. 

\begin{figure}
  (a)\\
  \includegraphics[width=\linewidth]{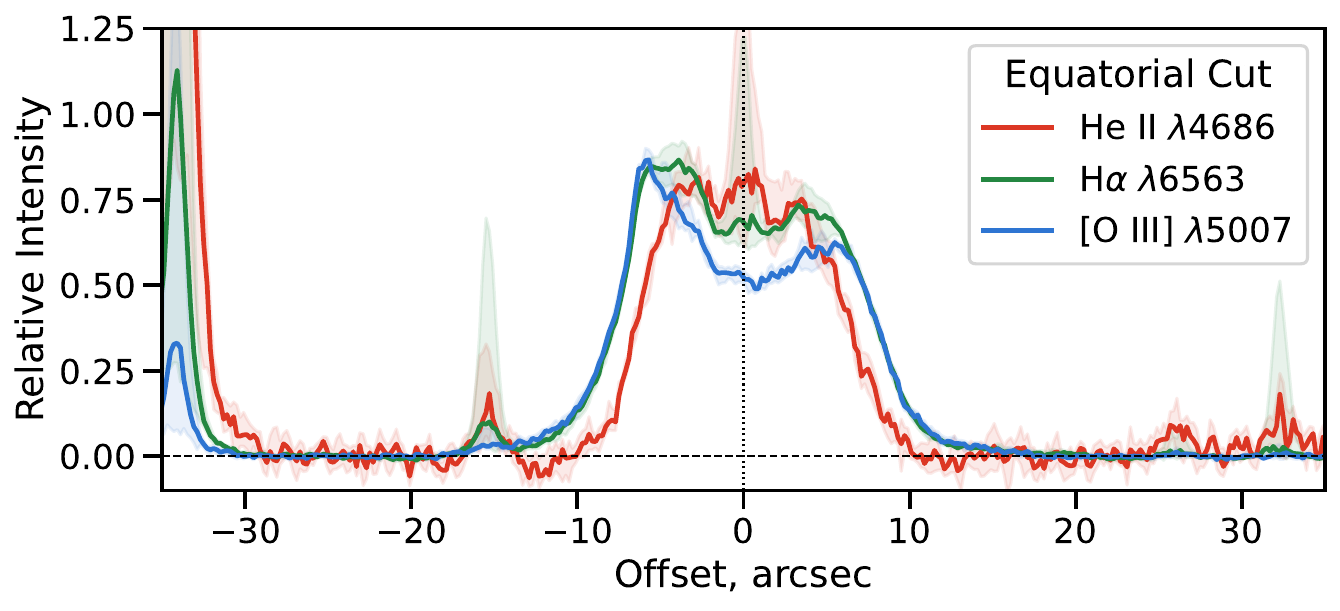}\\
  (b)\\
  \includegraphics[width=\linewidth]{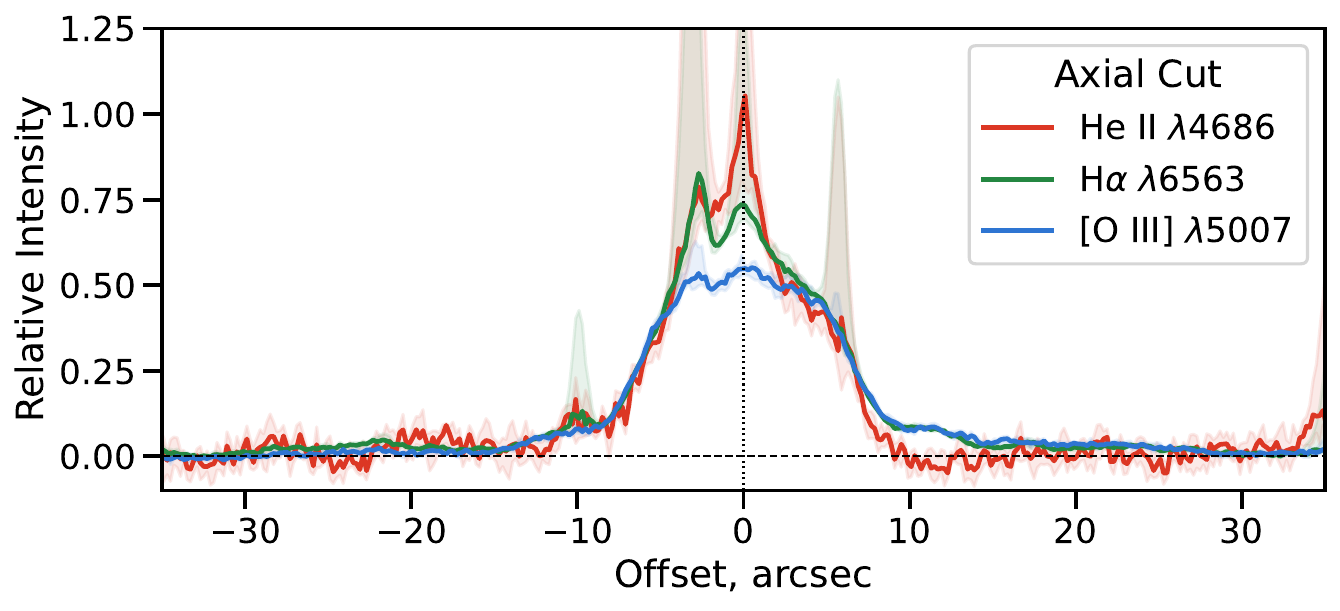}
  \caption{Profile cuts along the equator and polar axis of Ou~5
    from the ALFOSC-NOT images shown in Fig.~\ref{fig:rgb-image}.
    Colored lines show the median profile in each emission line filter
    for a strip of width \qty{4}{arcsec}, while gray shading shows the interquartile range (first to third quartile). Positions with a broad interquartile range correspond to stars that fall in the strip.
    (a)~Cuts along equator. The profile is affected by stars
    at offsets of \(-34\), \(-16\), \(0\), \(+27\), and \(+33\).
    (b)~Cuts along the nebular axis.
    The profile is affected by stars at offsets of  \(-10\), \(-4\), \(0\), and \(+6\).
  }
  \label{fig:image-cuts-equator}
  \label{fig:image-cuts-axis}
\end{figure}

\begin{figure}
  (a)\\
  \includegraphics[width=\linewidth]{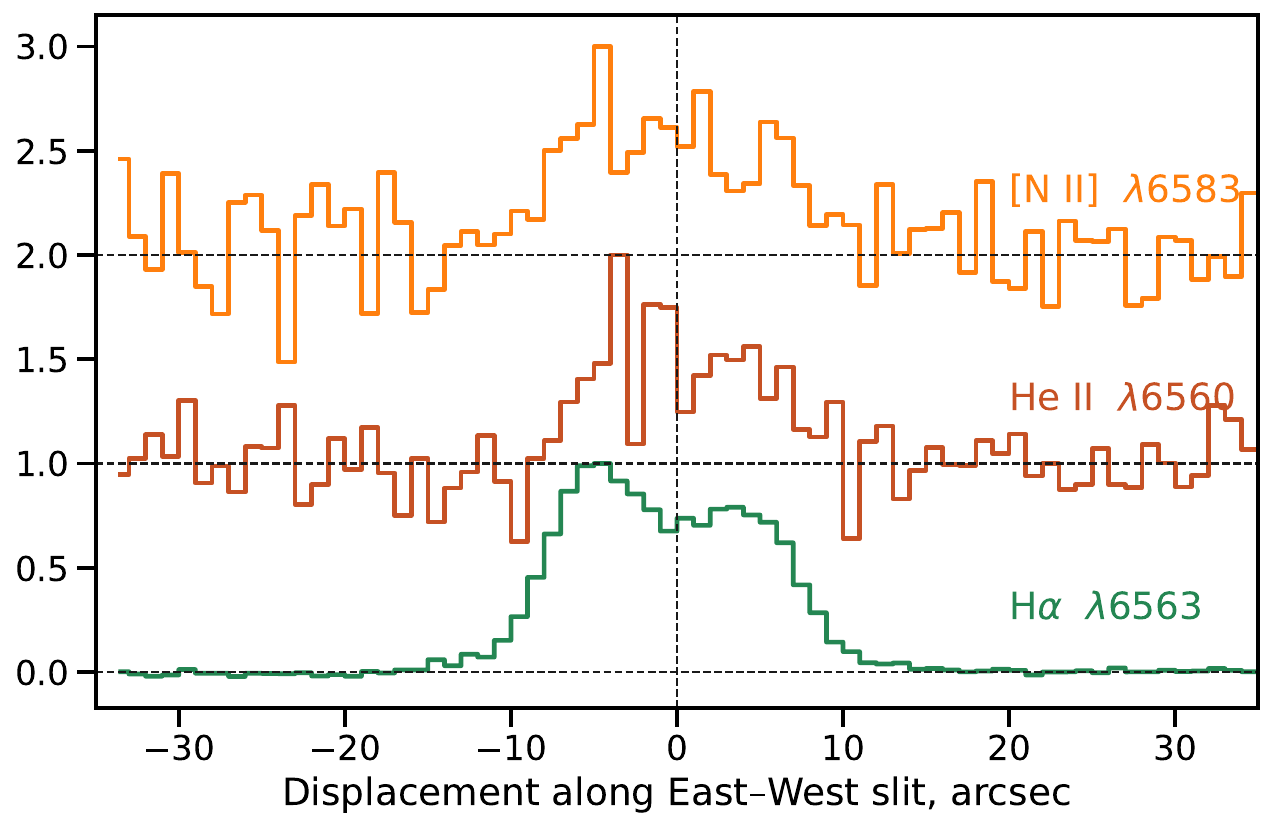}
  (b)\\
  \includegraphics[width=\linewidth]{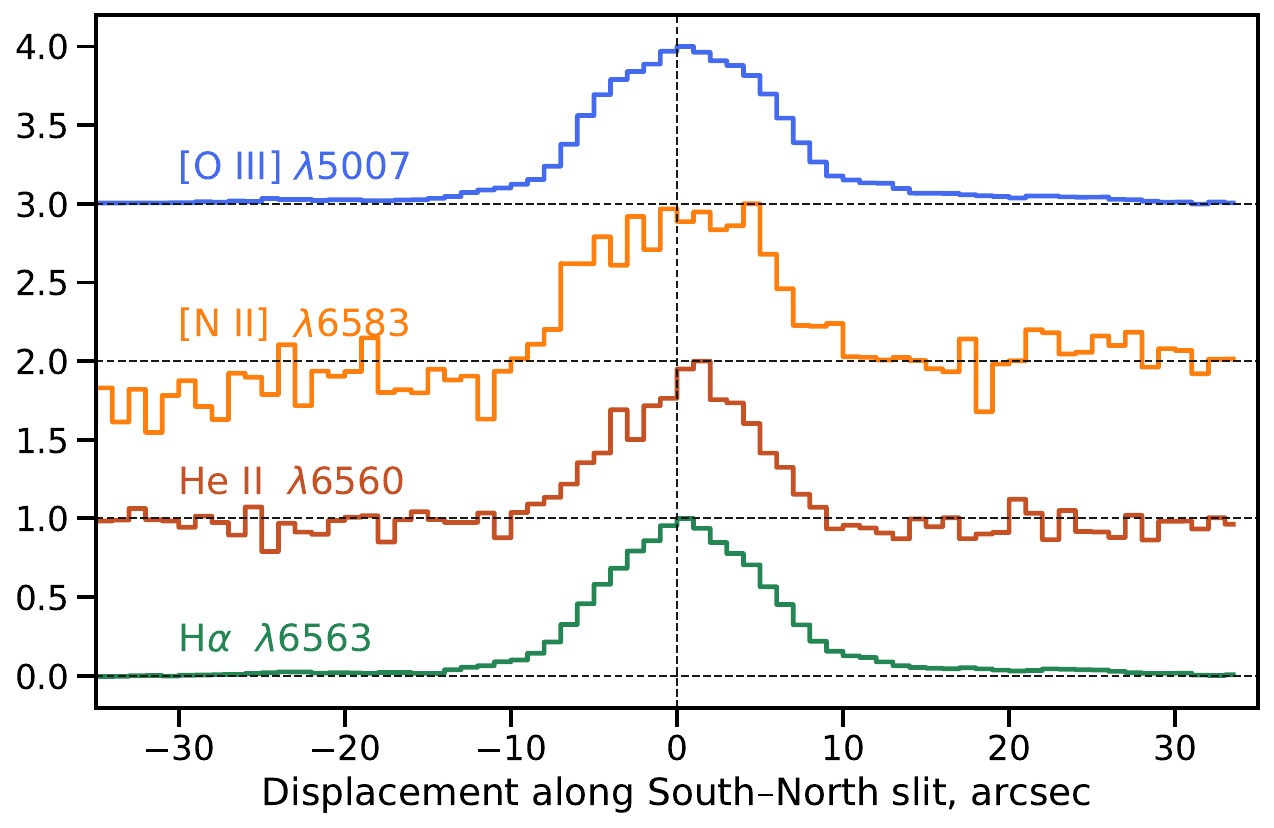}
  \caption{Spatial profiles extracted from the MES-SPM longslit spectra,
    integrated over all nebular velocities.
    (a)~Horizontal slit (position g), which can be compared with
    the image-extracted profiles in Fig.~\ref{fig:image-cuts-equator}a.
    Positive displacements are towards the West.
    (b)~Co-added axial slits (positions c, d, e), which can be compared with Fig.~\ref{fig:image-cuts-axis}b.
    Positive displacements are towards the North.
  }
  \label{fig:slit-spatial-profiles}
\end{figure}

Fig.~\ref{fig:image-cuts-equator} shows profile cuts
along the equator and polar axis.
The equatorial profile (Fig.~\ref{fig:image-cuts-equator}a) shows most clearly the difference
between the emission lines, with \heii{} peaking at \qty{\pm 4}{arcsec}
and \oiii{} peaking at \qty{\pm 7}{arcsec}.
The \Ha{} profile is approximately the average of the \oiii{} and \heii{} profiles.
The cuts along the polar axis (Fig.~\ref{fig:image-cuts-axis}b) are much
more similar between the different lines.
There is some indication that \oiii{} is more flat-topped than \Ha{} and \heii{},
but contamination by continuum emission from stars complicates the analysis.
To remedy this, we show corresponding profiles derived from our
high-resolution longslit spectra
in Fig.~\ref{fig:slit-spatial-profiles},
which allow the nebular emission to be cleanly separated from the
stellar continuum and foreground \hii{} region.
The profiles along the nebular axis (Fig.~\ref{fig:slit-spatial-profiles}b)
confirm that \Ha{} and \heii{} are more centrally peaked than \oiii{},
but the difference is very slight. 

\subsection{Kinematic analysis}
\label{sec:kinematics}

\begin{figure}
  \includegraphics[width=\linewidth, clip, trim=90 50 90 40]{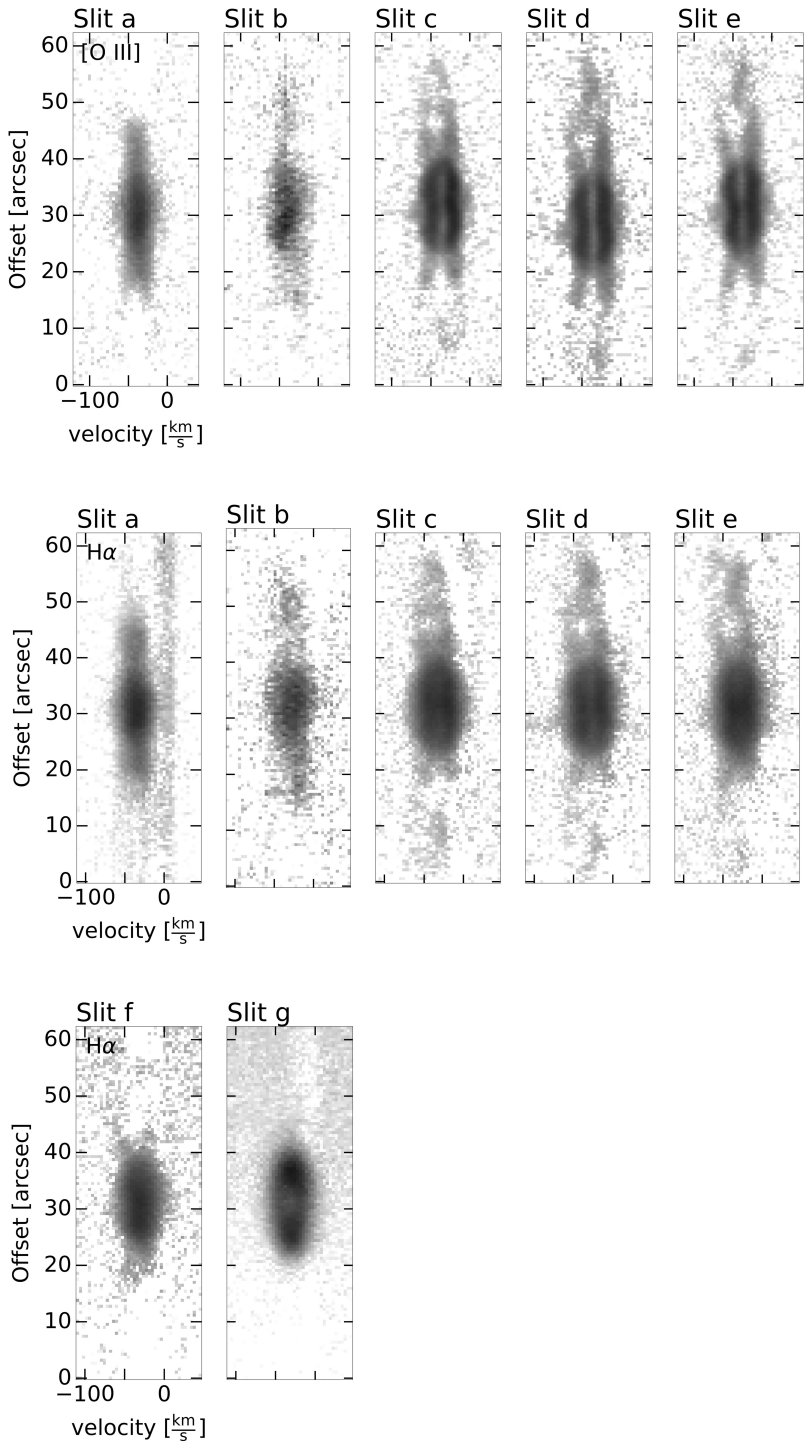}
  \caption{Position-velocity mosaics of the \oiii and  \Ha{} spectra, corrected for heliocentric velocity}
  \label{fig:spectro-ha}
  \label{fig:sp}
\end{figure}

\begin{figure}
  \centering
  \includegraphics[width=\linewidth]{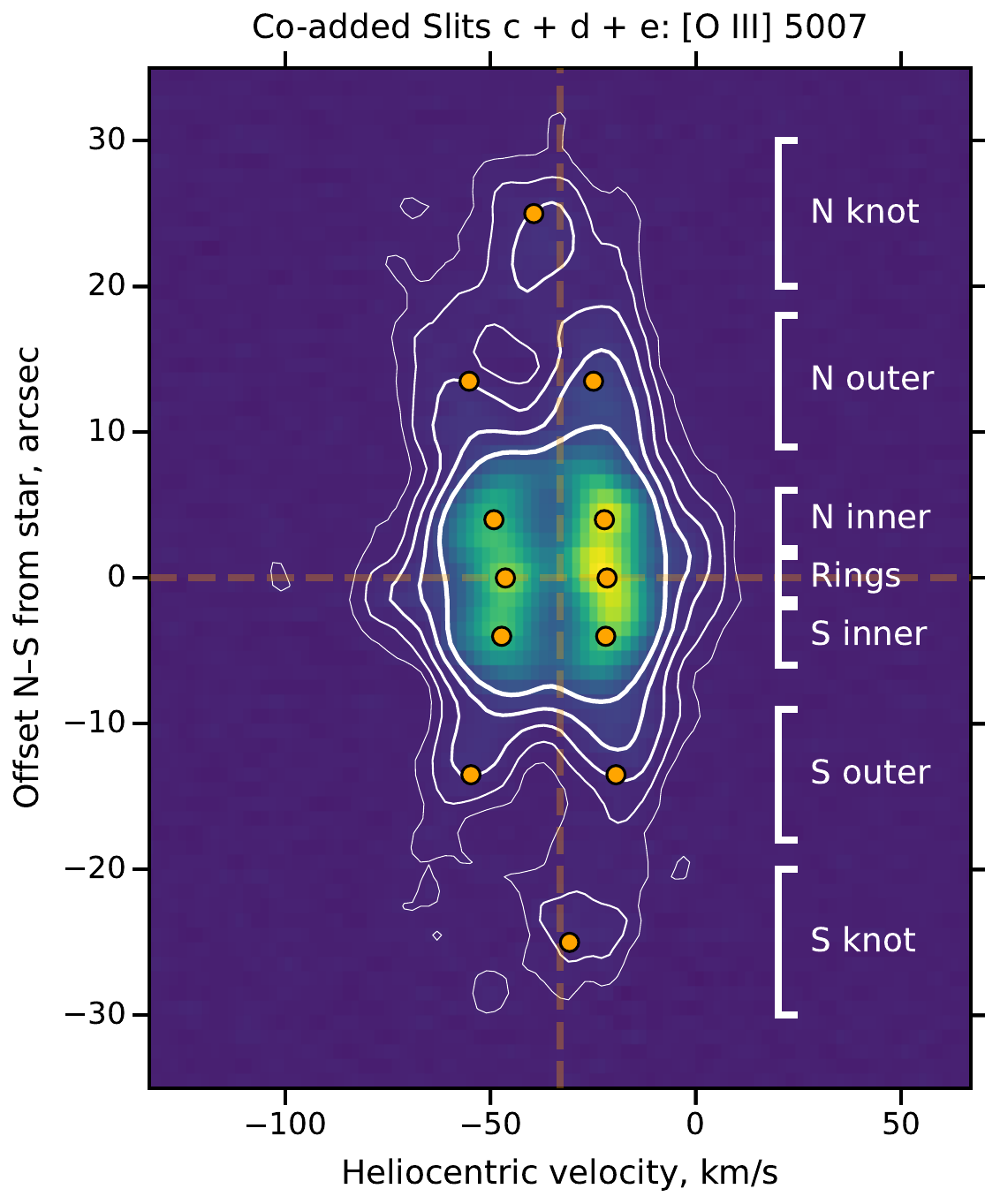}
  \caption{
    Co-added \oiiilam{} spectra from the three vertical 
    slits nearest the symmetry axis (positions c, d, and e), corrected to 
    heliocentric velocity. 
    A linear colour scale shows the bright inner regions at full resolution,
    while contours (levels: 1, 2, 4, 8, and 16\% of the peak)
    trace the fainter outer regions, Gaussian smoothed 
    with a kernel of \(\sigma = \qty{1}{pix}\) to enhance the signal-to-noise ratio.
    The position–velocity array exhibits the characteristic hourglass pattern 
    of the inner and outer lobes, together with the compact kinematic 
    signatures of the northern and southern polar knots. 
    The equatorial rings are also visible at small offsets. 
    Labels indicate the principal nebular components (N/S knots, inner and 
    outer lobes, and equatorial rings), and orange dots denote the centroids 
    of double-Gaussian fits to the integrated line profile of each component.
  }
  \label{fig:co-added-oii-pv}
\end{figure}

\begin{figure}
  \includegraphics[width=1.\columnwidth]{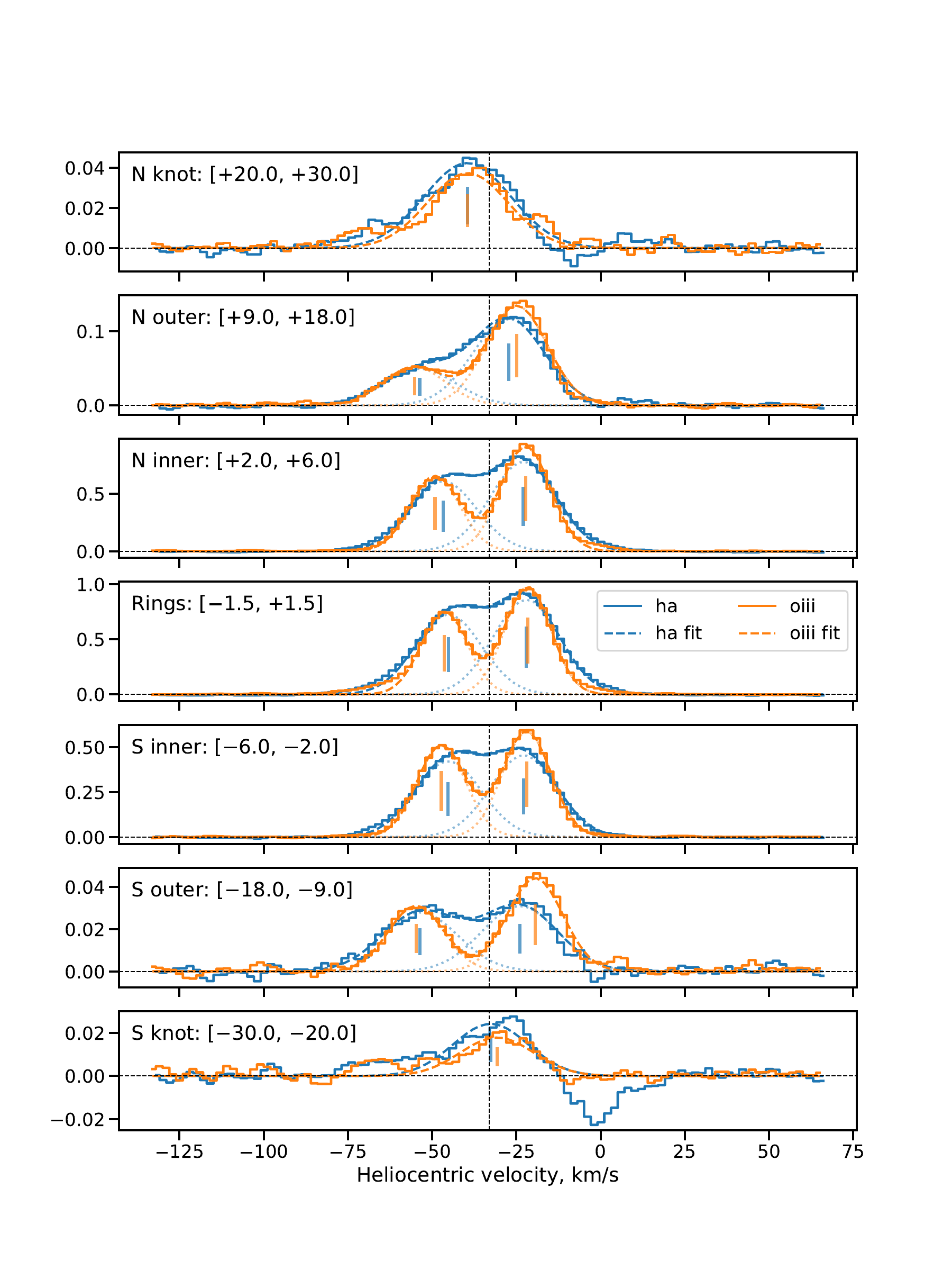}
  \caption{
    One-dimensional spectra of \oiii{} and \Ha{}
    for the different regions of interest along the nebular axis, extracted from the
    co-added position-velocity image of Fig.~\ref{fig:co-added-oii-pv}.
    Two-component Gaussian fits to each profile are also shown,
    except for the knots, where a single component is used.}
    \label{fig:spectro-oiii}
\end{figure}

Fig.~\ref{fig:02} shows the location of the central slit~d superimposed on the SPM \oiii{} image, which clearly reveals the bipolar axial symmetry of Ou5. Several regions of interest are indicated in this figure. Specifically, the blue squares indicate the positions of the polar knots, the  pink rectangles trace the approximate extent of the outer lobes,  and the yellow area highlights the brighter inner lobes. 
The P-V array to the right of the image and at the centre of Fig.~\ref{fig:02} corresponds to the bi-dimensional \oiii{} spectrum  from slit~d. The same regions of interest as in the image are indicated here.
Position-velocity images of all the slits are shown in Fig.~\ref{fig:sp},
which shows that there is very little variation
between the three positions closest to the nebular axis:
c, d and e.
We therefore coadd these three slits to improve the signal-to-noise,
with results shown in Fig.~\ref{fig:co-added-oii-pv},
with Fig.~\ref{fig:spectro-oiii} showing line profiles extracted
from each region of interest.

\begin{figure}
  \includegraphics[width=1.\columnwidth]{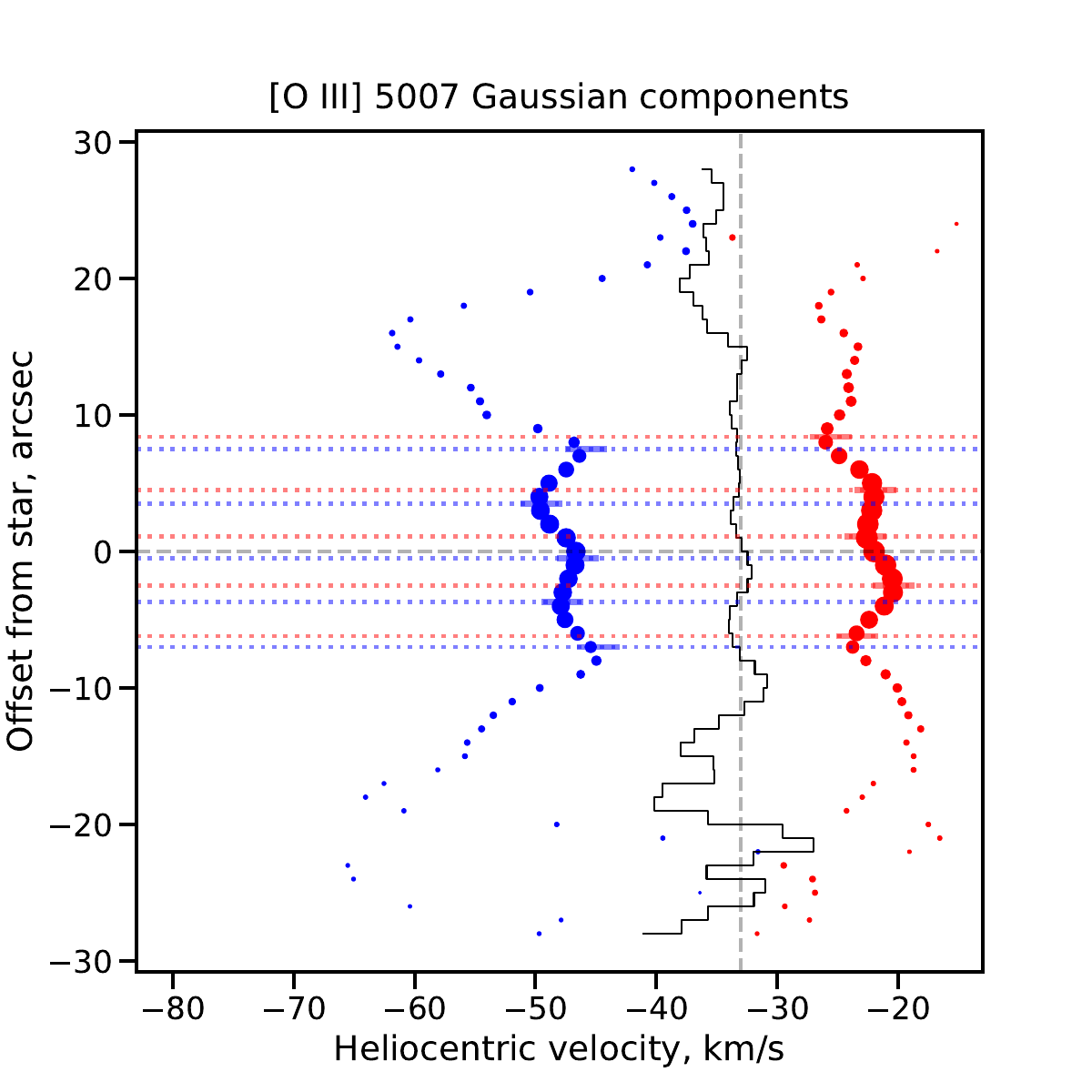}
  \caption{
    Velocities of components from two-gaussian fits to
    \oiiilam{} line profiles in the co-added N-S slits along the major axis of the nebula. 
    Symbol size is proportional to the flux of each component. 
    The flux-weighted mean velocity of the two components is shown by the black stepped line
    and the nominal systemic velocity of \qty{-33}{\kms}
    is indicated by the vertical dashed line. 
    Turning points in the velocity profiles of the red and blue components of the inner lobes are marked by horizontal dotted lines.
  }
  \label{fig:gaussian-components-velocities}
\end{figure}
  
\begin{figure}
  \includegraphics[width=1.\columnwidth]{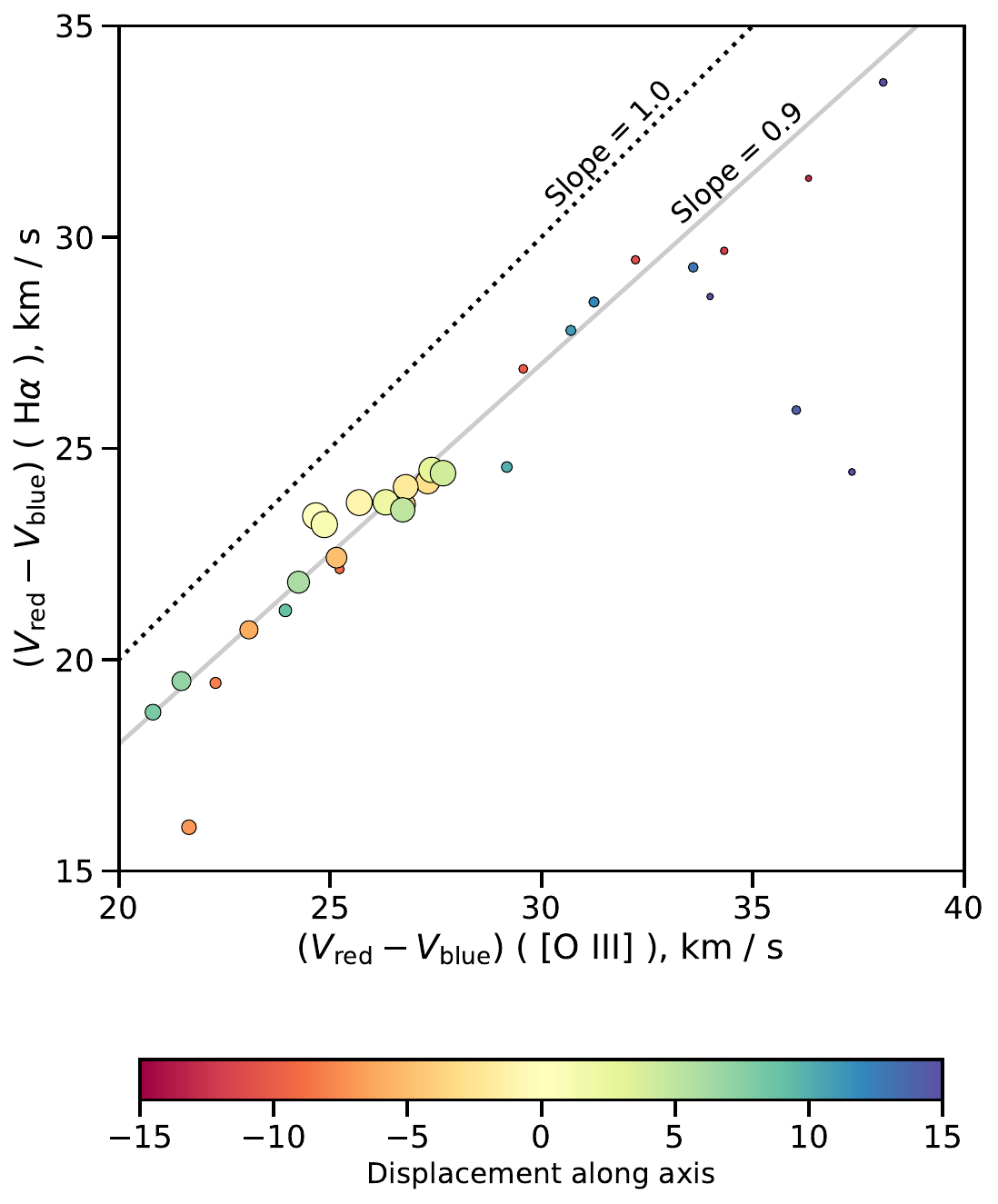}
  \caption{Comparison of velocity splitting
    (difference between red and blue gaussian components)
    of \oiii{} versus \Ha{}.
    Symbol size is proportional to brightness
    and color varies with position along the axis (see color bar).
    Dotted line shows the prediction for the case of equal splitting,
    while solid line the prediction for the case of \Ha{} splitting
    being 0.9 times \oiii{} splitting.
  }
  \label{fig:gauss-split}
\end{figure}

\begin{figure}
  \includegraphics[width=\columnwidth]{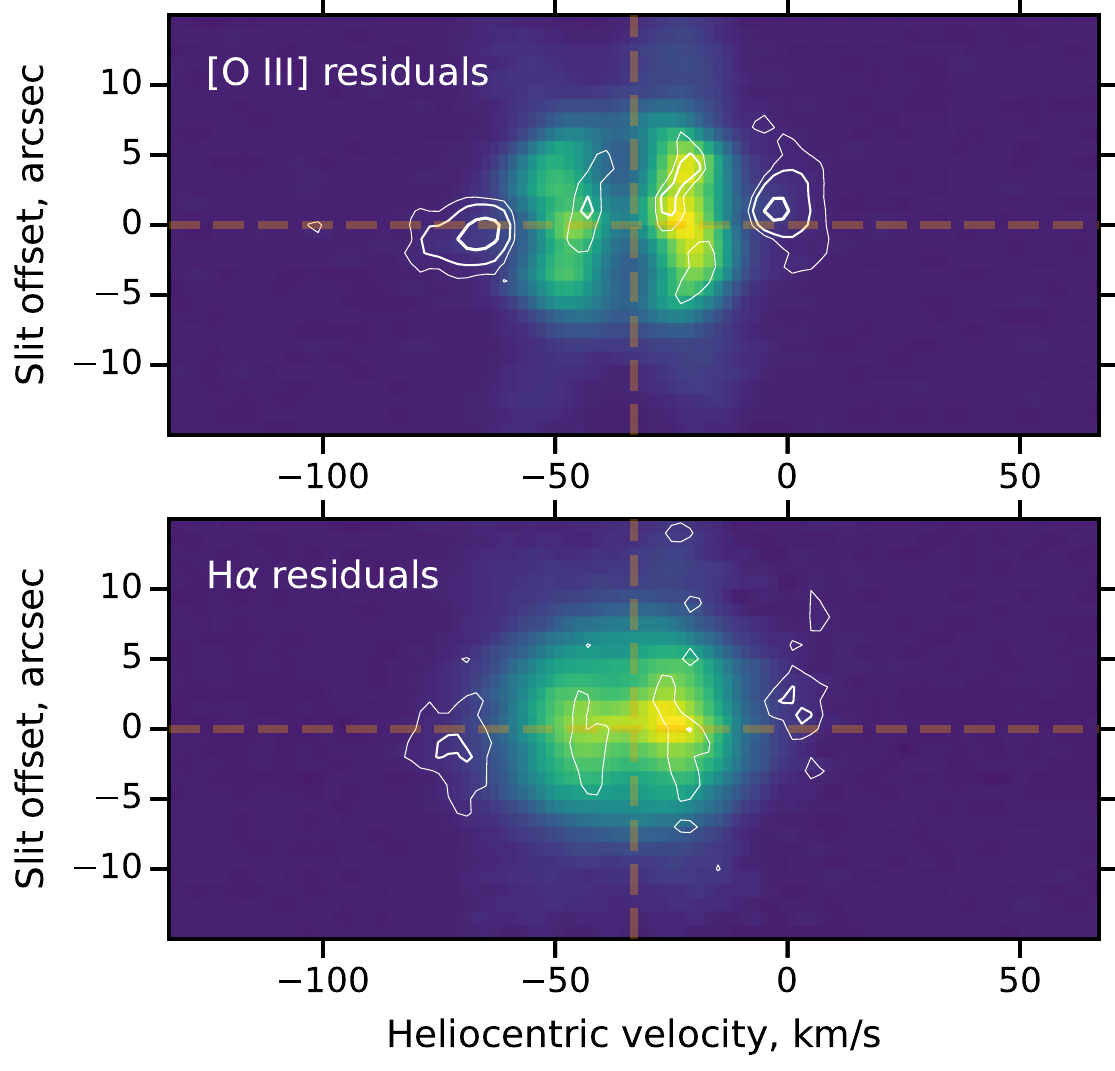}
  \caption{
    Residual emission after subtracting the
    two-gaussian fit from the observed profile at each position
    along the co-added axial slit.
    White countours of increasing thickness show excess emission
    of 1.5\%, 3\%, and 6\% of the peak for \oiii{} (upper)
    and \Ha{} (lower), superimposed on a color image of the
    observed profile.
    The excess emission in the wings can be ascribed to the
    outer equatorial ring.}
    \label{fig:residual-outer-ring}
\end{figure}

The systemic velocity of Ou~5 was determined using slit~d from the \Ha{} and \oiii{} lines. The flux-weighted heliocentric mean velocity at the projected position of the central star is \(V_\mathrm{sys} = \qty{-33 \pm 1}{\kms}\), which we adopt here as the systemic velocity of the nebular shell. 
Note that this systemic velocity differs from the \qty{-26}{\kms} center-of-mass velocity of the binary central star \citep{Jones:2022a}.

We performed 
gaussian fits to the different regions of interest along the axis, with resultant profiles shown in Fig.~\ref{fig:spectro-oiii}. 
Two gaussians were fitted for the core, inner, and outer lobes, with one gaussian fitted for the polar knots.
In general, there is a good agreement between the 
two emission lines in the derived velocities of the gaussian components,
although the splitting is consistently larger by about \qty{2}{\kms} in \oiii{}  compared with \Ha{}.

To investigate velocity variations along the slit on finer spatial scales,
we performed double-Gaussian fits to the \oiii{} profile
at a spatial resolution of \qty{1}{arcsec} along the co-added slits.
The results are shown in Fig.~\ref{fig:gaussian-components-velocities}.\footnote{%
The position angle of the nebular axis is \(\mathrm{PA} = \ang{356 \pm 1}\),
yielding a projection factor of \(\cos (\mathrm{PA}) = \num{0.998 \pm 0.001}\)
between the axis and the NS-oriented slits.
We ignore this correction since it is small compared with our measurement precision.}
The open hourglass pattern of the position--velocity structure in
the inner disk and lobes is readily apparent,
with the nebular morphology in direct images being closely mirrored in velocity space.
Close inspection also shows evidence for a spatial offset between the red and blue kinematic components,
probably due to the inclination of the nebular axis (see Section~\ref{sec:line-sight-incl}).

The \oiii{} velocity splitting \(\delta V\splitting = V\red - V\blue\) is
\qty{25 \pm 1}{\kms} at the equator, increasing to a local
maximum in the inner lobes of \qty{28 \pm 1}{\kms} at \(z = \qty{\pm 4}{arcsec}\),
then dropping to a local minimum of \qty{21 \pm 2}{\kms}
at the boundary with the outer lobes at \(z = \qty{\pm 8}{arcsec}\).
In the outer lobes themselves, the splitting increases once more,
reaching a maximum \(\delta V\splitting = \qty{35 \pm 3}{\kms}\)
at \(z = \qty{\pm 15}{arcsec}\).
Unlike the inner lobes, the outer lobes do not show antisymmetric skew,
but instead both bend towards negative velocities,
giving the appearance of two nested closing parentheses in the PV diagram.
Beyond displacements of \qty{20}{arcsec} from the equator,
the profiles are dominated by the polar knots, which show no local splitting within each knot but are once again globally antisymmetric,
with a small splitting between the N and S knots of \(\delta V\splitting = \qty{7 \pm 2}{\kms}\).


Figure~\ref{fig:gauss-split} shows the relation between line splitting in \oiii{} and \Ha{}.
There is a consistent linear trend across the inner and outer lobes,
with \(\delta V\splitting(\Ha) \approx 0.9 \,\delta V\splitting(\oiii)\).
Once again, the mirroring between nebular morphology and
kinematics is evident, since spatial profile cuts (Figs.~\ref{fig:image-cuts-equator}, \ref{fig:slit-spatial-profiles})
show that \Ha{} is slightly more centrally concentrated than \oiii{}.
The only significant deviation from the linear trend occurs at the equator
(yellow symbols in Fig.~\ref{fig:gauss-split}), where the constant of proportionality increases to \(0.95\).

Given the consistent mirroring between spatial and velocity structures,
one would expect the outer ring and outermost lobes (Fig.~\ref{fig:rgb-image}) to produce faint, high-velocity emission near the equator.
Such emission is indeed visible as equatorial protrusions
in the lowest contours of the position--velocity profile shown in Fig.~\ref{fig:co-added-oii-pv}.
To isolate these components,
we subtract the double-Gaussian fit from the observed line profile at each position
and show the residuals as white contours in Fig.~\ref{fig:residual-outer-ring}.
High-velocity components are revealed with a splitting
\(\delta V\splitting = \qty{70 \pm 5}{\kms}\),
together with a slight asymmetry (red component displaced north, blue south).
The relative intensity of this residual high-velocity emission
reaches 6\% of the \oiii{} main shell, but only 3\% in \Ha{}.
We also see lower-level residuals near the line peaks,
but these are almost certainly due to non-Gaussian line shapes in the inner lobes,
rather than a distinct kinematic component.

\subsection{Line-of-sight inclination of the nebula axis}
\label{sec:line-sight-incl}

The hour-glass pattern of the red and blue velocity components of the inner lobes
is slightly sheared along the slit
(Fig.~\ref{fig:gaussian-components-velocities}).
The simplest explanation for this spatio-kinematic anti-symmetry 
is that the nebula is approximately cylindrically symmetric,
but with an axis that is tilted from the plane of the sky.

If this is the case, then we can estimate the inclination angle by measuring
the displacement along the axis between equivalent points on the near (blue)
and far (red) sides of the nebula shell. 
For each component (red and blue)
we mark the turning points in the velocity profiles:
\(d V / d x = 0\), where \(x\) is the displacement along the slit, 
with the position of each indicated with red or blue dotted lines in the figure. 
There are five turning points for each kinematic component:
a minimum in the core region (the waist of the hourglass), 
a pair of maxima in the N and S inner lobes,
and a pair of minima at the boundaries between the outer and inner lobes
(maxima and minima refer to the absolute velocity difference of each component from the systemic velocity).
As can be seen in the figure,
all of the turning points in the velocity of the red kinematic component
are systematically displaced to the N
compared with the corresponding turning points in the blue kinematic component.
The mean and standard deviation of the displacements is 
\(\delta x = \qty{1.1 \pm 0.3}{arcsec}\).

Assuming that the inner lobes consist of a cylindrically symmetric shell,
with axis inclined at an angle \(i\) to the line of sight, 
then the projected displacement along the symmetry axis between any corresponding features of the near and far sides of the shell is given by 
\begin{equation}
    \label{eq:displacement}
    \delta x = D \cos i,
\end{equation}
where \(D\) is the cylindrical diameter of the shell.
Given the barrel-like shape of the inner lobes,
the diameter does not vary significantly along the axis,
so we take the average value of  \(D = \qty{11(1)}{arcsec}\),
which yields an inclination of \(i = \qty{84(2)}{\degree}\).

We can obtain a second independent estimate of the inclination 
from the high-velocity wings revealed after subtracting the inner shell
from the observed profiles (Fig.~\ref{fig:residual-outer-ring})
under the assumption that they are associated with the outer ring (Fig.~\ref{fig:rgb-image}b).
The spatial displacement between the residual peaks on the blue and red sides
is measured to be \(\delta x = \qty{2.5(0.9)}{arcsec}\),
while the diameter of the outer ring is \(D = \qty{33(1)}{arcsec}\),
which combine to yield an inclination of \(i = \qty{86(2)}{\degree}\).

\citet{Jones:2022a} estimate an inclination for the binary orbit of
\(i = \qty{82(1)}{\degree}\) which is in slight tension with our results
for the inclination of the nebular axis \NEW{(see further discussion in section \ref{sec:impl-our-kinem} below)}.

\begin{figure}
  \centering
  \captionof{table}{Correspondence between model components
    (Fig.~\ref{fig:geometrical-model})
    and morphological features (Fig.~\ref{fig:rgb-image}b)
    \label{tab:geometrical-model}
  }
  \begin{tabular}[b]{l l l }
    \toprule
    & Geometrical component
    & Morphological features\\
    \midrule
    \textbf{A} & Inner shell & Inner Ring, Inner Lobes \\
    \textbf{B} & Outer shell & Outer Lobes  \\
    \textbf{C} & Outermost shell & Outer Ring, Outermost Lobes\\
    \textbf{D} & Polar knots & N Knot, S Knot\\
    \bottomrule
  \end{tabular}
  \bigskip
  \null\\
  \includegraphics[width=0.5\linewidth]{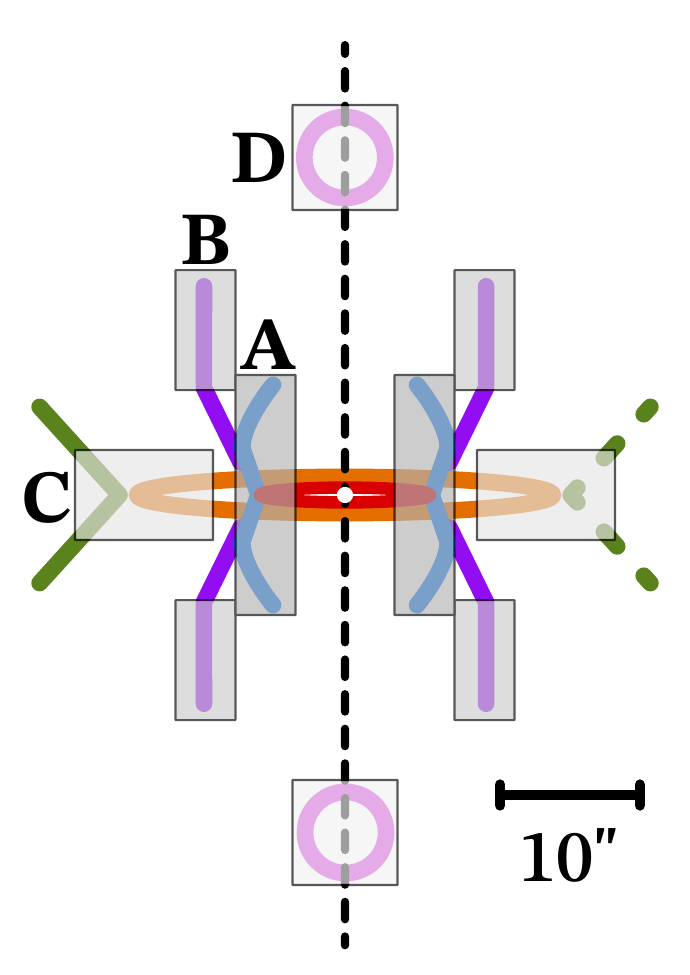}
  \captionof{figure}{Geometrical model for the nebula. Diagram shows cross-section through
    the four-component model,
    where all the components are approximated by cylinders that are
    indicated by gray boxes with dimensions detailed in Table~\ref{tab:shells}.
    Colored lines trace the morphological features identified in the images of the nebula
    (Fig.~\ref{fig:rgb-image}b),
    which map onto the model components as listed in Table~\ref{tab:geometrical-model}.
  }
  \label{fig:geometrical-model}
\end{figure}

\subsection{A geometrical model for Ou~5}
\label{sec:nebular-geometry}

In order to account for the imaging and spectroscopic observations of the nebula, at least four distinct geometrical components are required:
\begin{enumerate}[A.]
\item An inner, barrel-shaped shell to reproduce the inner lobes and inner equatorial ring
\item An outer shell to reproduce the outer lobes at high latitudes above and below the equator 
\item A third, outermost shell to reproduce the outer equatorial ring
  and outermost lobes, which is also responsible for the high-velocity wings in the line profiles near the equator (Fig.~\ref{fig:residual-outer-ring})
\item A pair of bipolar knots to reproduce the compact kinematic features observed along the axis (Fig.~\ref{fig:sp}) 
\end{enumerate}
The model is illustrated in Fig.~\ref{fig:geometrical-model} and Table~\ref{tab:geometrical-model}
summarises the correspondence between model components and the morphological features described in section~\ref{sec:morph-feat}.

Based on the observed images and position-velocity diagrams,
we determine approximate geometric parameters for the components,
which are listed in the first pair of rows of Table~\ref{tab:shells}.
Each component is idealised as a cylindrical shell,
extending from \(r_\mathrm{min}\) to \(r_\mathrm{max}\) in cylindrical radius
(distance from the axis)
and from \(z_\mathrm{min}\) to \(z_\mathrm{max}\) in height
(distance above and below the equator).

The second pair of rows in Table~\ref{tab:shells} gives the maximum velocity splitting
\(\delta V\splitting = V\red - V\blue\)
observed in each component,
together with the height \(z\splitting\) where this occurs.
The velocities \(V\red,  V\blue\) are the centroids of double-Gaussian fits to the line profiles
(Fig.~\ref{fig:gaussian-components-velocities}). 

The third pair of rows in Table~\ref{tab:shells} lists the corresponding average spherical radius \(R\splitting\) and
average latitude \(\theta\splitting\) (angle above and below the equator)
for each component.
The radius is given in parsecs, assuming a distance of \qty{4}{kpc}
(see section~\ref{sec:dist-determ}).
The \(\pm\) values denote the typical spread rather than the uncertainty.

The fourth pair of rows in Table~\ref{tab:shells} lists the radial expansion velocity, \(V\),
and dynamical age timescale, \(t_\mathrm{dyn}= R / V\),
for each component.
Assuming that there are no non-radial motions, then the expansion velocity can be found
from the splitting as
\begin{equation}
  \label{eq:velocity-splitting}
  V = \frac{\delta V\splitting}{2 \left\vert \sin(\theta\splitting + i) \right\vert},
\end{equation}
where we assume an inclination of \(i = \qty{84 \pm 2}{\degree}\) (see section \ref{sec:line-sight-incl}).
The dynamical age is found to be roughly \(t_\mathrm{dyn} \approx \qty{1e4}{year}\)
for all four components,
with slight evidence that the higher latitude components \textbf{B} and \textbf{D}
may be a few thousand years older than the lower latitude components \textbf{A} and \textbf{C}.

The remaining rows of the table concern physical properties of the nebula, which we discuss below in section~\ref{sec:phys-param-nest}.

\section{Physical parameters}
\label{sec:physical-parameters}
In this section we use our spectroscopic data to constrain the physical conditions of Ou~5.
We begin in Section~\ref{sec:estim-nebula-kinet} by deriving the mean kinetic temperature of the ionized gas from thermal line broadening, considering both single-temperature and two-phase interpretations.
We then estimate in Section~\ref{sec:phys-param-nest} the electron densities and ionized masses of the principal nebular components from the calibrated \Ha{} surface brightness.
Finally, in Section~\ref{sec:temp-lumin-centr} we apply the \heii{} Zanstra method to determine the effective temperature and luminosity of the central star.

\subsection{Estimate of nebula kinetic temperature}
\label{sec:estim-nebula-kinet}

\begin{table*}
  \caption{Geometrical and physical properties of the nebular components of Ou 5}
  \label{tab:shells}
  \centering
  \begin{tabular}{lLl RRRR}
    \toprule
    & & {Units}
    & \text{\textbf{A} Inner shell} & \text{\textbf{B} Outer shell} & \text{\textbf{C} Outermost shell} & \text{\textbf{D} Polar knots}\\
    \midrule
    Cylindrical radii & [r_\mathrm{min}, r_\mathrm{max}] & \unit{arcsec}
    & [3.3, 7.3] & [7.3, 11.3] & [8.8, 18] & [0, 4]\\
    Heights & [z_\mathrm{min}, z_\mathrm{max}] & \unit{arcsec}
    & [0, 8] & [7, 15] & [0, 3] & [19, 26]\\[\smallskipamount]
    Maximum splitting & \delta V\splitting & \unit{\kms}
    & 24 \pm 2 & 35 \pm 3 & 70 \pm 5 & 7 \pm 2 \\ 
    Height of splitting & z\splitting & \unit{arcsec}
    & 3.6 \pm 0.4   & 15 \pm 2      & 0  \pm 1      & 23 \pm 2  \\[\smallskipamount]
    Spherical radius & R\splitting & \unit{pc}
    & 0.12 \pm 0.01 & 0.34 \pm 0.03 & 0.26 \pm 0.03 & 0.45 \pm 0.04\\
    Latitude & \theta\splitting  & \unit{deg}
    & 34 \pm 3 & 58 \pm 3 & 0 \pm 9 & 90\\[\smallskipamount]
    Expansion velocity & V & \unit{\kms}
    & 14 \pm 1& 28 \pm 3 & 35\pm 3& 33\pm 15\\
    Dynamical age & t_\mathrm{dyn} & \qty{1000}{yr}
    & 8.7 \pm 0.7 & 11.9 \pm 1.7 & 7.3 \pm 0.6 & 13.3 \pm 6.1\\[\smallskipamount]
    Electron density & n_\mathrm{e} & \unit{cm^{-3}}
    & 150 \pm 10 & 40 \pm 10 & 30 \pm 10 & 30 \pm 10\\
    Ionized mass & M_\mathrm{i} & \qty{0.001}{\msun}
    & 28 \pm 3 & 26 \pm 7 & 12 \pm 4 & 3 \pm 1\\
    \bottomrule
  \end{tabular}
\end{table*}

Ou~5 is an example of a class of planetary nebulae with a high
\textit{abundance discrepancy factor} (ADF),
in which metal ion abundances derived from recombination lines
are systematically higher than those derived from collisional lines
\citetext{see \citealt{Garcia-Rojas:2019a} and references therein}.
One potential explanation for these discrepancies is the coexistence
of different ``phases'' within the ionized nebula with varying
metallicity and physical conditions \citep{Liu:2000a}.
The presence of warm and cool photoionized phases has been postulated
for Ou~5 \citep{Corradi:2015a},
with the warm-phase temperature determined from \oiii{} line ratios as
\qty{10150(300)}{K} after correction for a recombination contribution to the auroral line.
For some other nebulae with high ADF,
the Paschen jump or Balmer jump implies a lower mean temperature
for the H recombination spectrum than for collisional lines
\citep{Garcia-Rojas:2022a}, but no such measurements exist for Ou~5.
However, an independent method of estimating the mean temperature
of ionized hydrogen in the nebula is via
the excess thermal broadening of hydrogen emission lines,
as compared with lines from heavier ions.
Our high resolution spectra are well-suited to such a measurement.

Several different physical processes contribute to the observed
profile of an emission line \(I(v)\).
For lines from ionized nebulae, it is convenient to divide
the processes as follows (see \citealt{Garcia-Diaz:2008a}):
\newcommand\obs{\ensuremath{_\mathrm{obs}}}
\newcommand\ins{\ensuremath{_\mathrm{ins}}}
\newcommand\therm{\ensuremath{_\mathrm{th}}}
\newcommand\expand{\ensuremath{_\mathrm{exp}}}
\newcommand\fine{\ensuremath{_\mathrm{fs}}}
\newcommand\nontherm{\ensuremath{_\mathrm{nt}}}
\newcommand\noise{\ensuremath{_\mathrm{noise}}}

\begin{enumerate}[1.]
\item Instrumental profile, \(I\ins(v)\), assumed to be Gaussian with FWHM given in Table~\ref{table:kn61}.
\item Thermal broadening profile, \(I\therm(v)\), assumed to be gaussian with RMS width \(\sigma\therm = (\qty{9.1}{\kms}) (T_4/A)^{1/2}\), where \(T_4 = T / \qty{10000}{K}\) and \(A\) is the atomic mass of the emitting species.
\item Kinematic profile, \(I\expand(v)\), due to the systematic expansion of the nebula.
\item Fine-structure profile, \(I\fine(v)\), due to closely spaced atomic sub-levels.
\item Non-thermal profile, \(I\nontherm(v)\), which is a catch-all term for any other physical processes that might broaden the lines, such as turbulence or scattering by dust.
\end{enumerate}
The observed profile can be regarded
as the convolution the individual profiles:
\newcommand\CONV{\star}
\begin{equation}
  \label{eq:convolution}
  I\obs(v) = I\ins(v) \CONV I\therm(v) \CONV
  I\expand(v) \CONV I\fine(v) \CONV I\nontherm(v) .
\end{equation}
In the case that all the profiles are gaussian, then the RMS widths
add in quadrature:
\begin{equation}
  \label{eq:quadrature-widths}
  \sigma^2\obs = \sigma^2\ins + \sigma^2\therm +
  \sigma^2\expand + \sigma^2\fine + \sigma^2\nontherm ,
\end{equation}
but for non-gaussian profiles this relation is only approximate.

\subsubsection{One-phase analysis}
\label{sec:one-phase-analysis}

\begin{figure}
  \textbf{a}\\[\medskipamount]
  \null\hfill
  \includegraphics[width=0.93\linewidth]{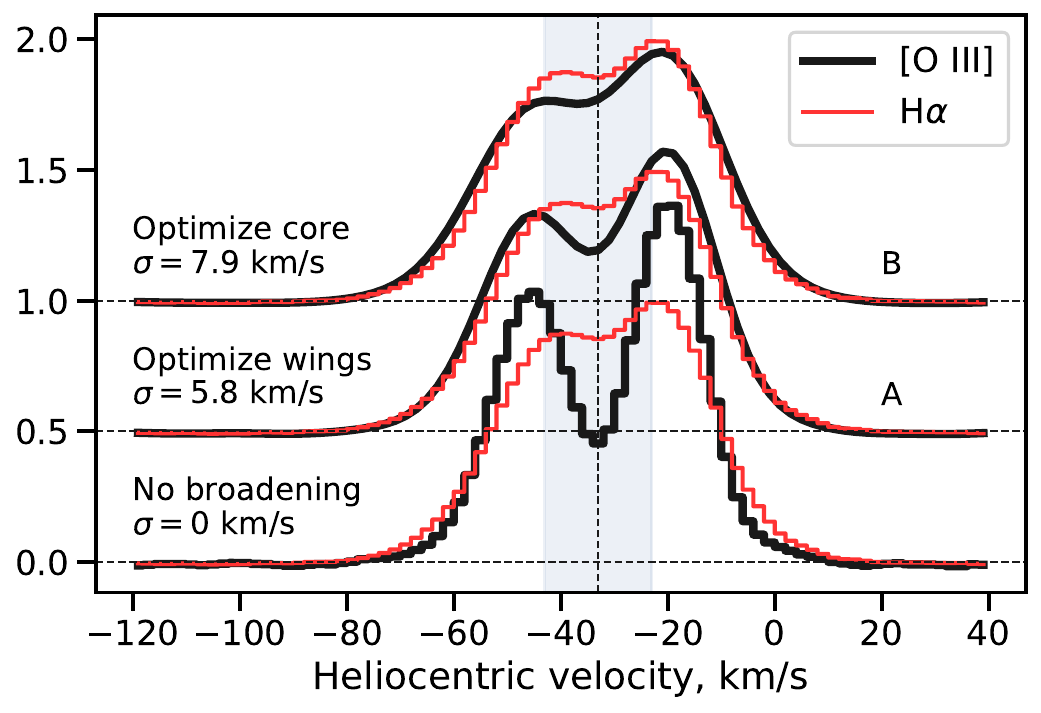}\\
  \textbf{b}\\[\medskipamount]
  \includegraphics[width=\linewidth]{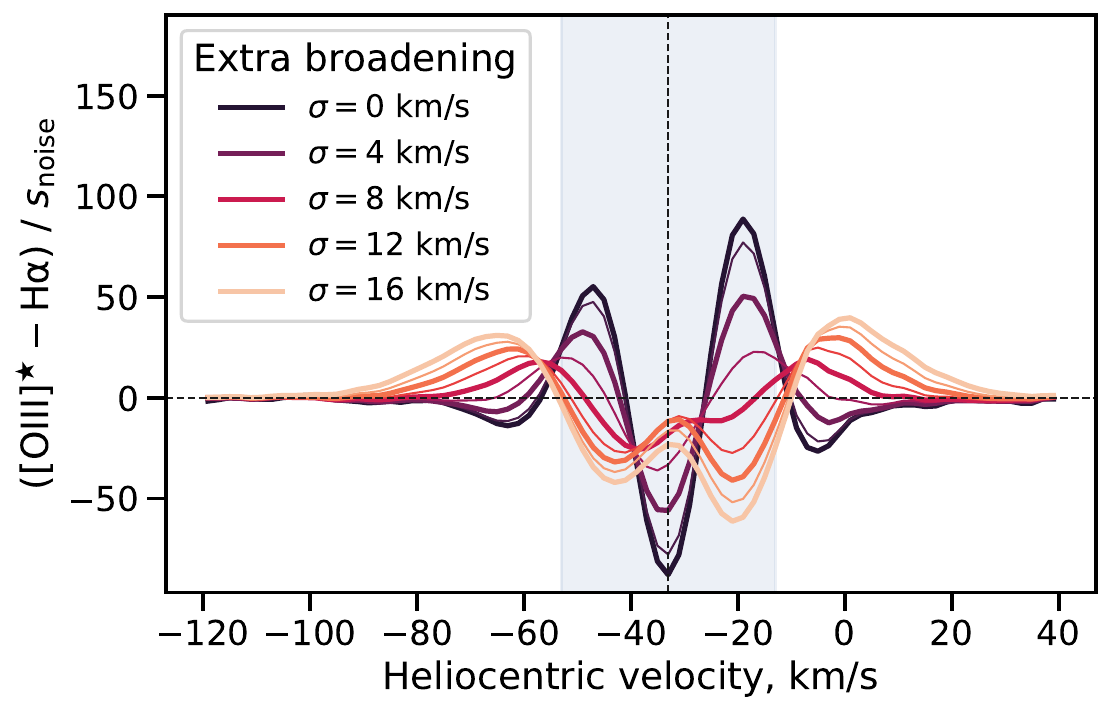}\\
  \textbf{c}\\[\medskipamount]
  \includegraphics[width=\linewidth]{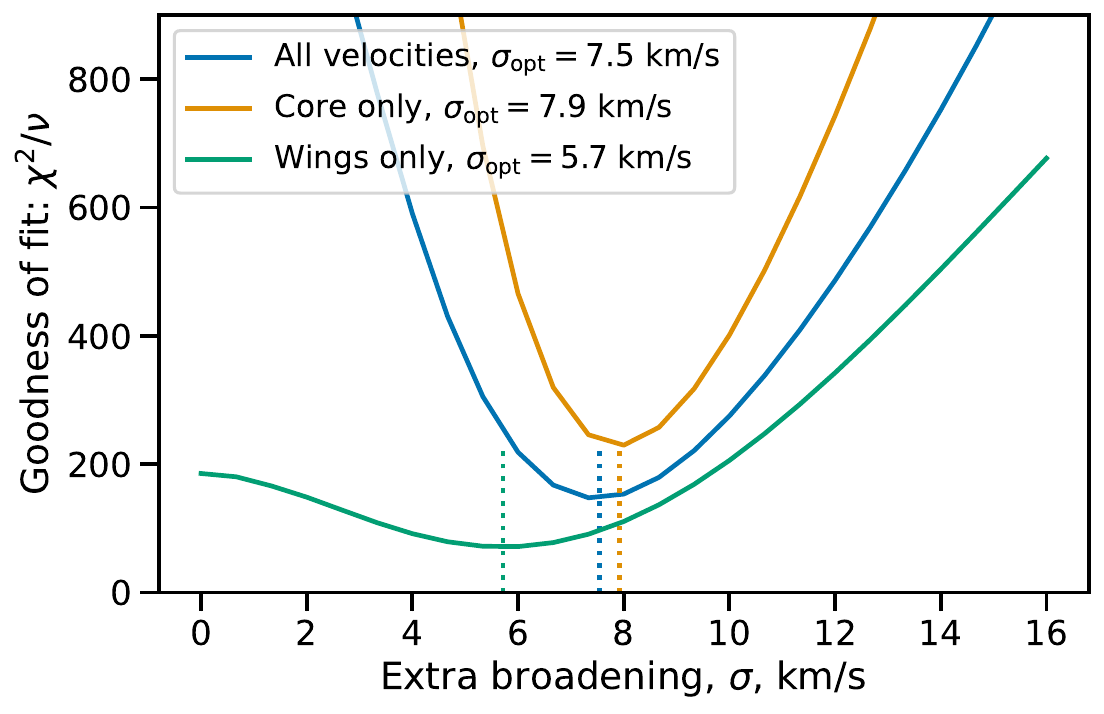}\\
  \caption{
    Direct convolution method for determining the kinetic temperature.
    (a)~Comparison of integrated line profiles of the inner lobes
    for \Ha{} and \oiii{}. Bottom row shows the original profiles,
    while upper rows show the effect of convolving the \oiii{}
    profile with a gaussian kernel of increasing rms width, \(\sigma\).
    (b)~Residual profiles of convolved \oiii{} minus \Ha{}
    for kernel widths of \(\sigma = 0\) to \qty{16}{\kms},
    normalized by the rms noise of the profiles. 
    (c)~Reduced \(\chi^2\) measure of ``goodness of fit''
    as a function of \(\sigma\) for
    the entire profile (blue line), core only (orange line),
    and wings only (green line).
    The core is defined as velocities with \qty{\pm 20}{\kms}
    of the systemic velocity, indicated by blue shading
    in panels a and b. The optimum values of \(\sigma\) for
    each case are indicated by vertical dotted lines.
  }
  \label{fig:T-convolution}
\end{figure}

\begin{figure}
  \includegraphics[width=\linewidth]{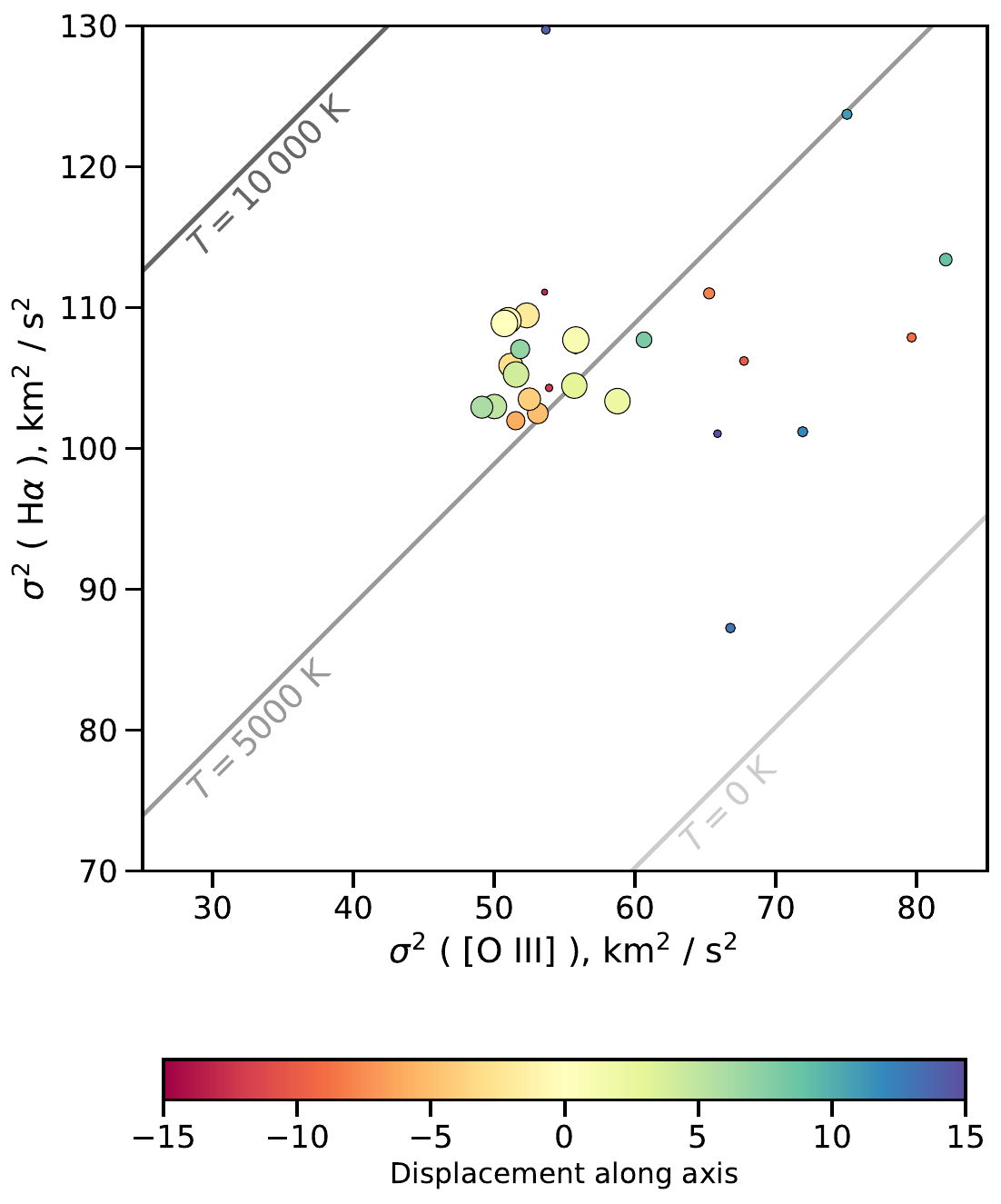}
  \caption{
    Gaussian decomposition method for determining the kinetic temperature.
    Comparison of squared rms component widths
    \(\sigma^2\) of two-gaussian fits to the line profiles of
    \oiii{} (horizontal axis)
    and \Ha{} (vertical axis)
    Colored symbols show results for different positions
    within the inner lobes along the coadded axial slits.
    Symbol size is proportional to brightness,
    while color represents displacement along the axis
    from south (dark colors) to north (light colors).
    Gray diagonal lines show predictions for kinetic temperatures
    of \(T = \qtylist{0;5000;10000}{K}\).
  }
  \label{fig:T-gaussian-fits}
\end{figure}

We estimate the mean kinetic temperature of the \Ha{}-emitting
gas by comparing the \Ha{} and \oiii{} line profiles.
Due to the difference in atomic masses
(\(Z = 1\) and \(Z = 16\), respectively),
the thermal broadening is much larger for \Ha{} than for \oiii{}:
\begin{equation}
  \label{eq:thermal-ha-oiii}
  \sigma^2\therm(\Ha) - \sigma^2\therm(\oiii) = 77.34 \, T_4 \, \unit{km^2.s^{-2}}
\end{equation}
For \oiii{}, the \(\lambda5007\) line is a single fine-structure component, but \Ha{} consists of multiple unresolved components \citep{Clegg:1999a}. For typical nebular conditions,
this can be approximated \citep{Garcia-Diaz:2008a}
as a gaussian profile with
\begin{equation}
  \label{eq:fs-ha-oiii}
  \sigma^2\fine(\Ha) - \sigma^2\fine(\oiii) \approx \qty{10.2}{km^2.s^{-2}} .
\end{equation}
The instrumental broadening is dominated by the finite slit width,
yielding \(\sigma\ins = \qty{5.0}{\kms}\) for both lines for the \qty{150}{\micron} slit. 

If the emitting volumes of the two lines are
homogeneous and coincide exactly,
then the expansion and non-thermal profiles,
\(I\expand(v)\) and \(I\nontherm(v)\),
will be identical between the \Ha{} and \oiii{} lines, in which case
the \Ha{} profile should be equal to the convolution of the \oiii{} profile with a gaussian kernel whose variance \(\sigma^2\) is equal to the sum of equations~\eqref{eq:thermal-ha-oiii} and~\eqref{eq:fs-ha-oiii}.
This is the basis of the method illustrated in Fig.~\ref{fig:T-convolution}.
See \citet{Richer:2022a} for further details and caveats concerning this method.

The \oiii{} and \Ha{} profiles are integrated over
the entire inner lobes and are normalized to the same total flux.
The peak signal-to-noise ratio (\SN) is
\(\SN = 310\) for \Ha{} and \(\SN = 425\) for \oiii{}.
To bring the mean heliocentric velocities into perfect agreement
we make a slight shift to the \Ha{} profile of \qty{+1.1}{\kms},
which is within the
precision of our absolute wavelength calibrations.
The respective profiles are shown in Fig.~\ref{fig:T-convolution}~a,
with the original profiles below and results of two different smoothing kernels (A and B) above.
For different values of the extra smoothing width \(\sigma\),
Fig.~\ref{fig:T-convolution}~b shows the residual difference between the \oiii{} and \Ha{} line,
normalized by the rms noise of the spectra.
It can be seen that the residuals are dominated by systematic effects
for these high-\SN{} spectra.
Fig.~\ref{fig:T-convolution}~c shows the reduced \(\chi^2 / \nu\) goodness-of-fit measure
as a function of \(\sigma\),
where  \(\chi^2\) is the summed squared normalized residuals
and \(\nu\) is the number of degrees of freedom
(number of data points minus number of model parameters).
Considering the entire line profile, the best fit (minimum \(\chi^2\))
is obtained for a smoothing kernel with rms width
\(\sigma = \qty{7.5}{\kms}\).

However, there is a tension between the behavior of the core
and the wings of the line, with the wings favoring a smaller value
of \(\sigma = \qty{5.8}{\kms}\) (profile~A in Fig.~\ref{fig:T-convolution}a)
and the core a slightly larger value of \(\sigma = \qty{7.9}{\kms}\)
(profile~B).
Given the values of \(\chi^2 / \nu \gg 1\), none of the fits are formally acceptable,
indicating that this one-parameter model is too simplified to
fully reproduce the observations, with the discrepancies being greatest
in the core and flanks of the line.
One possible reason for this is apparent from close inspection of the
``No broadening'' profiles in the figure:
the FWHM of the observed \oiii{} line is
already larger than the \Ha{} line,
which indicates a larger mean expansion velocity in \oiii{},
due to radial gradients in physical conditions
within the nebular shell.
Convolution with a gaussian can only ever increase the FWHM,
so this violation of the assumption of equal \(I\expand(v)\)
means that the fit to the flanks of the profile can never be good.
Taking the range of best-fit values between the wings and the core as indicative of the uncertainty in the derived \(\sigma\),
then this method implies a mean \Ha{} temperature of
\(T = \qty{6000(600:3000)}{K}\).

\newcommand\comp{\ensuremath{_\mathrm{comp}}}
In order to check that the tension mentioned above is not biasing
the results, we use a second method based on the two-gaussian decomposition of the line profiles
(see section~\ref{sec:kinematics} and Fig.~\ref{fig:gaussian-components-velocities}).
By analyzing the individual widths, \(\sigma\comp\), of the red-shifted and blue-shifted components,
this allows us to relax the assumption that the expansion profile \(I\expand(v)\)
need be the same between \Ha{} and \oiii{}.
In Fig.~\ref{fig:T-gaussian-fits}, we plot \(\sigma^2\comp(\Ha)\) against \(\sigma^2\comp(\oiii)\)
for all positions within the inner lobes with brightness exceeding 30\% of the peak.
The diagonal lines show the predictions for different temperatures under the assumption that
both \(\sigma\nontherm\) and \(\sigma\ins\) are constant between \Ha{} and \oiii{}.
We find \(T = \qty{5700(1300)}{K}\) by this method,
with no clear trends with respect to position or brightness.
This is fully consistent with the results of the convolution method above,
which strengthens our confidence in the robustness of this mean temperature measurement.

\subsubsection{Two-phase analysis}
\label{sec:two-phase-analysis}
\newcommand\warm{\ensuremath{_\mathrm{warm}}}
\newcommand\cool{\ensuremath{_\mathrm{cool}}}

\begin{figure}
  \textbf{a}\\[\medskipamount]
  \null\hfill
  \includegraphics[width=\linewidth]{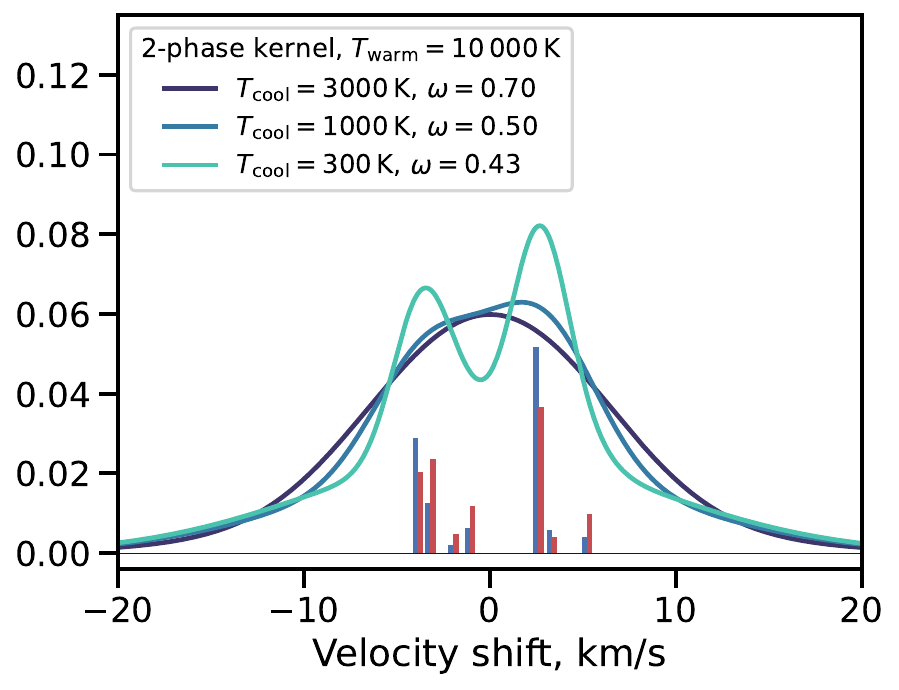}\\
  \textbf{b}\\[\medskipamount]
  \includegraphics[width=\linewidth]{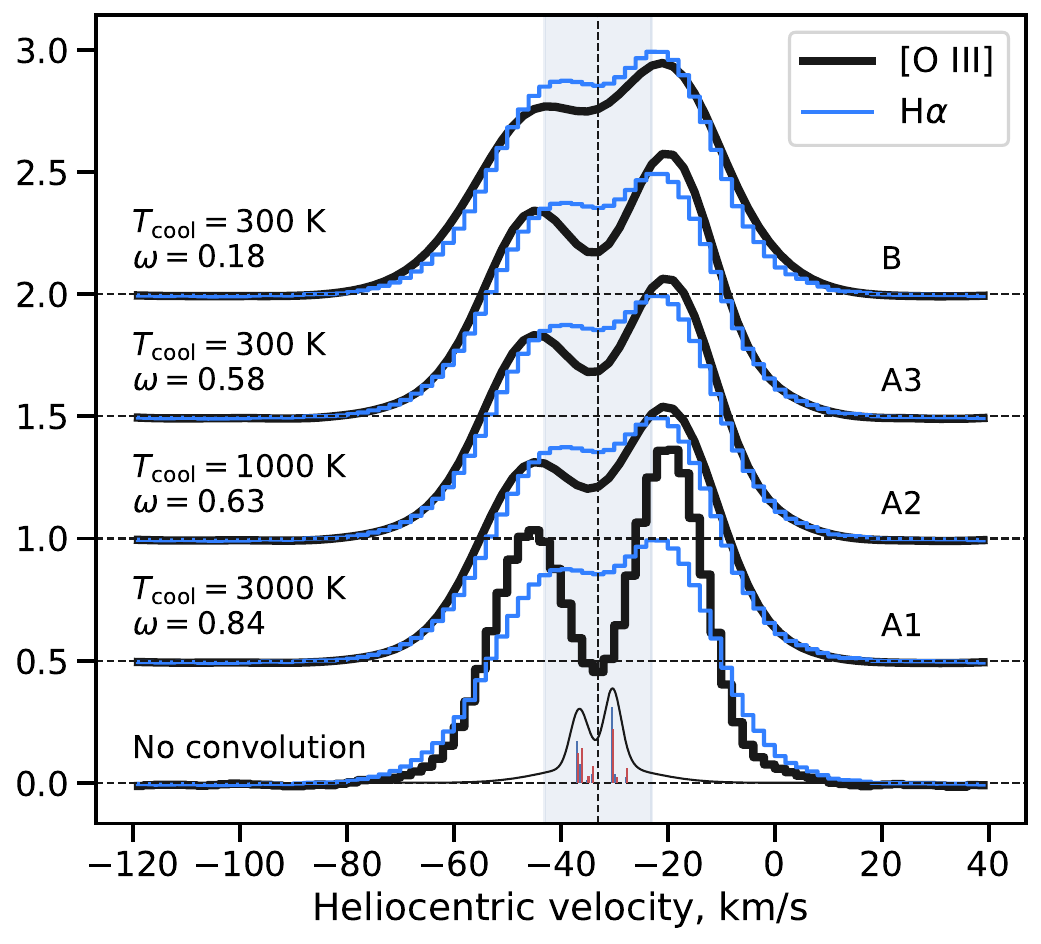}\\
  \caption{
    Two-phase model and fine-structure components \NEW{of the \Ha{} line}.
    (a)~Predicted thermal plus fine-structure profile of the two-phase model with \(T\warm = \qty{10000}{K}\) for different values of \(T\cool\).
    \NEW{The shape of each kernel reflects the combined thermal broadening and fine-structure splitting of the \Ha{} components weighted by the cool- and warm-phase emissivities.}
    The cool-phase emission fraction \(\omega\) is adjusted to give a mean temperature of \(T = \qty{5500}{K}\) in each case, so that all model \NEW{kernel profiles} have approximately the same rms width.
    \NEW{The seven individual fine-structure components of \Ha{}
      \citep{Clegg:1999a} are indicated
      by colored vertical bars below the profiles, with relative intensities for
      \(T = \qty{10000}{K}\) shown in red and \(T = \qty{300}{K}\) in blue
      (assuming Case~B and \(n = \qty{100}{\pcc}\) for both).
      The zero point of the velocity axis corresponds to the mean \Ha{} rest wavelength,
      \(\lambda_0\), which varies slightly with temperature as the relative intensities
      of the components change
      (from \(\lambda_0 = \qty{6562.791}{\angstrom}\) at \qty{10000}{K}
      to  \(\lambda_0 = \qty{6562.797}{\angstrom}\) at \qty{300}{K}).
      This is what causes the small horizontal shift of \qty{0.25}{\kms}
      between the red and blue bars,
      even though the absolute wavelengths of the components are fixed.
    }
    (b)~As Fig~\ref{fig:T-convolution}a, but for \NEW{the} two-phase model.
    Model-convolved profiles A1, A2, and A3 are fitted to the line wings,
    while B is fitted to the line core.
    \NEW{Note that the horizontal velocity scale is different between the two panels.
      Despite the apparent similarity, the double-peaked profiles in panel~b
      are not related to the double-peaked convolution kernel shown in panel~a.
      An example kernel is shown on the same scale at the bottom of panel~b
      (thin line and red/blue vertical bars)
      to emphasize this point. 
    }
  }
  \label{fig:2-phase-convolution}
\end{figure}

The analysis of the previous section implies a mean H temperature of
\(T \lesssim \qty{6000}{K}\), which is significantly lower than the values of \(T \ge \qty{10000}{K}\)
derived from collisional lines \citep{Corradi:2014a, Corradi:2015a}.
In addition, temperature diagnostics that involve metal recombination lines
\citep{Peimbert:2013a, Peimbert:2014a} imply \(T \le \qty{5000}{K}\).
All this implies the existence of significant variation in the temperature
within the nebula, which can be most simply modeled by the co-existence of two phases,
with temperatures \(T\warm\) and \(T\cool\),
as discussed above.
The steep temperature dependence of collisional line emissivities means
that they will be emitted almost exclusively by the warm phase,
so we fix \(T\warm =  \qty{10000}{K}\).
The model is then specified by two parameters:
\(T\cool\) and \(\omega\), defined as the fraction
of the \Ha{} emission that is emitted by the cool phase
\citetext{e.g., \citealp{Garcia-Rojas:2022a}}.
The mean H temperature of the model is therefore
\begin{equation}
  \label{eq:two-phase-Tmean}
  T = \omega T\cool + (1 - \omega) T\warm .
\end{equation}

For sufficiently low temperatures,
the thermal broadening of \Ha{} becomes smaller than the fine-structure splitting,
so that the line profile is highly non-gaussian
and equation~\eqref{eq:fs-ha-oiii} is no longer accurate.
We therefore explicitly calculate the convolution kernel for the two-phase
model, using the fine-structure component intensities from \citet{Clegg:1999a},
which we interpolate on \(\log_{10} T\) from their Table~3,
assuming Case~B emissivities and \(n = \qty{100}{\pcc}\).
Model kernels are shown in Fig.~\ref{fig:2-phase-convolution}a
for different combinations of \(T\cool\) and \(\omega\),
which all correspond to the same mean temperature of \(T = \qty{5000}{K}\).
\NEW{Further technical details are given in the figure caption.}
It can be seen that the profile is approximately gaussian for \(T\cool = \qty{3000}{K}\)
but an asymmetry in the core becomes apparent for \(T\cool = \qty{1000}{K}\),
while for \(T\cool = \qty{300}{K}\) there is a clear division into
extended wings from the warm phase and a narrow double-peaked core
from the cool phase, corresponding to two groups of fine-structure components
separated by \(\approx \qty{6}{\kms}\) \NEW{(indicated by colored bars in the figure)}.

The results from fitting this model to the observed line profiles are illustrated
in Fig.~\ref{fig:2-phase-convolution}b.
The observed profiles are shown at bottom,
\NEW{together with an example model smoothing kernel on the same scale
  (thin line)},
while the convolution of the \oiii{}
profile with different smoothing kernels are shown above.
Models A1, A2, and A3 are fitted to the wings,
using the different values of \(T\cool\) shown in panel~a and optimizing for \(\omega\),
while Model~B is fitted to the core.
It can be seen that the respective profiles are almost identical to the
one-phase models A and B from Fig.~\ref{fig:T-convolution}a,
showing very similar patterns in the residuals.
It is not possible to individually constrain \(T\cool\) and \(\omega\)
from this comparison because different models with the same mean \(T\)
from equation~\eqref{eq:two-phase-Tmean} give very similar results
when convolved with the observed \oiii{} profile.
The expansion, non-thermal, and instrumental broadening completely washes out
the distinctions that are apparent in Fig.~\ref{fig:2-phase-convolution}a.

\subsubsection{High-resolution slit}
\label{sec:high-resolution-slit}

\begin{figure}
  \textbf{a}\\[\medskipamount]
  \null\hfill
  \includegraphics[width=0.91\linewidth]{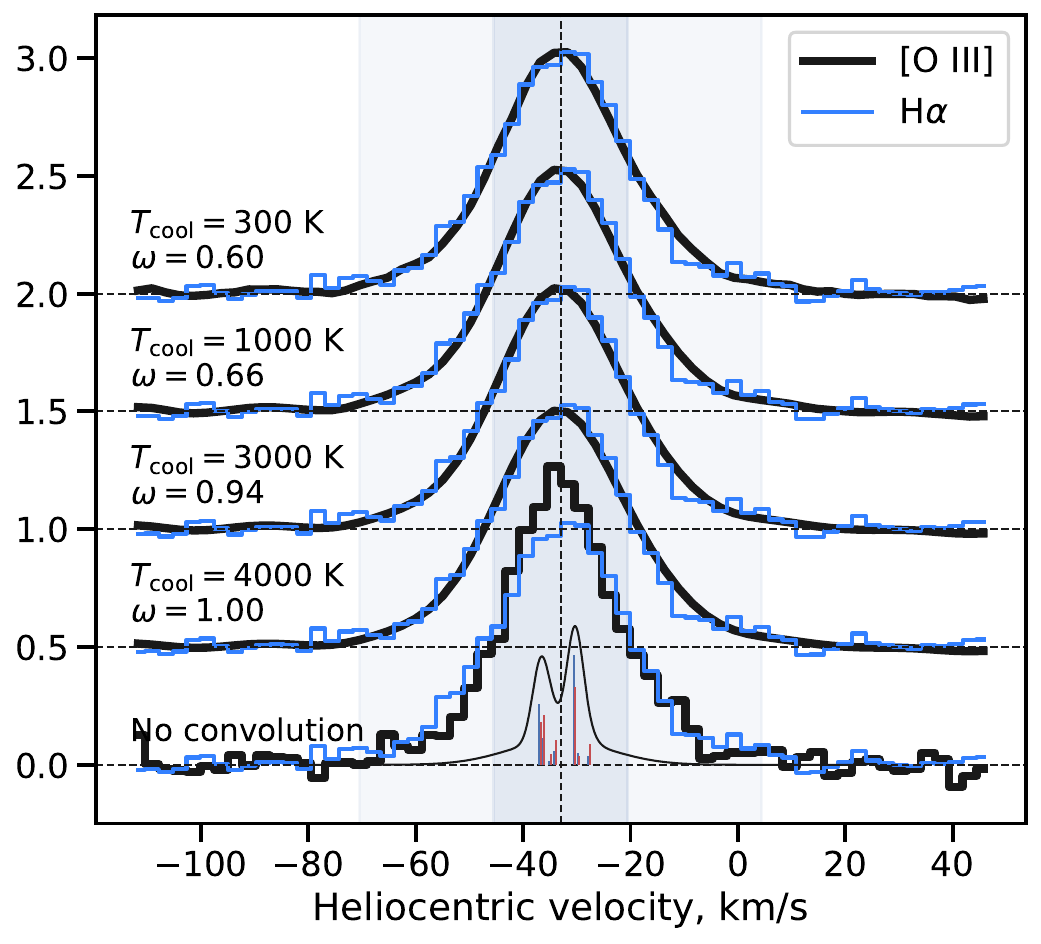}\\
  \textbf{b}\\[\medskipamount]
  \null\hfill
  \includegraphics[width=0.93\linewidth]{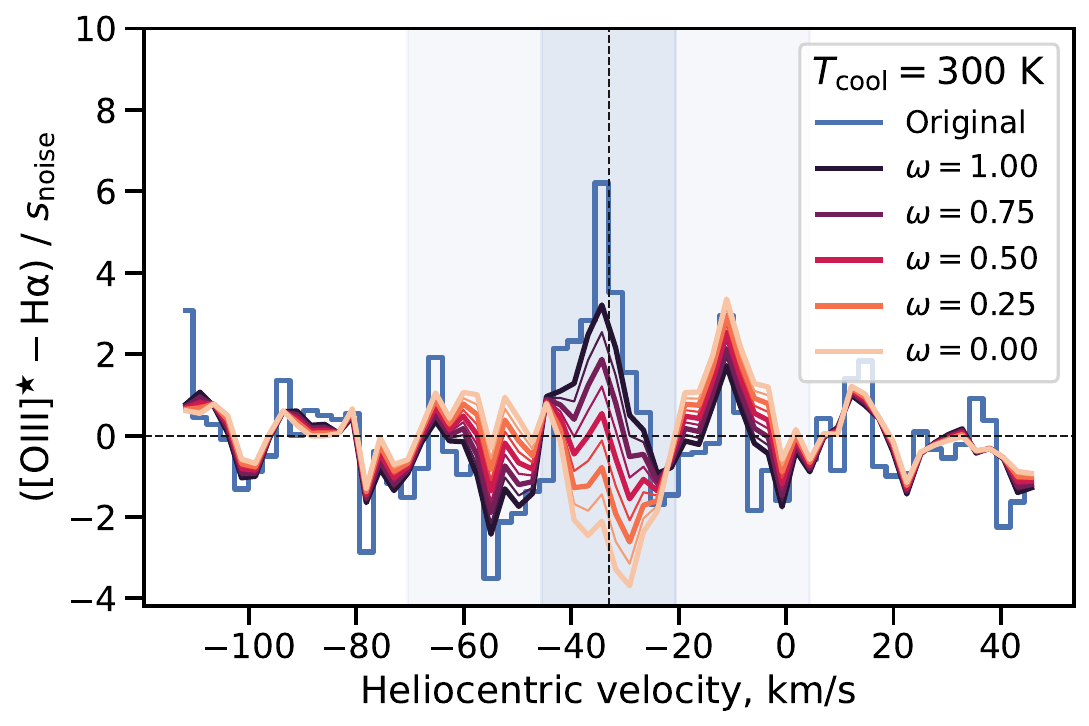}\\
  \textbf{c}\\[\medskipamount]
  \null\hfill
  \includegraphics[width=0.94\linewidth]{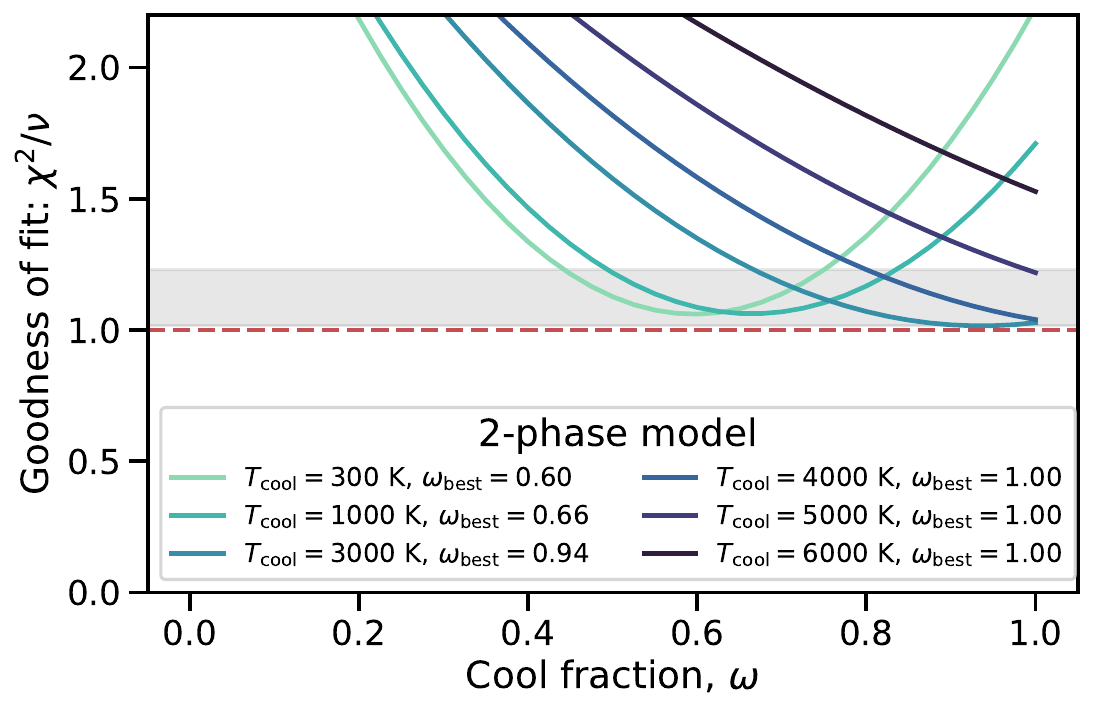}\\
  \caption{
    Application of the two-phase convolution analysis to
    the narrowest observed line profiles.
    (a)~As Fig.~\ref{fig:2-phase-convolution}b but for
    single-peaked profile from slit position~b.
    Light blue colored background shading indicates the core and wings of the line.
    Fits are shown that optimize \(\omega\) for distinct values of \(T\cool\).
    \NEW{The individual fine-structure components (vertical bars) and
    an example smoothing kernel (thin line) are shown at bottom.}
    (b)~Normalized residuals between \Ha{} and model-convolved \oiii{}
    (see Fig.~\ref{fig:T-convolution}b).
    Blue histogram-style line shows \((\oiii - \Ha) / s\noise\)
    for the observed profiles,
    while black-red-yellow continuous lines show the result of convolving
    \oiii{} with two-phase model kernels with \(T\cool = \qty{300}{K}\)
    and decreasing cool-phase fractions \(\omega\).
    (c)~Reduced \(\chi^2\) goodness-of-fit measure as a function of \(\omega\)
    for distinct values of \(T\cool\).
    Gray shaded box indicates the range of acceptable models at
    the 95\% confidence level.
  }
  \label{fig:70-micron}
\end{figure}

\begin{figure}
  \includegraphics[width=\linewidth]{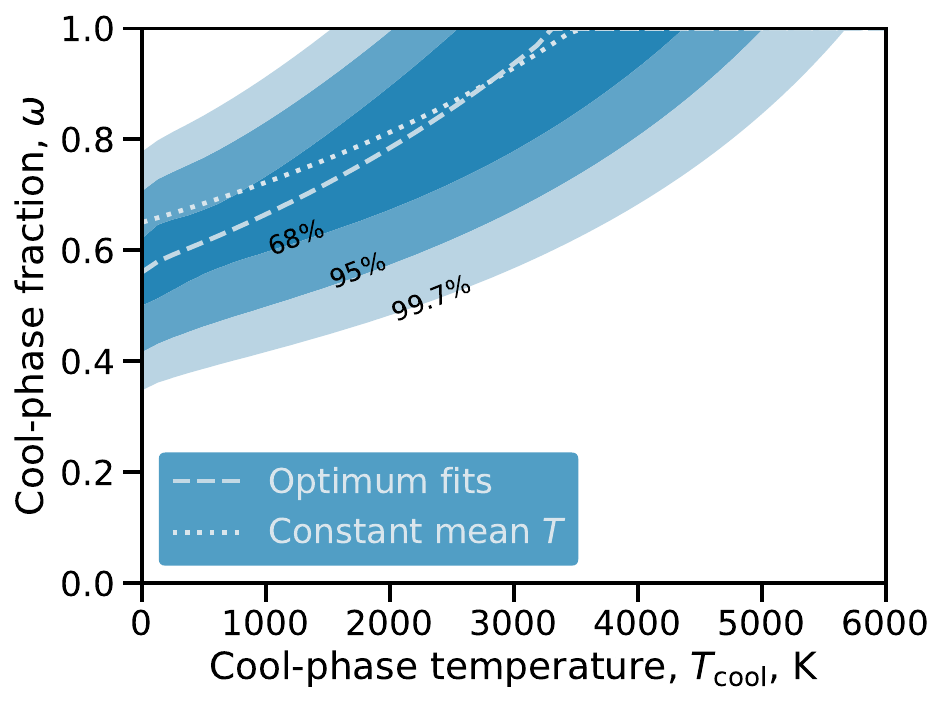}\\
  \caption{
    Confidence intervals for the two-phase model parameters based on fits to the
    \qty{70}{\micron} spectra (data degrees of freedom: \(\nu = 29\)).
    Contours of difference in \(\chi^2\) statistic with respect to best-fit value
    (\(\chi^2 = 29.3\) at \(T\cool = \qty{3430}{K}\), \(\omega = 1\)).
    From dark to light:
    \(\Delta\chi^2 = 2.3\) (1-\(\sigma\), 68\% confidence);
    \(\Delta\chi^2 = 6.2\) (2-\(\sigma\), 95\% confidence);
    \(\Delta\chi^2 = 11.8\) (3-\(\sigma\), 99.7\% confidence).
    Dashed line indicates the maximum-likelihood values of \(\omega\)
    as a function of \(T\cool\).
    Dotted line shows a naive application of equation~\eqref{eq:two-phase-Tmean}, assuming a mean temperature
    \(T = \qty{3500}{K}\).
  }
  \label{fig:2-phase-ci}
\end{figure}

In order to address the issues raised in the previous section and to better
constrain the two-phase model parameters,
it is important to minimize the contributions from additional broadening
mechanisms: non-thermal motions, expansion, and instrumental effects.
To that end, we analyze the spectra from slit position~b
(see Fig.~\ref{fig:sp}), which were obtained with a narrower \qty{70}{\micron} slit.
In addition to having a sharper instrumental profile (\(\sigma\ins \approx \qty{2.5}{\kms}\)),
the slit's position at the edge of the inner lobe means that
the expansion profile \(I\expand(v)\) is single-peaked, since the shell expansion there is predominantly in the plane of the sky.
Consequently, \(\sigma\expand\) is greatly reduced.
The drawback, however, is that these spectra are significantly noisier
than those from the co-added \qty{150}{\micron} slits analyzed previously.
The peak \(\SN = 44\) and \(33\) for \Ha{} and \oiii{}, respectively.

Fig.~\ref{fig:70-micron} presents the results of a convolution analysis
of the two-phase model applied to slit~b,
paralleling those shown in the previous sections.
Unlike the broad double-peaked profiles from the entire inner nebula
(Figs.~\ref{fig:T-convolution}ab and \ref{fig:2-phase-convolution}b),
the narrower, single-peaked profile from slit~b
shows no significant systematic mismatch between the \Ha{} and \oiii{} kinematics.
As a result, the convolved \oiii{} profiles provide good fits to the \Ha{} profile
(Fig.~\ref{fig:70-micron}a)
with small residuals (e.g., red-brown lines in Fig.~\ref{fig:70-micron}b)
and the reduced-\(\chi^2\) statistic is of order unity (Fig.~\ref{fig:70-micron}c).

In Fig.~\ref{fig:2-phase-ci} we show joint confidence intervals for
the two-phase model parameters \(T\cool\) and \(\omega\),
with contours representing the \mbox{1-\(\sigma\)}, \mbox{2-\(\sigma\)}, and \mbox{3-\(\sigma\)} confidence levels.
These contours correspond to specific values of \(\Delta \chi^2\)
(\(2.3\), \(6.2\), and \(11.8\) respectively),
where \(\Delta \chi^2\) is equal to \(-2\) times the logarithm of the likelihood ratio between each parameter set and the best-fit model
\citetext{see, for example, section~11.5 of \citealp{Bevington:2003a}}.
The dashed line shows the best-fit \(\omega\) for each value of \(T\cool\),
while the dotted line shows the naive prediction of equation~\eqref{eq:two-phase-Tmean},
assuming that the mean temperature is constant at \(T = \qty{3500}{K}\).
The prediction approximates the full results,
but tends to overestimate \(\omega\) at low values of \(T\cool\),
implying that the mean temperature of the best-fit model increases
slightly as \(T\cool\) decreases.
Such a disagreement with equation~\eqref{eq:two-phase-Tmean}
is expected due to
(i)~the non-gaussian smoothing kernel, and 
(ii)~the slight temperature dependence in the relative intensity of the
fine-structure components \citep{Clegg:1999a}.

Although a range of different models
are consistent with the line profile data,
the contours allow us to place constraints
at the 95\% confidence level of 
\(T\cool < \qty{5000}{K}\) and \(\omega > 0.4\).

\subsection{Electron densities and ionized masses of the nebular components}
\label{sec:phys-param-nest}

The fourth pair of rows in Table~\ref{tab:shells} lists the root-mean-square electron density, \(n\elec\),
and ionized mass, \(M_\mathrm{i}\), of each component.
These are derived from the \Ha{} surface brightnesses, measured from the images
and position-velocity diagrams,
which are calibrated to physical units using the total
extinction-corrected nebular flux (see section~\ref{sec:dist-determ}).
The \Ha{} surface brightness is converted to an emission measure (\(\mathrm{EM} = \int n_\mathrm{e} n_\mathrm{p} \,dz\))
using the emission coefficient for \(T = \qty{6000}{K}\) \citep{Osterbrock:2006a}.
Assuming \(\mathrm{He}/\mathrm{H} = 0.18\) and \(\mathrm{He^{++}/He} = \mathrm{He^{+}/He} = 0.5\)
\citep{Corradi:2014a}, then \(n_\mathrm{e} = 1.27 n_\mathrm{p}\) and we find
\begin{equation}
  \label{eq:density-from-em}
  n\elec =  1.13 \left (\frac{\mathrm{EM}}{z_\mathrm{los}} \right)^{1/2}
\end{equation}
where \(z\Sub{los} =(2 / \sin i) \sqrt{r\Sub{max}^2 - r\Sub{min}^2}  \)
is the peak line-of-sight depth through the component. 
It is noteworthy that the electron density that we derive for the inner shell from this method,
\(n\elec(\Ha) = \qty{150(10)}{\pcc}\), is very similar to the values derived from the \sii{} nebular lines:
\(n\elec(\sii) = \qty{150(40)}{\pcc}\) \citep{Corradi:2014a} or \qty{150(300:100)}{\pcc} \citep{Corradi:2015a}.
Collisional lines from other ions are mostly consistent with this: \(n\elec([\ion{Ar}{IV}]) = \qty{300(800:200)}{\pcc}\),
with the exception of \(n\elec([\ion{O}{II}]) = \qty{1200(550:350)}{\pcc}\), 
but that may be contaminated by a recombination contribution to the lines
\cite[section~5.5.1 of][]{Wesson:2018a}.


Ionized masses for each component are found by multiplying the derived proton density by the mean mass per proton and by the cylindrical volume of each component.
Interestingly, despite its faintness, the outer shell is a comparable mass to the inner shell.

\subsection{Temperature and luminosity of the central star}
\label{sec:temp-lumin-centr}

We perform a classical Zanstra analysis of the nebula and its central star \citep{Zanstra:1931a}.
\newcommand\Fstar{\ensuremath{\lambda F_{\lambda,\star}}}
This makes use of 3 observed quantities: the continuum flux of the ionizing star,
\(\Fstar \),
together with the total flux of the nebula in a \heii{} recombination line,
e.g., \(F(\Wav{4686})\),
and an \hi{} recombination line,
e.g., \(F(\Hb)\),
(see \citealp{Osterbrock:2002a} for an historical overview).
The analysis proceeds by comparing different estimates of the stellar luminosity parameter
\citep{Harman:1966a}
\begin{equation}
  \label{eq:big-lambda}
  \Lambda \equiv \frac{L_\star / \,\unit{\lsun}}{(D / \,\unit{kpc})^2},
\end{equation}
which are derived from each independent observed quantity as
a function of the assumed stellar effective temperature \(T_\star\).
Under the assumption that the \chem{He^+}-ionizing continuum is optically thick,
we first compare the estimates
\begin{equation*}
  \label{eq:2}
  \Lambda_\star\left[ \Fstar ;\ T_\star \right]
  \quad \text{versus} \quad
  \Lambda_\heii\left[F(\Wav{4686});\ T_\star \right]
\end{equation*}
in order to find the value of \(T_\star\) where the two estimates are equal.
We determine the continuum flux of the CSPN from the photometric light curves around primary total eclipse
(\citealp{Jones:2022a}, Fig.~2),\footnote{\label{fn:photometry}%
   We use the \(g\)-band observations since the central wavelength
  (\(\lambda \sim \qty{4686}{\angstrom}\)) is close to the \heii{} line. 
  We assume that the flat minimum of the primary eclipse
  (\(|\phi| < 0.03; m_g = 23.3\))
  corresponds to the light from the secondary star alone,
  whereas the phase immediately before/after the eclipse
  (\(|\phi| \approx 0.05; m_g = 19.7\))
  corresponds to the sum of the secondary and the primary.
  This implies a primary magnitude of \(m_g = \num{19.75(5)}\). 
  Since the secondary is in front of the primary in this part of the orbit,
  the illuminated face of the secondary is hidden and irradiation effects are minimized.
  We adopt the AB magnitude zero point of \qty{3631}{Jy} \citep{Oke:1983a} to derive
  \(\log_{10} \Fstar = -0.4 (m_g - A_g) - 4.6340\), where the \(g\)-band extinction
  is estimated as \(A_g = 2.5 c(\Hb)\), with \(c(\Hb) = \num{0.94(10)}\) \citep{Corradi:2015a}.
}
yielding \(\Fstar = \qty[separate-uncertainty-units = bracket]{2.57(12)(58)e-12}{\cgsflux}\).
Note that two uncertainties are given, with the first being purely photometric random error
and the second being the systematic uncertainty in the extinction correction.

From the spectrophotometry of \citet{Corradi:2015a},
combined with the extinction-corrected \Ha{} flux from \citet{Corradi:2014a},
We determine \(F(\Wav{4686}) = \qty[separate-uncertainty-units = bracket]{9.0(1.3)(2.1)e-13}{\cgsflux}\). 
We then calculate \(\Lambda_\star\left[ \Fstar ;\ T_\star \right]\) assuming a black-body stellar spectrum
and \(\Lambda_\heii\left[F(\Wav{4686});\ T_\star \right]\),
assuming an area covering fraction\footnote{Estimated from the geometry of the inner shell.}
of \chem{He^{++}}-thick gas of \(\xi = \num{0.83(8)}\)
and that a fraction \(f_{4686} = 0.26\) of \chem{He^{++}} recombinations
produce a \Wav{4686} photon.\footnote{%
  Calculated using pyneb \citep{Luridiana:2015a} for Case~B with \(n = \qty{150}{\pcc}\), \(T = \qty{6000}{K}\)
  using atomic data from \citet{Storey:1995a}.}
The two estimates coincide for \(T_\star = \qty{103(4:3)}{\kilo\kelvin}\), with
\(\Lambda = 24.9\), which implies a stellar luminosity
\(L_\star = \qty{400(92)}{\lsun}\) at \(D = \qty{4}{kpc}\).
\NEW{This can then be combined with the \Hb{} flux to estimate the fraction, \(\eta\), of
  H-ionizing photons that are absorbed in the nebula, yielding \(\eta \approx 0.20\)
  for Ou~5, confirming that the nebula is optically thin at \qty{1}{Rydberg}.}

To check how sensitive \NEW{the derived temperature and luminosity are}
to the assumption of a black-body spectral shape,
we repeated the analysis using metal-line blanketed NLTE model atmospheres \citep{Rauch:2003a}.
The model spectra at \(T_\star \sim \qty{1e5}{\kelvin}\) are more strongly peaked below
\(h\nu \lesssim \qty{50}{eV}\) than a black body, which reduces the flux both in the
visual range (\(h\nu \approx \qty{2.5}{eV}\)) and above the \chem{He^+}-ionization edge
(\(h\nu > \qty{54.4}{eV}\)) for a given bolometric luminosity.
As a result, the Zanstra analysis for Ou~5 requires a slightly higher stellar temperature
and luminosity, \(T_\star \approx \qty{115}{kK}\) and \(L_\star \approx \qty{660}{L_\odot}\).
These results firmly place the central star above \(10^5\)~K, favouring a post-AGB interpretation and contrasting with the much lower \(T_\star \simeq 67\)~kK found by \citet{Jones:2022a}; we return to this issue in Section~\ref{sec:evol-stat-centr}.

\NEW{Although the \heii{} Zanstra method is generally considered to be a reliable
  way of determining the central star temperature in high-excitation PNe
  \citep{Villaver:2003a, Moreno-Ibanez:2016a},
  it is worth considering ways in which it might go wrong.
  \citet{Gruenwald:2000a} calibrated the method against nebula photoionization models
  and provide two-dimensional diagnostic diagrams of \(T\Sub{Z}(\heii)\) versus
  \(\heiilam / \ion{He}{I}\ \Wav{5876}\),
  which confirm that Ou~5 is in the regime where \(T\Sub{Z}(\heii) = \Teff\) to a very good approximation. 
  Another concern is that the line ratios from the spectrograph slits of
  \citet{Corradi:2014a, Corradi:2015a} may not be representative of the entire nebula.
  We therefore repeated the analysis using the much weaker \heiiredlam{} line
  that is detected near \Ha{} in our high resolution spectra (see section~\ref{sec:nebular-morphology}),
  finding an identical result within the observational uncertainties. 
  }

\section{Cloudy photoionization models}
\label{sec:cloudy-phot-models}

To further explore the physical conditions in Ou~5, we computed a series of photoionization models using the \textsc{cloudy} plasma physics and spectral synthesis code
\citep{Ferland:1998a, Ferland:2013a, Ferland:2017a, Chatzikos:2023a, Gunasekera:2025a}. 
The models are intended as illustrative rather than exhaustive, and are constrained to reproduce the observed \Ha{} flux of the nebula,
with input parameters that are summarised in Table~\ref{tab:cloudy-input}.
They assume spherical symmetry with inner/outer radii matched to the inner shell, and they adopt the stellar effective temperature and luminosity derived above in section~\ref{sec:temp-lumin-centr}. 
The gas-phase abundances are based on the empirical determinations of \citet[][Table~8]{Corradi:2015a}, with the warm-phase composition (derived from collisional lines) used as the reference set
for He, C,\footnote{
  Direct carbon abundances from collisional lines requires UV observations,
  which are not available for Ou~5.
  Therefore we use the ionic \(\chem{C^{+2}} / \chem{H^+}\)
  derived from recombination lines \citep[][Table~7]{Corradi:2015a},
  together with an ionization correction factor of \(1 \pm 0.2\)
  \citep[][Fig.~14, given that \(\chem{O^{+2}} / (\chem{O^+} + \chem{O^{+2}}) \approx 0.9\) for Ou~5]{Delgado-Inglada:2014b}
  and assume that \chem{C / O} is the same in the cool and warm phases
  to derive \(12 + \log_{10}(\chem{C / H}) = 8.13 \pm 0.14\) for the warm phase.
}
N, O, Ne, S, Cl, and Ar,\footnote{
  In summary, these are roughly \(2\, \times\) solar for \chem{He} and \(0.5\, \times\) solar
  for most metals (referenced to \citealp{Asplund:2009a} values).
}
supplemented by ISM values (scaled to the empirically measured \chem{O/H}) for other elements not directly measured.  

As a baseline, we first constructed chemically homogeneous models with constant gas pressure. 
These provide a useful reference but are not expected to be realistic given the strong evidence for thermal inhomogeneities (Section~\ref{sec:estim-nebula-kinet}). 
The homogeneous models predict electron temperatures
(green line in Fig.~\ref{fig:cloudy-p-Z06}b)
that are \numrange{1.5}{2} times higher than the empirically derived
warm-phase temperature of \(\approx \qty{1e4}{K}\).
As a result, such models over-predict the intensity of strong forbidden lines such as \oiiilam{}.
At the same time, they systematically underpredict the strengths of optical recombination lines (ORLs) such as \oii\ \Wavv{4651}.  

We therefore considered models in which the metal abundances vary with radius
\citep{Kingdon:1998a}, motivated by the two-phase interpretation of the line profiles. 
At each radius \(r\), the abundance of all metals is multiplied by a factor
\begin{equation}
  \label{eq:Z-fluctuation}
  Z(r) = \tfrac12 (Z_0 + 1) + \tfrac12 (Z_0 - 1)
  \cos \Bigl[2 \pi \bigl(\tfrac{r - r_{\mathrm{min}}}{r_0} \bigr)\Bigr] ,
\end{equation}
where \(r_{\mathrm{min}} = \qty{0.06}{pc}\) is the inside radius of the inner shell,
and \(r_0\) is the spatial period of the fluctuation. 
Here \(Z_0\) denotes the \emph{metallicity contrast}, i.e.~the amplitude of the 
abundance variation relative to the baseline value,
which is similar to the Abundance Contrast Factor (ACF) defined by \citet{Gomez-Llanos:2020a}.
Two prescriptions were explored for the spatial period:
in the ``large-scale fluctuation'' models (\(r_0 = \qty{0.16}{pc}\))
the metallicity varies gradually over radial scales comparable to the shell thickness,
whereas in the ``small-scale fluctuation'' models
(\(r_0 = \qty{0.016}{pc}\),
the abundances oscillate more rapidly with depth
(five spatial periods over the width of the inner shell).

\begin{table}
  \centering
  \caption{Input parameters adopted for the \textsc{cloudy} photoionization models of the inner shell of Ou~5. See text for details.}
  \label{tab:cloudy-input}
  \renewcommand\arraystretch{1.4}
  \begin{tabular}{@{\ } P{0.2\linewidth} P{0.73\linewidth} @{\ }}
    \toprule
    Parameter & Value / Description \\
    \midrule
    Geometry & Spherical shell; inner radius \(r_{\min} = \qty{0.06}{pc}\); outer radius \(r_{\max} = \qty{0.14}{pc}\) (Section~\ref{sec:nebular-geometry}) \\
    Central star & \(T_\star = \qty{115}{kK}\), \(L_\star = \qty{660}{L_\odot}\), \(\log g = 6.66\), Rauch stellar atmosphere (Section~\ref{sec:temp-lumin-centr}) \\
    Density law  & Constant density (\(n_{\rm H}\)) or constant thermal pressure (\(P/k = (n\elec + \sum_{i \in \text{ions}} n_i) T\)) \\
    Normalization & All models are tuned to reproduce the fraction
                    \num{0.74} of the total nebular
                    \Ha{} flux that we ascribe to the inner shell,
                    assuming an area covering fraction for the shell of \num{0.83}. \\
    Abundances   & Warm-phase values from \citet{Corradi:2015a}; ISM-scaled values for unmeasured elements; \chem{C/H} via \chem{C/O} of ORLs with ICF (see text); standard ISM dust recipe \\
    Metallicity\newline fluctuations & Homogeneous (\(Z(r)=1\)); large-scale (\(r_0 = \qty{0.16}{pc}\)); small-scale (\(r_0 = \qty{0.016}{pc}\)) \\
    Fluctuation\newline amplitude & Metallicity contrast of \(Z_0 = 6, 15\) (moderate), \(30, 60\) (extreme) \\
    \bottomrule
  \end{tabular}
\end{table}

\subsection{Model results: physical structure}
\label{sec:model-results-phys}

\begin{figure}
  \centering
  \includegraphics[width=\linewidth]{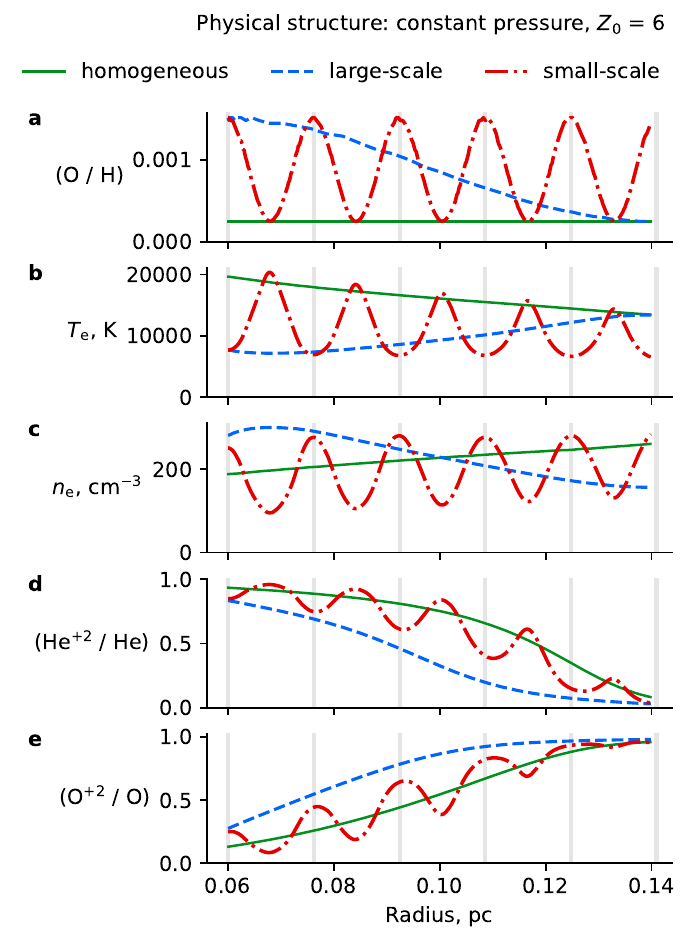}
  \caption{Physical structure of constant-pressure Cloudy models for the
    inner shell of Ou~5:
    constant metallicity (solid green lines),
    large-scale fluctuating metallicity (dashed blue lines),
    and small-scale fluctuating metallicity (dot-dashed red lines).
    For the latter two cases, results are shown for a moderate metallicity contrast of 
    \(Z_0 = 6\).
    (a)~Relative gas-phase oxygen abundance \(\chem{O} / \chem{H}\)
    by number with respect to hydrogen.
    Gray vertical lines in all panels indicate the positions of metallicity peaks in the
    small-scale fluctuation model.
    (b)~Gas temperature, \(T\elec\), which anti-correlates with \(\chem{O} / \chem{H}\).
    (c)~Electron density, \(n\elec\), which anti-correlates with
    temperature within each model, while 
    variation between models is due to all being tuned 
    to reproduce the observed \Ha{} flux of the nebula. 
    (d)~Fraction of doubly ionized helium \(\chem{He^{+2}} / \chem{He}\).
    (e)~Fraction of doubly ionized oxygen \(\chem{O^{+2}} / \chem{O}\).
  }
  \label{fig:cloudy-p-Z06}
\end{figure}

\begin{figure}
  \centering
  \includegraphics[width=\linewidth]{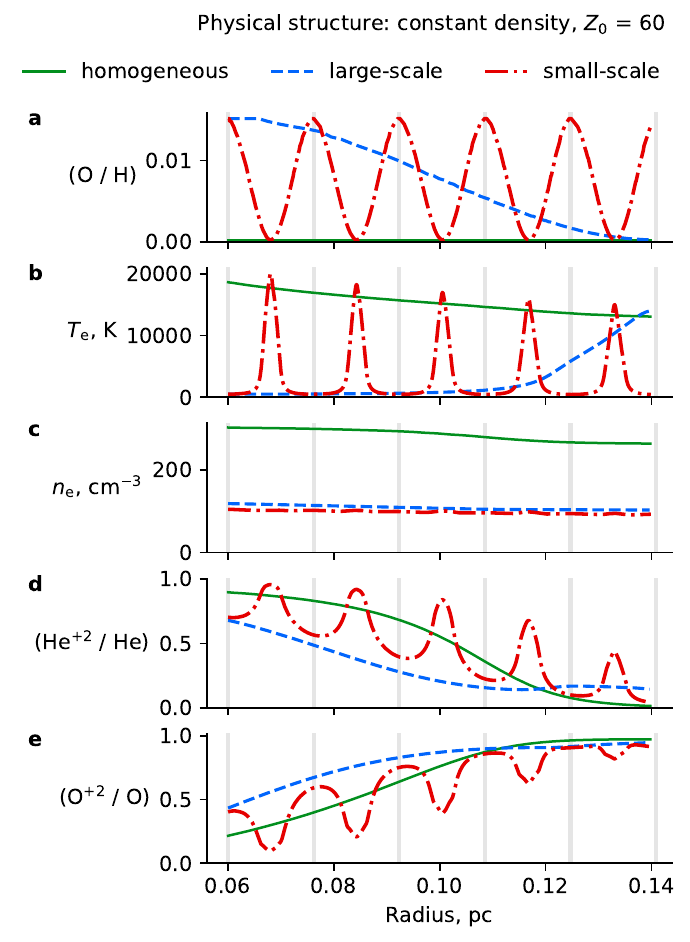}
  \caption{As Fig.~\ref{fig:cloudy-p-Z06} but for physical structure of constant-density Cloudy models with more extreme metallicity fluctuations: \(Z_0 = 60\).}
  \label{fig:cloudy-n-Z60}
\end{figure}

The physical structure of one such pair of models,
calculated at constant gas pressure,
is illustrated in Figure~\ref{fig:cloudy-p-Z06} for \(Z_0 = 6\).
The large-scale fluctuations are shown in blue (dashed line)
and the small-scale fluctuations in red (dash--dotted line),
in comparison with the homogeneous model in green (solid line).
The functional form of the oscillations (panel~a) is such that the highest metallicity occurs at the inner edge,
\(Z(r_{\mathrm{min}}) = Z_0\), while the lowest metallicity, \(Z(r) = 1\),
occurs at \(r = r_{\mathrm{min}} + (k + 1/2)\,r_0\) for \(k = 0, 1, 2, \dots\).
In both fluctuating cases,
the varying metallicity induces an opposite modulation of the temperature
(panel~b),
although in this case the changes are relatively modest:
from \(T \approx \qtyrange{15000}{20000}{K}\) (low metallicity regions)
to \(T \approx \qtyrange{6000}{8000}{K}\) (high metallicity regions).
Because the total pressure is fixed, the electron density varies in anti-correlation with the temperature (panel~c),
\(n_{\mathrm{e}} \propto T^{-1}\), so that the per-ion recombination rate rises steeply with falling temperature,
scaling as \(\alpha(T)\,n_{\mathrm{e}} \sim T^{-1.7}\).
The ion fractions of \chem{He^{+2}} (panel~d) and \chem{O^{+2}} (panel~e) both show modulations as a result,
but with opposite sense.
The enhanced recombination rates at lower temperatures reduce the \chem{He^{+2}} fraction since it is the highest ionization stage.
In contrast, \chem{O^{+2}} is primarily competing with the higher stage \chem{O^{+3}} (at least in the inner half of the shell),
so enhanced recombination in the cooler zones leads to an \emph{increase} in the \chem{O^{+2}} fraction.

Figure~\ref{fig:cloudy-n-Z60} illustrates models with
a more extreme metallicity contrast of \(Z_0 = 60\),
but this time at constant hydrogen density.\footnote{%
  It is unclear whether constant-density or constant-pressure assumptions are more appropriate for the nebula.
  For an age of \(\qty{1e4}{yr}\) (section~\ref{sec:nebular-geometry}),
  the sonic scale -- defined as the size of a region with sound-crossing time equal to the nebular age --
  is \(\qtyrange{0.03}{0.10}{pc}\) for \(T = \qtyrange{1e3}{1e4}{K}\).
  This is comparable to the shell thickness but smaller than its full diameter,
  which suggests that constant-pressure models are applicable to small-scale metallicity oscillations,
  whereas constant-density models remain plausible for larger-scale oscillations.}
In this case the temperature fluctuations (panel~b) are much stronger than in the previous case, with minima reaching \(< \qty{1000}{K}\).
Since all models, regardless of their internal structure, are tuned to reproduce the observed total \Ha{} flux,
the density of these highly fluctuating models is significantly lower than in the homogeneous case (panel~c),
in order to offset the enhanced recombination efficiency.
The ionization variations (panels~d and~e) are similar to
in Figure~\ref{fig:cloudy-p-Z06} since the larger temperature fluctuations
are offset by the shallower temperature-dependence of recombination
rates at constant density compared with constant pressure.

For simplicity our models use the standard ISM grain recipe with constant abundance,
and we find that grain photoelectric heating
is most important in the low-metallicity warm phase
(contributing \(\sim 10\)–40\% of the heating rate)
and negligible in the cool phase.
If the dust abundance were instead to scale with metallicity,
then the cool phase would experience strong dust heating at high-\(Z\),
altering its thermal balance \citep[see][]{Ercolano:2003a},
but we do not explore this idea further here. 

\subsection{Model results: emission lines}
\label{sec:model-results-emiss}

\begin{figure}
  \centering
  \includegraphics[width=\linewidth]{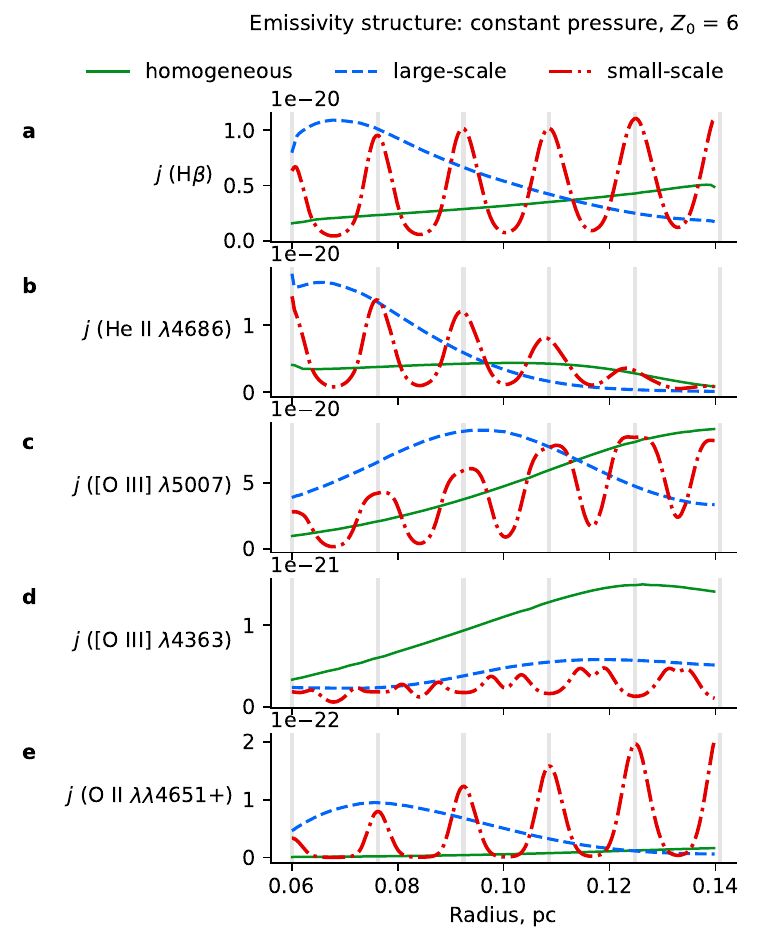}
  \caption{
    Line emissivity structure of the constant-pressure,
    moderate-fluctuation Cloudy models
    shown in Fig.~\ref{fig:cloudy-p-Z06}.
    (a)~\Hb{} hydrogen recombination line.
    (b)~\heii{} \Wav{4686} recombination line.
    (c)~\oiii{} \Wav{5007} forbidden nebular line,
    which is primarily excited by collisions.
    (d)~\oiii{} \Wav{4363} forbidden auroral line,
    which is excited by electron collisions with \chem{O^{+2}}
    at high temperatures, but
    primarily by recombination of \chem{O^{+3}} at low temperatures.
    (e)~\oii{} \Wavv{4651+} permitted line multiplet,
    which is primarily excited by recombination of \chem{O^{+2}}.
  }
  \label{fig:cloudy-p-Z06-emiss}
\end{figure}

\begin{figure}
  \centering
  \includegraphics[width=\linewidth]{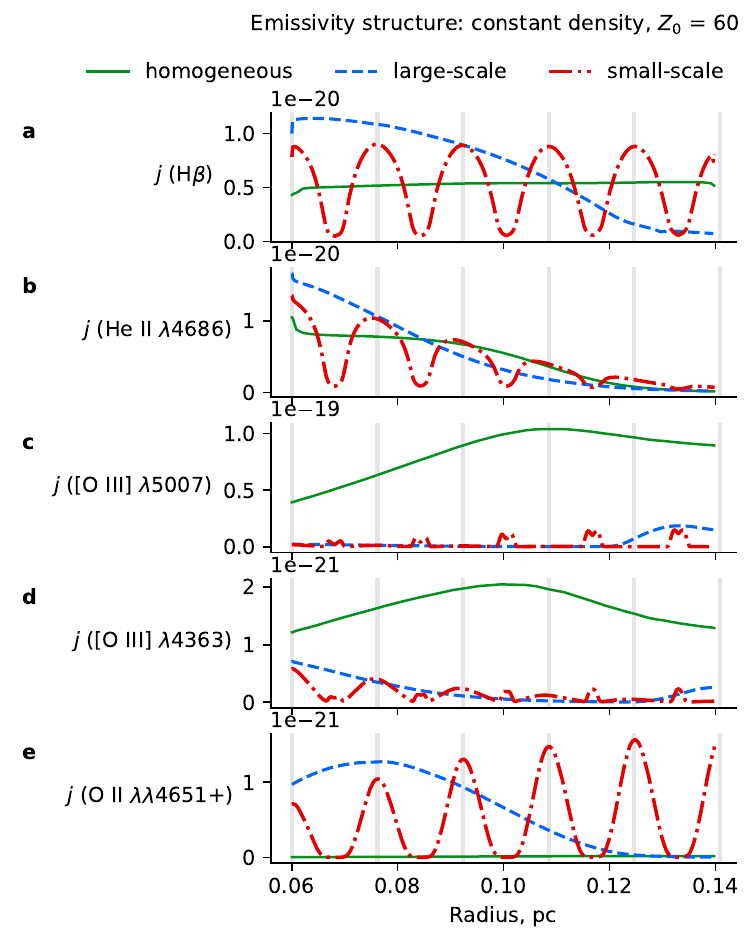}
  \caption{
    As Fig.~\ref{fig:cloudy-p-Z06-emiss} but for emission structure
    of constant-density, extreme-fluctuation Cloudy models shown in Fig.~\ref{fig:cloudy-n-Z60}
    with more extreme metallicity fluctuations: \(Z_0 = 60\).}
  \label{fig:cloudy-n-Z60-emiss}
\end{figure}

The emission structures of the same two sets of models are shown in
Figure~\ref{fig:cloudy-p-Z06-emiss} (moderate fluctuations: constant
pressure, \(Z_0 = 6\)) and Figure~\ref{fig:cloudy-n-Z60-emiss}
(extreme fluctuations: constant density, \(Z_0 = 60\)).
The \Hb{} emissivity (panel~a in each figure) is dominated by recombination,
and so peaks in the cool phase.
\heiilam{} behaves similarly
but is more concentrated toward the inside of the shell
due to the behavior of the \(\chem{He^{+2}}/\chem{He}\) fraction.
Both recombination lines show qualitatively similar behavior in the two sets of models.
In contrast, the forbidden \oiii{} lines differ markedly
between the moderate fluctuation and extreme fluctuation cases.
In the moderate case (Fig.~\ref{fig:cloudy-p-Z06-emiss}c),
the \oiiilam{} nebular line emissivity profile results from a complex combination
of the density, temperature, and ionization variations.\footnote{%
  For the small-scale fluctuations,
  the increase in \(n\elec\) and \chem{O^{+2}} fraction
  in the high-metallicity peaks more than compensate for the reduced temperature,
  resulting in emissivity peaks that coincide with those in \Hb{}.
  For the large-scale fluctuations, a gradually declining \(n\elec\)
  combines with a rising \chem{O^{+2}} fraction and temperature
  to produce an emissivity peak at an intermediate radius.
}
In the case of extreme fluctuations (Fig.~\ref{fig:cloudy-n-Z60-emiss}c)
the \oiiilam{} emission is greatly suppressed throughout the shell:
The high-metallicity zones are now too cool to collisionally excite the line,
while the low-metallicity zones have too low a \chem{O^{+2}} fraction.
The auroral line \oiiialam{} is also suppressed by the metallicity fluctuations,
but to a lesser extent since the recombination of \chem{O^{+3}}
makes an important contribution in the cool phase (Fig.~\ref{fig:cloudy-n-Z60-emiss}d).
Finally, the permitted metal line \(\oii\ \Wavv{4651+}\) (panel~e) is dominated by recombination
and therefore closely follows the oxygen abundance profile,
modulated by the varying \chem{O^{+2}} fraction.

The behavior of forbidden lines from singly-ionized metals is problematic in \emph{all} of our models.
The predicted ratios \(\niilam / \Hb\) and \([\ion{O}{ii}]\ \Wav{3726} / \Hb\) (not illustrated)
for the homogeneous models are lower than the observations by factors of two to three,
and this discrepancy only worsens in the metallicity fluctuation models.
The ratios can be brought into agreement with the observations
by reducing the stellar luminosity,
but the predicted \(\heiilam / \Hb\) ratio then becomes too low. 

\subsection{Model results: trends with metallicity contrast, \boldmath\(Z_0\)}
\label{sec:model-results-trend}

\begin{figure}
  \centering
  \includegraphics[width=\linewidth]{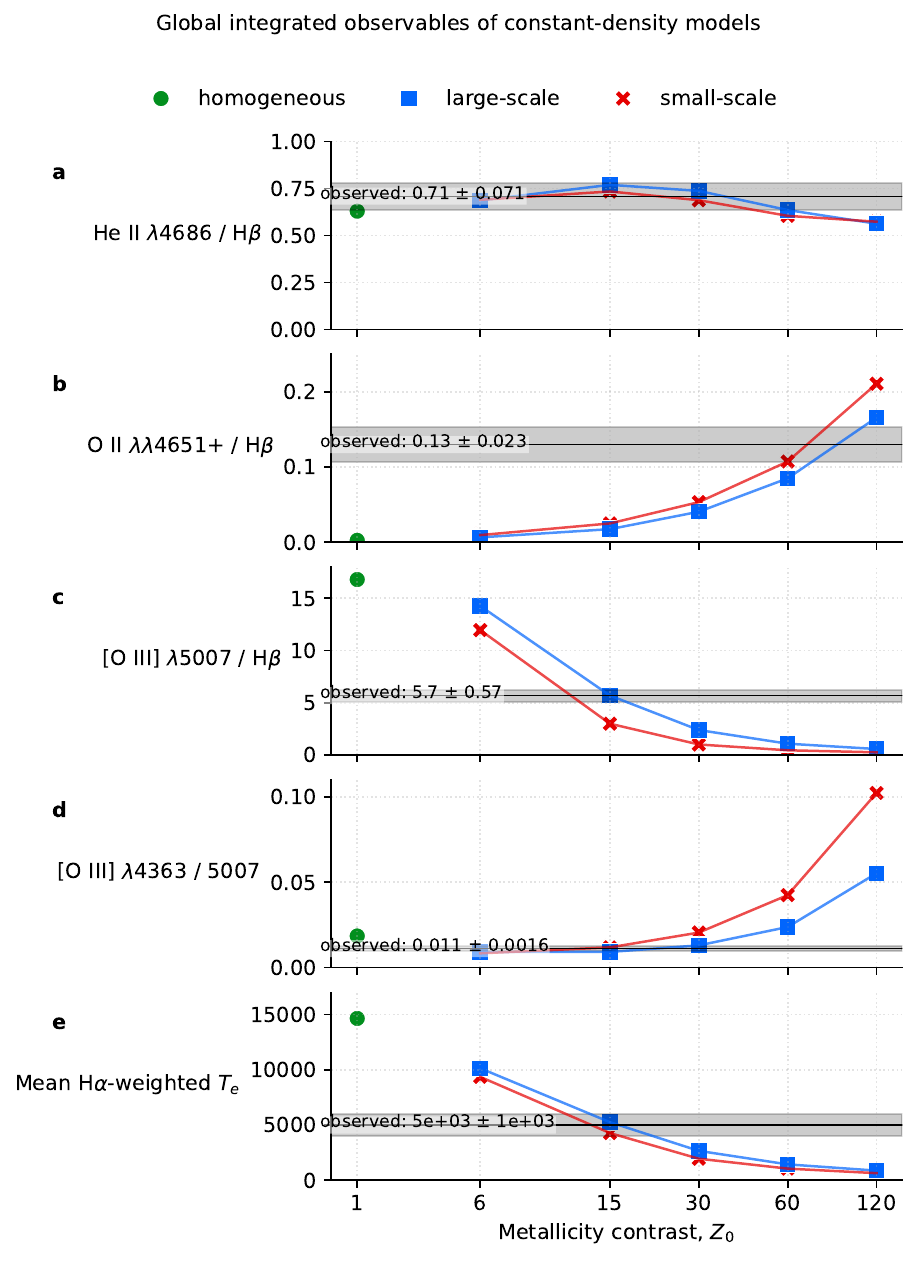}
  \caption{
    Selected predicted emission-line ratios from constant-density Cloudy models of Ou~5 as a function of the metallicity contrast \(Z_0\). 
    Panels show (a)~\(\oii\ \Wav{4651+}/\Hb\), (b)~\(\oiii\ \Wav{5007}/\Hb\),  (c)~\(\oiii\ \Wav{4363}/\Hb\), and (d)~mean \Ha{} temperature. 
    Colours and symbols distinguish the fluctuation scale (homogeneous, short, long). 
    Solid curves join models of the same scale, and the dashed curves in panel~c show results without
    the recombination or charge-exchange contributions to \(\oiii\ \Wav{4363}\). 
    Grey bands mark the observed values with uncertainties (including an additional 10\% allowance for spatial variation across the nebula).}
  \label{fig:cloudy-line-ratios-vs-observed-n}
\end{figure}
\begin{figure}
  \centering
  \includegraphics[width=\linewidth]{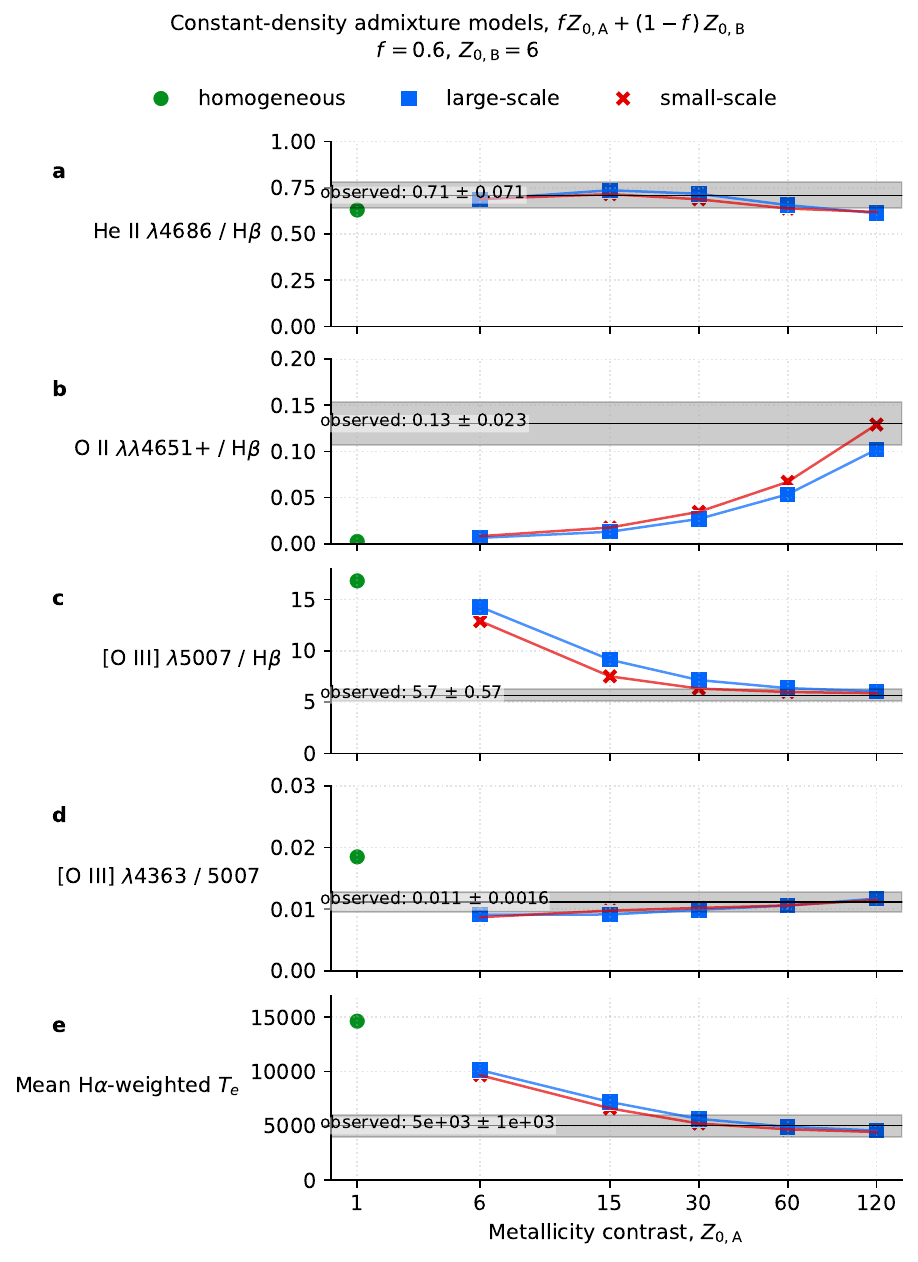}
  \caption{Same as Fig.~\ref{fig:cloudy-line-ratios-vs-observed-n}
    but for a mixture of two different radial fluctuation models.
  }
  \label{fig:cloudy-line-ratios-vs-observed-admixture}
\end{figure}

Figure~\ref{fig:cloudy-line-ratios-vs-observed-n}
summarizes the dependence of selected global diagnostics
on the amplitude of the imposed metallicity fluctuations,
parameterized by the contrast \(Z_\mathrm{max}/Z_\mathrm{min} = Z_0\),
and compare the model predictions with observations.
Each panel shows the integrated value of a diagnostic line ratio
(or mean temperature) for models with homogeneous abundances,
and with large- or small-scale fluctuations of increasing amplitude.
The figure shows constant-density models only,
but constant-pressure models give qualitatively similar results.

The behaviour of the optical recombination line
\(\oii\ \Wavv{4651+}/\Hb{}\) (Fig.~\ref{fig:cloudy-line-ratios-vs-observed-n}~b)
is monotonic with \(Z_0\).
Homogeneous models fall far below the observed ratio,
whereas the degree of enhancement grows steadily with \(Z_0\),
reaching agreement with observations for the most extreme models
(\(Z_0 \gtrsim 60\)),
and being only weakly sensitive to the density law and the spatial scale of the fluctuations.

For the forbidden nebular line ratio \(\oiii\ \Wav{5007}/\Hb{}\) (Fig.~\ref{fig:cloudy-line-ratios-vs-observed-n}~c),
the situation is reversed:
homogeneous models overpredict the observed ratio,
but increasing \(Z_0\) suppresses the \oiii{} line emission.
Agreement with the observed value is reached for
intermediate fluctuation amplitudes
(\(Z_0 \sim \numrange{10}{15}\) for constant density and \(Z_0 \sim \numrange{8}{10}\) for constant-pressure).
Fluctuations more extreme than that greatly suppress \(\oiii\ \Wav{5007}\) as described above (Fig.~\ref{fig:cloudy-n-Z60-emiss}),
resulting in a predicted ratio that is many orders of magnitude below what is observed.

The auroral-to-nebular line ratio \(\oiii\ \Wav{4363}/\Wav{5007}\)
(Fig.~\ref{fig:cloudy-line-ratios-vs-observed-n}~d) is also over-predicted by the homogeneous models,
and as with \Wav{5007} initially falls with \(Z_0\), coming into agreement with
the observations for \(Z_0 = \numrange{6}{15}\).
However, for higher values of \(Z_0\) the ratio rises again,
reaching values as high as \num{0.1} in the constant-density models
and \num{0.3} in the constant-pressure models.
This is primarily due to the increased contribution of radiative recombination of \chem{O^{+3}}
to both lines \citep{Pequignot:1991a} as the cool-phase temperature decreases.
There are also minor contributions from charge exchange excitation and continuum starlight fluorescent pumping  of \chem{O^{+2}} \citep{Morisset:2026a}.

The final panel (Fig.~\ref{fig:cloudy-line-ratios-vs-observed-n}d)
presents the mean \Ha{}-weighted electron temperature,
which decreases systematically with \(Z_0\).
This reflects both the higher fraction of the total recombinations
that occur in cool, metal-rich gas in high-contrast models,
together with the lower temperature of that gas.
Homogeneous models yield temperatures greater than \(\qty{1e4}{K}\),
whereas \(Z_0 = \numrange{10}{15}\) brings the mean down to \(\sim \qty{5000}{K}\),
which is the value we infer from our line-profile analysis (Section~\ref{sec:estim-nebula-kinet}).
Higher values of \(Z_0\) than this produce even lower values
of the \Ha{} temperature, \(\lesssim \qty{1000}{K}\) for \(Z_0 \ge 60\),
which is in conflict with our measurements.

Thus, no single model of radial metallicity fluctuations can simultaneously reproduce all the observations. 

\subsection{Model results: mixture of two models}
\label{sec:model-results-admix}

\newcommand\Za{\ensuremath{Z_{\mathrm{0,A}}}}
\newcommand\Zb{\ensuremath{Z_{\mathrm{0,B}}}}
It would be more realistic to consider 3-dimensional models with both radial
and non-radial variations in metallicity.
Such models are beyond the scope of this paper, but we can approximately simulate the
effect by combining two models with different metallicity contrasts.
We suppose that a covering fraction \(f\) of the nebular shell has radial metallicity
fluctuations with amplitude \(\Za\), while the remaining covering fraction \(1 - f\)
has fluctuations with amplitude \(\Zb\).
The flux of any emission line is then given by a linear combination
of the flux from the two models: \(F = f F\Sub{A} + (1 - f) F\Sub{B}\).

We first tried \(\Zb = 1\) (homogeneous metallicity),
but all combinations of \Za{} and \(f\) yield line ratios
\(\oiiilam / \Hb\) and \(\oiii \Wav{4363} / \Wav{5007}\)
that are at least two times too large.
On the other hand, with \(\Zb = 6\), we find an acceptable fit to all
the line ratios with \(f = \num{0.60(5)}\) and \(\Za \approx 120\)
(see Fig.~\ref{fig:cloudy-line-ratios-vs-observed-admixture}).

\section{Discussion}
\label{sec:discussion}

In this section we place our results in the broader context of previous studies of Ou~5
and of high-abundance-discrepancy nebulae in general.
Table~\ref{tab:param-comparison} summarizes the principal geometrical,
kinematic, and physical parameters derived in this paper
and compares them with those from earlier works.
Here we highlight the implications of our new kinematic analysis for the dynamical history of the nebula,
examine the evidence for a two-phase temperature structure and its relation to the abundance discrepancy,
assess the successes and limitations of the Cloudy photoionization models,
and finally consider Ou~5 in the context of other post-common-envelope and high-ADF planetary nebulae.

\subsection{Implications of our kinematic results}
\label{sec:impl-our-kinem}
The kinematic analysis of Section~\ref{sec:kinematics} shows that all of the principal components of the nebula
(inner shell, outer shell, outermost shell, and polar knots) share a common kinematic age of
\(\sim \qty{1e4}{yr}\) (Table~\ref{tab:shells}).
This might seem to argue against a sequence of distinct ejection events,
since the factor of three difference in radius between components is not accompanied by a difference in age.
However, the nebula is rather old, so we cannot rule out a sequence of ejections
over a period of \(\lesssim \qty{1e3}{yr}\).
Indeed, we find weak evidence
that the high-latitude nebular components
(outer shell and polar knots) are slightly older kinematically (\qty{1.2(2)e4}{yr})
than the low-latitude inner and outermost shells (\qty{8.0(5)e3}{yr}).

Morphologically, the nebula bears considerable resemblance to
nested bipolar nebulae around symbiotic stars,
such as Ced~211 around R~Aquarii \citep{Solf:1985a, Henney:1992a}
and the Southern Crab (PN~G315.4+09.4) around Henize~2-104 \citep{Schwarz:1989a, Corradi:2001a}.
The Southern Crab has a kinematic age of \qty{4000(1000)}{yr}
and the expansion velocities are similar to Ou~5 at the equator (\qtyrange{10}{15}{\kms})
but much larger at the poles (\qtyrange{50}{250}{\kms}).
As with Ou~5 the expansion follows a Hubble-type law (\(V \propto r\))
such that the kinematic age of all nebular components is approximately equal
\citep{Corradi:2001a, Santander-Garcia:2008a}.
R~Aquarii is much younger, with the outer nebula, inner bipolar nebula,
and helical filament having kinematic ages of
\qty{450(25)}{yr}, \qty{270(20)}{yr}, and \qty{285(20)}{yr}, respectively  \citep{Santamaria:2024a}.
This is suggestive of two different outburst events, separated
by \qty{170(25)}{yr},
but such a small time difference will be very difficult to perceive once the nebula is \qty{10000}{yr} old.
This provides a natural explanation for the globally Hubble-like velocity laws seen in older nebulae.

Ou~5 is not a symbiotic system (white dwarf accreting from a cool giant),
but it has passed through a common-envelope stage (see below) and the shell ejections
may be related to enhanced mass transfer to the low-mass main-sequence secondary
during or just before this stage
\citep{Chamandy:2018a, Khouri:2021a, Lopez-Camara:2022a}.
Similar nested hourglass shells are seen in some young planetary nebulae such as
Hb~12 \citep{Clark:2014d, Hsia:2021a}, which may be a closer analogue to Ou~5 since
it is also possibly hosts a short-period post-common-envelope binary \citep{Hsia:2006a}\footnote{
  Although the evidence for eclipses in this system has been questioned  \citep[App.~A.3.2 of][]{De-Marco:2008a, Louie:2010a}.
}

If this scenario of closely spaced multiple outbursts is true, one might ask why the outer shells (which were ejected first)
are always expanding faster than the inner shells (which were ejected slightly later).
But this can be simply explained as a selection effect: any shell that is ejected at a faster velocity
than the shells from earlier outbursts will eventually catch up with them and merge.
Therefore, in the long run, the only shells that survive to be observed as
evolved nested bipolar nebulae are those with the ``correct'' ordering of expansion velocities.

Our measured systemic velocity of the nebula, \(\vsys = \qty{-33}{\kms}\),
is offset from the center-of-mass velocity of the binary, \(\qty{-26}{\kms}\) \citep{Jones:2022a}.
Although the difference is only \(\sim \qty{7}{\kms}\), it is much larger than the  velocity precision of our spectroscopy
and may indicate the imprint of asymmetric mass loss during envelope ejection.
The surface brightness of the nebula is also slightly lop-sided,
being 10 to 20\% brighter on the east side (see Fig.~\ref{fig:image-cuts-equator}),
which supports this interpretation.

Similarly, we find a small but statistically significant difference
between the inclination of the nebular axis, \(i = \qty{85 \pm 2}{\degree}\) (Sec.~\ref{sec:line-sight-incl}),
and the binary orbital inclination that is inferred from eclipse fitting,
\(\qty{82 \pm 1}{\degree}\) \citep{Jones:2022a}.
This is hard to explain, especially since we find consistent inclinations
for both the inner and outermost shells.
Given the small mass and low expansion velocity of the nebula,
it is not possible for an asymmetric outburst to have caused such a shift in a close binary orbit.\footnote{%
  The recoil velocity of a star of mass \(M_\star = \qty{0.6}{\msun}\) due to
  the ejection of a nebula of mass \(M\Sub{neb} = \qty{0.1}{\msun}\)
  at a speed of \(\vexp = \qty{30}{\kms}\) with an asymmetry of \(f\Sub{asym} = 0.2\)
  is \(\delta V = f\Sub{asym} \vexp M_\mathrm{neb} / M_\star = \qty{1}{\kms}\),
  whereas the relative velocity of the binary is
  \(V\Sub{orb} \ge \qty{260}{\kms}\) \citep{Jones:2022a}.
  The resultant tilt of the binary orbit is thus \(\delta i \le (\delta V / V\Sub{orb}) \, \unit{\radian} \le \ang{0.2}\).
}
The most likely explanation for the \ang{3} difference
is that the uncertainty of
at least one of the measurements has been underestimated.

\subsection{Evolutionary status of the central star}
\label{sec:evol-stat-centr-1}
A robust result of our nebular analysis (section~\ref{sec:temp-lumin-centr})
is that the ionizing central star must have an effective temperature \(\gtrsim \qty{1e5}{K}\).\footnote{%
  \(T_\star = \qty{103(4:3)}{\kilo\kelvin}\)  assuming a black-body spectrum,
  or \(T_\star \approx \qty{115}{\kilo\kelvin}\) assuming a model atmosphere from \citet{Rauch:2003a}.  
}
This contradicts the finding of a significantly cooler temperature (\qty{67(5)}{K}) from
analysis of the binary light curve and radial velocity variations \citep{Jones:2022a}.
Together with a low derived luminosity (Table~\ref{tab:param-comparison}),
this led those authors to propose that the primary is a post--red-giant-branch (RGB)
star with a \chem{He} core,
rather than a post--asymptotic-giant-branch (AGB) star with a \chem{C/O} core,
as is more usual in planetary nebulae.
However, we believe that our higher temperature determination from
the \heii{} Zanstra method is much more reliable,
since it is (1)~dependent on the high-energy tail of the stellar SED,
which is a steep function of \(T_\star\),
and (2)~based on a photon counting argument, and thus insensitive
to assumptions about the nebular density or \chem{He} abundance.

For these reasons,
we strongly disfavor the post-RGB hypothesis since models of post-RGB evolution
\citep{Hall:2013a, Istrate:2016a} show that the maximum effective temperature is always \(< \qty{1e5}{K}\).
The dynamical stellar mass \(M_\star = \qty{0.50(6)}{\msun}\) derived by \citet{Jones:2022a}
is very close to the threshold for core \chem{He} burning
of \qtyrange{0.48}{0.50}{\msun} \citep[e.g.,][p.~148]{Salaris:2005a},
so is consistent with either hypothesis.
We therefore proceed to analyze the central star in the light of
post-AGB evolution, for which
we employ interpolated tracks from the MESA Isochrones and Stellar Tracks (MIST)
project \citep{Dotter:2016a, Choi:2016a},
which are calculated using the stellar evolution code
Modules for Experiments in Stellar Astrophysics
\citep[MESA:][]{Paxton:2011a, Jermyn:2023a}.

\label{sec:evol-stat-centr}
\begin{figure}
  \centering
  \includegraphics[width=\linewidth]{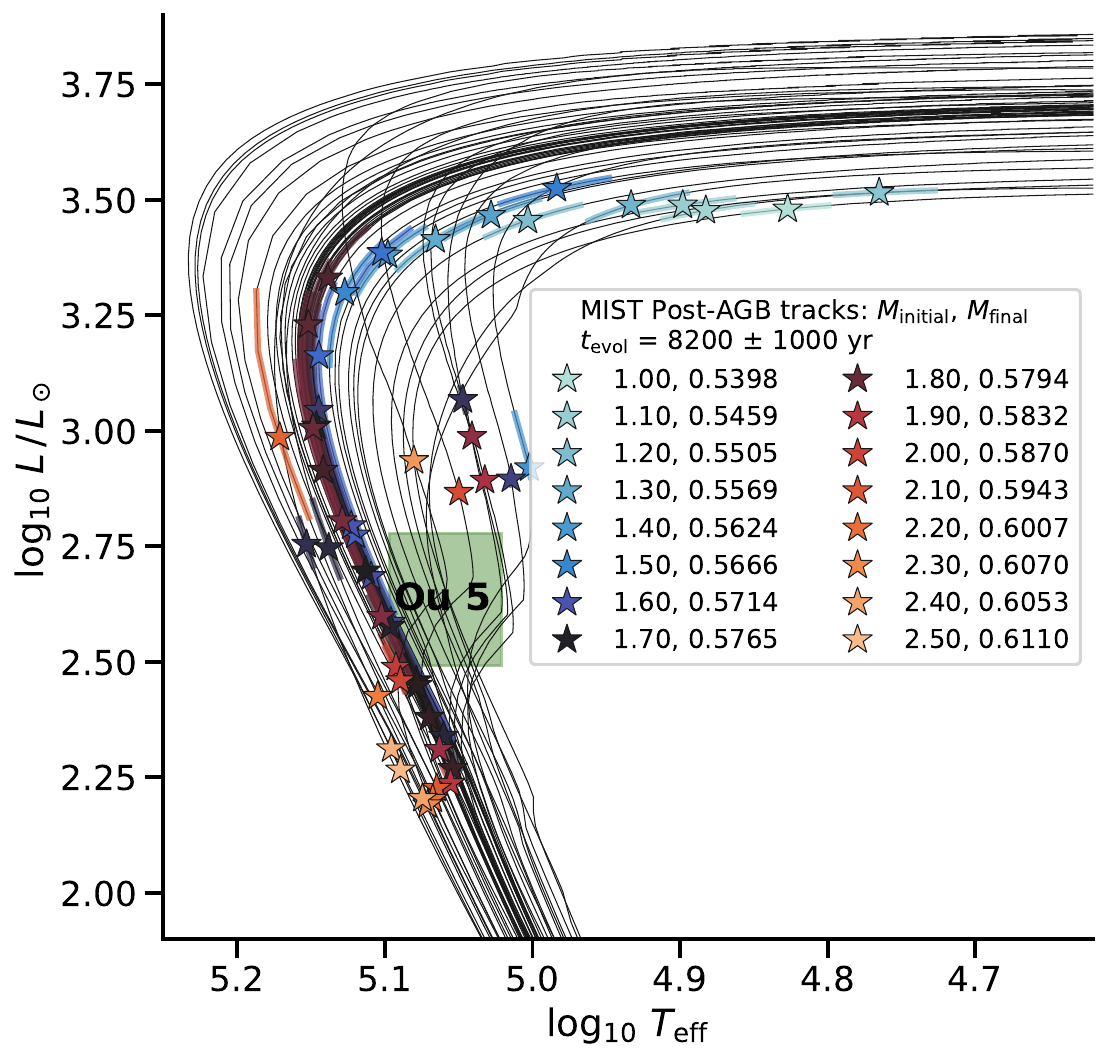}
  \caption{Tracks on the Hertsprung--Russell diagram for
    Post-AGB single-star evolutionary models from the MIST project, using a fine
    grid of initial (main-sequence) masses, \(M_\mathrm{initial}\)
    from \qtyrange{1.0}{2.5}{\msun}.
    Colored stars mark the point on the evolutionary track
    with a post-AGB age of \qty{8200}{yr}, corresponding to the
    kinematic age that we derive for the nebula, with heavy colored lines
    marking a range of \qty{\pm 1000}{yr}.
    The green rectangle shows the derived luminosity and effective temperature
    of the central star of Ou~5.
    The models that are consistent with the age and position of Ou~5 on this diagram
    all lie in the initial mass range \(M_\mathrm{initial} = \qtyrange{1.6}{2.0}{\msun}\),
    corresponding to final (white dwarf) masses of
    \(M_\mathrm{final} = \qtyrange{0.57}{0.59}{\msun}\).
  }
  \label{fig:post-agb}
\end{figure}

Results are shown on the Hertsprung--Russell diagram
in Fig.~\ref{fig:post-agb}, where we use the nebular
age that we derive from kinematic-geometric analysis (section~\ref{sec:nebular-geometry}) to fix the
post-AGB evolutionary age of the central star,\footnote{
  We follow \citet{Miller-Bertolami:2016a}
  by taking the zero-point for the post-AGB evolution as the time
  when the effective temperature first exceeds \qty{1e4}{K}.
  Varying this threshold by a factor of two in either direction makes no
  noticeable difference to the results.
}
finding the best agreement for a final mass\footnote{%
  For single stars, there is a well-defined relation between the
  initial and final mass \citep[e.g.,][]{Cunningham:2024a},
  implying an initial mass of \qty{1.8(2)}{\msun} for the Ou~5 primary,
  but binary evolution may modify this.
}
of \(M_\star = \qty{0.58(1)}{\msun}\)
(dark blue/black symbols in Fig.~\ref{fig:post-agb}),
which is higher than the binary-derived mass by \(1.3\,\sigma\).
It could be argued that the uncertainty in our result is underestimated, since it does
not include the uncertainty in the distance to the nebula: \(D = \qty{4(1)}{kpc}\).
Changes in \(D\) affect both the stellar luminosity
(\(\sim D^2\)) and the nebular kinematic age (\(\sim D\)).
However, it would be necessary to increase the distance to \(D \ge \qty{6}{kpc}\)
in order to bring our stellar-evolution-derived mass
into 1-\(\sigma\) agreement (\(M_\star < \qty{0.56}{\msun}\)) with the \citeauthor{Jones:2022a} value.
\startNEW
On the other hand, we note that the posterior corner plots shown in Fig.~5 of \citet{Jones:2022a}
show a positive correlation between the inferred primary mass and distance,
such that increasing the distance from \(D = \qty{3}{kpc}\)
to \qty{4}{kpc} would yield a primary mass consistent with our determination.
At the same time, the inferred secondary mass and radius would increase
by 40\% and 7\%, respectively, to \(M_2 \approx \qty{0.32}{\msun} \), \(R_2 \approx \qty{0.60}{\rsun} \).
In addition, our Zanstra solution for the luminosity and effective temperature of the primary star
implies a radius for the primary that is almost identical to the value derived by \citeauthor{Jones:2022a}
(see Table~\ref{tab:param-comparison}).
We have used the modeling tool of \citet[see also \citealt{Subebekova:2020a}]{Zharikov:2013a}
to investigate the feasibility of reproducing the observed light curve with the parameters derived in this paper
(\(D = \qty{4}{kpc}\); \(M_1 = \qty{0.58}{\msun}\); \(T_1 = \qty{115}{kK}\)).
We find an acceptable fit with a highly distorted illuminated secondary that nearly fills its Roche lobe:
\(M_2 = \qty{0.33}{\msun}\); \(R_2 = \qtyrange{0.62}{0.84}{\rsun}\).
This solution has a higher secondary temperature (\(T_2 = \qty{5250}{K}\))
and a smaller inclination (\(i = \ang{78}\)) than the \citet{Jones:2022a} solution,
which increases the tension with the observed colors during primary eclipse and our determination of the nebular axis
(Table~\ref{tab:param-comparison} and section~\ref{sec:line-sight-incl}).
Further investigation is required in order to construct a model that is consistent with all observations of the Ou~5 system.\footnote{%
  In the case of the very similar system UU~Sge \citep[central star of the PN Abell~63,][]{Afsar:2008a},
  the derived primary temperature is significantly higher if limb brightening is assumed
  for the secondary instead of limb darkening \citep{Bell:1994a}.
  For Ou~5, both \citet{Jones:2022a} the \citet{Zharikov:2013a} model consider only limb darkening,
  although limb brightening
  may be more appropriate given the strong irradiation of the secondary by the primary
  (but see also \citealp{Wawrzyn:2009a} for further discussion of the complexities of such a system).}
\stopNEW
 
\begin{table*}
\centering
\caption{Key parameters of Ou~5 compared with prior work}
\label{tab:param-comparison}
\begin{tabular}{
  l
  c
  R
  l
  l
  R
  l
  l
}
\toprule
{Parameter} & {Units} &
\text{This paper} &
{Method} &
{Section} &
\text{Prior studies} &
{Source\textsuperscript{a}} &
{Prior study method}\\
\midrule
Distance \(D\) & \unit{kpc} & \num{4.0(7)} & Statistical (adopted) & \ref{sec:dist-determ} & \num{3.1(3)} & J22 & Eclipsing-binary modeling \\
\addlinespace[2pt]
\multicolumn{8}{l}{\textit{Kinematics}}\\
Systemic velocity (nebula) \(V_{\mathrm{sys}}\) & \unit{\kilo\metre\per\second} & \num{-33(1)} & Flux-weighted mean& \ref{sec:kinematics} & {} &  &  \\
Systemic velocity (binary) \(V_{\mathrm{com}}\) & \unit{\kilo\metre\per\second} & {} &  &  & \num{-26(1)} & J22 & Eclipsing-binary modeling \\
  \addlinespace[2pt]
  Expansion velocity \(V\expand\) & \unit{\kms} & \text{14\textsuperscript{b}\ to\ 35\textsuperscript{c}} & PV analysis & \ref{sec:kinematics} & & & \\
Kinematic age & \qty{1000}{yr} & \num{8.2(4)} & PV analysis & \ref{sec:nebular-geometry} & {} &  &  \\
\addlinespace[2pt]
\multicolumn{8}{l}{\textit{Geometry}}\\
Axis inclination (nebula) \(i\) & \unit{\degree} & \num{85(2)} & PV shear & \ref{sec:line-sight-incl} & {} &  &  \\
Axis inclination (binary) \(i_{\star}\) & \unit{\degree} & {} &  &  & \num{82(1)} & J22 & Eclipsing-binary modeling \\
\addlinespace[2pt]
\multicolumn{8}{l}{\textit{Primary central star}}\\
Effective temperature \(T_{\mathrm{eff}}\) & \unit{\kilo\kelvin} & \numrange{100}{115} & \heii{} Zanstra & \ref{sec:temp-lumin-centr} & \num{67(5)} & J22 & Eclipsing-binary modeling \\
Luminosity \(L_\star\) & \unit{\lsun} & \numrange{300}{660} & \heii{} Zanstra & \ref{sec:temp-lumin-centr} & \num{214(70)}\textsuperscript{d} & J22 & Eclipsing-binary modeling \\
Mass \(M_\star\) & \unit{\msun} & \num{0.58(1)} & Post-AGB tracks & \ref{sec:evol-stat-centr} & \num{0.50(6)} & J22 & Eclipsing-binary modeling \\
Radius \(R_\star\) & \unit{\rsun} & \num{0.075(6)} & Stefan-Boltzmann & \ref{sec:evol-stat-centr} & \num{0.078(6)} & J22 & Eclipsing-binary modeling \\
\addlinespace[2pt]
\multicolumn{8}{l}{\textit{Nebular physical}}\\
Inner-shell \(n_{\mathrm e}\) & \unit{\per\centi\metre\cubed} & \num{150(10)} & \Ha{} EM & \ref{sec:phys-param-nest} & \num{560(550:300)} & C15 & [\ion{S}{ii}], [\ion{O}{ii}], [\ion{Ar}{iv}] ratios \\
\(\langle T (\Ha)\rangle\) & \qty{e4}{\kelvin} & \numrange{0.35}{0.6} & Line broadening & \ref{sec:estim-nebula-kinet} & {} &  &  \\
\addlinespace[2pt]
Warm-phase \(T\) & \qty{e4}{\kelvin} & \numrange{1}{2} & Cloudy models & \ref{sec:model-results-phys} & \num{1.01(3)} & C15 & \oiii{} line ratio \\
Cool-phase \(T\) & \qty{e4}{\kelvin} & \numrange{0.05}{0.2} & Cloudy models & \ref{sec:model-results-phys} & {} &  &  \\
\addlinespace[2pt]
Metallicity boost \(\langle Z \rangle\)\textsuperscript{e} &  & \num{45(10)} & Cloudy models & \ref{sec:model-results-admix} & \num{56(12)} & C15 & \chem{O^{+2}} ADF \\
\addlinespace[2pt]
Total ionized mass \(M_\mathrm{i}\) & \unit{\msun} & \num{0.069(9)} & Geometrical model & \ref{sec:phys-param-nest} & \numrange{0.005}{0.2} & C15 & \Ha{} flux \\
Inner shell ionized mass \(M_\mathrm{i}\) & \unit{\msun} & \numrange{0.03}{0.07} & Cloudy models & \ref{sec:cloudy-phot-models} &  &  & \\
\bottomrule
\end{tabular}
\begin{flushleft}
  \textsuperscript{a}~J22 = \citet{Jones:2022a}; 
  C15 = \citet{Corradi:2015a}.\\
  \textsuperscript{b}~Inner shell.\\
  \textsuperscript{c}~Outermost shell and polar knots.\\
  \textsuperscript{d}~Adjusted to \(D = \qty{4}{kpc}\).\\
  \textsuperscript{e}~\Ha-weighted mean metallicity with respect to baseline value (see eq.~\ref{eq:Z-fluctuation}).
\end{flushleft}
\end{table*}

\subsection{Two-phase temperature structure and the abundance discrepancy}
\label{sec:two-phase-temp}
Our measurements of thermal line broadening (Sec.~\ref{sec:estim-nebula-kinet})
imply a mean \Ha{} temperature of \(T \simeq \qtyrange{3500}{6000}{\kelvin}\),
significantly below the \(\sim \qty{1e4}{\kelvin}\) implied by collisionally excited line (CEL) diagnostics
\citep[cf.][]{Corradi:2015a}.
This robustly confirms that internal temperature fluctuations must be at least
a part of the explanation for the observed abundance discrepancy in this object. 
In a two-phase framework,
the fraction of \Ha{} emission arising in the cool phase in Ou~5 must be \(\omega > 0.4\),
which is larger than the typical values \(\omega \lesssim 0.1\) derived from other
nebulae with more moderate abundance discrepancies
\citep{Richer:2022a, Garcia-Rojas:2022a, Gomez-Llanos:2024a}.
However, it is comparable to the emission fraction of \(\omega = \numrange{0.2}{0.7}\)
estimated for Hf~2-2 \citep{Garcia-Rojas:2022a, Castaneda-Carlos:2025a},
which is another extreme-ADF nebula.

Spatially, the cool phase seems more centrally concentrated.
The NOT narrow-band images (Sec.~\ref{sec:morph-feat})
show that recombination lines (e.g., \heiilam, \Ha{})
are relatively brighter in the inner ring,
while CELs such as \oiiilam{} show limb-brightening in the lobes.
This agrees with the slit spectroscopy of \citet{Corradi:2015a},
whose ORL spatial profiles peak more centrally than the CELs.
Such differences argue for radial gradients in the two-phase parameters:
either the cool fraction \(\omega\), the cool-phase temperature \(T\cool\),
or both, must vary with radius. There is now a large body of evidence
that metallicity enhancements tend to be concentrated towards the inner edge of
planetary nebulae, both in high-ADF and moderate-ADF objects
\citep{Barker:1982a, Garnett:2001a, Tsamis:2008a, Garcia-Rojas:2016b, Richer:2017a}.

\subsection{Constraints from Cloudy modeling}
\label{sec:constr-from-cloudy}
The photoionization models of Section~\ref{sec:cloudy-phot-models} all
reproduce the global \Ha{} flux
and \heiilam/\Hblam{} ratio,
regardless of whether metallicity fluctuations are included,
but chemically homogeneous models are incapable of reproducing
either the forbidden or permitted metal lines.
Models with sinusoidal abundance fluctuations can fit the
observations better, and by combining moderate fluctuations (\(Z_0 \simeq 6\))
with more extreme fluctuations (\(Z_0 \simeq 120\)),
it is possible to simultaneously fit a broad range of the
observed line ratios, together with the inferred mean temperature.
The derived fraction of the nebular shell containing extreme fluctuations,
\(f = \num{0.60(5)}\) is of the same order as the cool-phase fraction \(\omega\)
derived from the two-phase analysis above.
We have not attempted to directly fit the observed spatial distribution
of different emission lines, which would require more sophisticated modeling
such as in \citet{Ercolano:2003a, Gomez-Llanos:2020a},
but the general trends are qualitatively reproduced by our
large-scale fluctuation models, while the small-scale fluctuation models
predict that the optical recombination line emission is biased towards
the outside of the shell, contrary to the observations.

The ionized mass derived from Cloudy ranges from 
\(M\Sub{i} \simeq \qty{0.10}{\msun}\) for the constant-metallicity
models to \qty{0.04}{\msun} for the most extreme metallicity fluctuations.
These are in reasonable agreement with the values derived from \Ha{} emission measures using the mean temperature (Sec.~\ref{sec:phys-param-nest}).
This reinforces the robustness of our simple geometrical estimates,
but note that the Cloudy model includes the inner shell only, whereas
we find (Table~\ref{tab:shells}) that this represents only 40\% of the total ionized
mass of the nebula.
We therefore find that the total ionized mass is at least \qty{0.10}{\msun},
which is very close to the median value for all planetary nebulae
\citep[e.g., Fig.~6 of][]{Stanghellini:2020a},
contradicting the suggestion of \citet{Corradi:2014a} that Ou~5 might have a
particularly low nebular mass.

\subsection{Ou~5 in the context of other high-ADF planetary nebulae}
\label{sec:ou-5-context}
\newcommand\adf{\ensuremath{\mathrm{ADF}}}
The most extreme abundance discrepancies (\(\adf \gtrsim 100\)) are
seen in localized \chem{H}-deficient knots inside the planetary nebulae
Abell~30 \citep{Wesson:2003a}, Abell~58 \citep{Wesson:2008a},
and Abell~46 \citep{Corradi:2015a}.
All but the last of these are ``born-again'' nebulae,
which may be the result of a very late thermal pulse (VLTP) that mixes the outer layers of the envelope 
and ejects \chem{H}-deficient material inside a pre-existing nebula
\citep{Iben:1983a}. 
Other examples of this class are Abell~78 \citep{Fang:2014a, Montoro-Molina:2023a}
and WR~72 \citep{Gvaramadze:2020a},
which also contain \chem{H}-deficient knots,
but which do not have detections of the metal recombination lines necessary to formally determine an ADF. 

Ou~5, with its measured \(\adf = 56\),
is unambiguously an \emph{extreme-ADF} nebula,
yet it belongs to the group of about ten objects with
\(\adf \simeq \numrange{20}{80}\) that form a distinct tier below
the ultra-extreme cases (\(\adf \gtrsim 100\)).
The properties of this group were studied in detail by \citet{Wesson:2018a},
who found a clear association with binarity:
\num{9} out of the \num{14} nebulae with measured
\(\adf > 20\) host close-binary nuclei,\footnote{%
  Based on data from \url{https://nebulousresearch.org/adfs/},
  last updated 2024-01-09.}
suggesting that Ou~5, like its peers, was ejected during
a common-envelope (CE) phase \citep{Jones:2025a}.
The details of the physical processes that gave
rise to both the nebula ejection and the formation of the high-metallicity phase
are both unknown,
although the empirical correlations suggest these must be linked \citep{Wesson:2018a}.
One possibility is explosive CE ejection initiated by an in-spiraling secondary
that overflows its Roche lobe.
This can cause a stream of \chem{H}-rich material
to penetrate deep into the AGB core, where it triggers the explosive destruction
of the \chem{He}-burning shell,
ejecting the envelope together with some \chem{He}-burning products.
This scenario has mainly been explored in the context of more massive binaries
\citep{Ivanova:2002a}, but we speculate that a similar process
\citep[section~3.3.4 of][]{Ivanova:2013a}
may give rise to high-ADF post-CE nebulae such as Ou~5.
However, in order to explain the observed ionized mass in the Ou~5 nebula
(\(\approx \qty{0.1}{\msun}\), sec.~\ref{sec:constr-from-cloudy}),
it seems necessary that a large part of the AGB envelope had already been lost prior to this event
(see \citealp{Santander-Garcia:2022a} for the general problem of
``missing mass'' in planetary nebulae).

An alternative mechanism for producing metallicity \NEW{inhomogeneities}
is the evaporation of solid bodies such as comets or planets
\citep{Liu:2003a, Liu:2006b}.
However, quantitative calculations \citep{Henney:2010a} show that
evaporation during the planetary nebula phase itself is
incapable of generating sufficient quantity of gas-phase metals.
If the evaporation began during the previous AGB phase,
then the mechanism becomes more feasible,
but that cannot explain the concentration of high-metallicity
towards the inner regions of the nebula, as observed in Ou~5.

\section{Conclusions}
\label{sec:conclusions}

We have presented new imaging and high-resolution spectroscopy of the
extreme-abundance-discrepancy planetary nebula Ou~5, together with photoionization
models that explore its unusual thermal and chemical structure.
Our principal conclusions are as follows:

\begin{enumerate}[1.]
\item The nebula has a complex, nested bipolar morphology consisting of
  inner and outer shells, faint outermost lobes, and polar knots.
  Despite their differing scales, all components share a common
  dynamical age of order \(\qty{1e4}{yr}\),
  implying that if they were ejected in discrete events,
  this must have occurred over only \(\sim \qty{1e3}{yr}\),
  much shorter than their present age.
\item Thermal line broadening implies a mean hydrogen temperature
  \(T \lesssim \qty{6000}{\kelvin}\) for the entire nebula
  and \(T \approx \qty{3500}{K}\) for the best-studied region,
  much lower than the \(\sim \qty{1e4}{\kelvin}\) derived from diagnostic line ratios of collisionally excited lines (CELs).
  This provides strong, independent evidence for the coexistence of
  multiple temperature/metallicity phases within the ionized gas,
  as previously proposed to explain the extreme abundance discrepancies
  observed in this nebula.
\item Simple \textsc{cloudy} models with sinusoidally varying metallicity
  are very successful at reproducing the observed emission line spectrum.
  In order to simultaneously fit the CEL and ORL diagnostics,
  a mixture of fluctuations with large amplitude
  (metallicity contrast \(Z_0 \approx 120\)) and small amplitude
  (\(Z_0 \approx 6\)) is required.
\item The relatively bright \heii{} emission from the nebula requires a higher
  stellar temperature and luminosity than previously claimed, which rules out
  the possibility that Ou~5 is a low-mass post-RGB object.
  Instead, the primary central star is consistent with post-AGB evolution
  toward a CO white dwarf of mass \(\approx \qty{0.58}{\msun}\).
\end{enumerate}
Ou~5 strengthens the emerging link between high abundance
discrepancies and close-binary central stars,
since most nebulae with \(\adf > 20\) host post-CE binaries.
It thus provides an important test case for understanding how the
common-envelope interaction shapes planetary nebulae and for probing
the physical origin of the abundance discrepancy problem.

\section*{Acknowledgements}
We gratefully acknowledge financial support provided by
\foreignlanguage{spanish}{%
  Dirección General de Asuntos del Personal Académico,
  Universidad Nacional Autónoma de México},
through grants
``\foreignlanguage{spanish}{%
  Programa de Apoyo a Proyectos de Investigación
  e Inovación Tecnológica IG101223, IN111124,  IN109823 and AG101223}''.
FPN gratefully acknowledges the generous and invaluable support of the Klaus Tschira Foundation.
We thank the daytime and night support staff at the OAN-SPM for facilitating and helping obtain our observations.  
We thank Felipe Montalvo, Salvador Monroy and Lester Fox for
assistance with observations at OAN-SPM. 
Some of the data analyzed here were obtained with ALFOSC, which is provided by the Instituto de Astrofisica de Andalucia (IAA) under a joint agreement with the University of Copenhagen and NOT.\@ We are grateful to
Sergio Armas Pérez for assistance in accessing the data archive.
We thank Jane Arthur, José Eduardo Méndez-Delgado and Christophe Morisset for useful discussions.
\NEW{We also thank the anonymous referee for a helpful report, which led to improvements in the presentation of our results.}


\section*{Data Availability}
\label{sec:data-avail}
All observational data, analysis and modelling scripts, and Jupyter notebooks 
used in this paper, together with extensive documentation and notes, are 
currently available from the GitHub repository 
\url{https://github.com/will-henney/teresa-pn-ou5}. 
A permanent Zenodo archive with DOI will be provided upon publication.



\bibliographystyle{mnras}
\bibliography{ou5-references} 

\bsp

\label{lastpage}

\end{document}